\documentclass{SciPost}

\hypersetup{
    colorlinks,
    linkcolor={red!50!black},
    citecolor={blue!50!black},
    urlcolor={blue!80!black}
}

\usepackage{comment}
\usepackage{physics}
\usepackage[bitstream-charter]{mathdesign}
\usepackage{mathtools} % Provides \mathllap, \mathrlap, \mathclap for safe math overlays.
\usepackage{graphicx}
\usepackage{subcaption}
\usepackage{stackengine}
\usepackage[utf8]{inputenc}
\usepackage{xcolor}

\stackMath

\DeclareSymbolFont{usualmathcal}{OMS}{cmsy}{m}{n}
\DeclareSymbolFontAlphabet{\mathcal}{usualmathcal}

\fancypagestyle{SPstyle}{
\fancyhf{}
\lhead{\colorbox{scipostblue}{\bf \color{white} ~SciPost Physics }}
\rhead{{\bf \color{scipostdeepblue} ~Submission }}

\fancyfoot[C]{\textit{\thepage}}
}

\begin{document}

\pagestyle{empty}

\begin{center}{\Large \textit{\color{scipostdeepblue}{
%%%%%%%%%% TODO: Write your article's title here
Eigenstate chaos in the presence of non-Abelian symmetries
}}}\end{center}

\begin{center}\textit{
% Use (full) first name (+ middle name initials) + surname format.
% Separate subsequent authors by a comma, omit comma and use "and" for the last author.
% Mark the corresponding author(s) with a superscript symbol in this order
% \star, \dagger, \ddagger, \circ, \S, \P, \parallel, ...
Siddharth Jindal\textsuperscript{1$\star$} and
Pavan Hosur\textsuperscript{1}
}\end{center}

\begin{center}
{\bf 1} Department of Physics and Texas Center for Superconductivity, University of Houston
%%%%%%%%%% END TODO: AFFILIATIONS
% Provide email address of corresponding author(s)
\\[\baselineskip]
$\star$ \href{mailto:siddharth@sidjindal.dev}{\small siddharth@sidjindal.dev}
\end{center}

\section*{\color{scipostdeepblue}{Abstract}}
\textit{\boldmath{
The eigenstate thermalization hypothesis (ETH) posits that energy eigenstates encode local properties of the microcanonical ensemble. Motivated by recent interest in the physics of non-commuting conserved charges and the non-Abelian ETH, we study chaotic eigenstates in the presence of symmetries described by general compact Lie groups, such as SU(2). By applying non-Abelian symmetry resolution, we develop a \textit{non-Abelian microcanonical entropy} and relate this entropy to the entanglement entropy of chaotic eigenstates. We find that microcanonical entropy is closely related to the symmetry-resolved entanglement entropy, which differs from conventional entanglement entropy by a universal logarithmic correction. Our results depend on the global Casimir charge, e.g. total spin. At finite charge density, we find a logarithmic enhancement to conventional entanglement entropy. At zero density, we find no such correction to entanglement entropy, but a logarithmic reduction to microcanonical entropy and symmetry-resolved entanglement entropy. We discuss the implications of our approach for non-Abelian eigenstate thermalization.}
}
\vspace{\baselineskip}

%%%%%%%%%% BLOCK: Copyright information
% This block will be filled during the proof stage, and finilized just before publication.
% It exists here only as a placeholder, and should not be modified by authors.
%\noindent\textcolor{white!90!black}{%
%\fbox{\parbox{0.975\linewidth}{%
%\textcolor{white!40!black}{\begin{tabular}{lr}%
%  \begin{minipage}{0.6\textwidth}%
%    {\small Copyright attribution to authors. \newline
%    This work is a submission to SciPost Physics. \newline
%    License information to appear upon publication. \newline
%    Publication information to appear upon publication.}
%  \end{minipage} & \begin{minipage}{0.4\textwidth}
%    {\small Received Date \newline Accepted Date \newline Published Date}%
%  \end{minipage}
%\end{tabular}}
%}}
%}
%%%%%%%%%% BLOCK: Copyright information

% For convenience during refereeing we turn on line numbers:
%\linenumbers

\vspace{10pt}
\noindent\rule{\textwidth}{1pt}
\tableofcontents
\noindent\rule{\textwidth}{1pt}
\vspace{10pt}

% Write your article contents here, starting from first \section.
% An example structure is given below.

\section{Introduction}
\label{sec:intro}
% TODO: write your article here.

A core postulate of quantum many-body chaos is that simple observables perceive high-energy eigenstates of chaotic Hamiltonians as random vectors up to constraints placed by symmetry. This heuristic, which has been recently referred to as the many-body Berry's conjecture (MBBC)~\cite{lu_renyi_2019, jindal_generalized_2024}, underlies the eigenstate thermalization hypothesis (ETH) and its corollary the ergodic bipartition (EB)\footnote{Ref.~\cite{goldstein_canonical_2006} contains a related idea known as canonical typicality.}~\cite{deutsch_quantum_1991, srednicki_chaos_1994, srednicki_approach_1999, deutsch_thermodynamic_2010}. The EB in particular is used to study entanglement in isolated systems~\cite{deutsch_thermodynamic_2010, lu_renyi_2019, murthy_structure_2019, shi_local_2023, kudler-flam_distinguishing_2021, jindal_generalized_2024}, which carries a close relationship to chaos and thermodynamics~\cite{popescu_entanglement_2006, deutsch_thermodynamic_2010}. Symmetries play a central role in physics and are known to impede some aspects of chaos and thermalization~\cite{kudler-flam_information_2022, majidy_noncommuting_2023}. Such impediments appear especially pronounced for non-Abelian symmetries~\cite{manzano_non-abelian_2022, majidy_noncommuting_2023, murthy_non-abelian_2023}. In the context of Abelian symmetries, the imposed constraint is merely a restriction of the randomness of eigenstates to charge-conservation sectors~\cite{murthy_structure_2019, belin_charged_2022}. 

Non-Abelian symmetries, central to myriad quantum many-body phenomena~\cite{wigner_group_1959, fradkin_field_2013, zee_group_2016}, entail some difficulties. For example, the Hamiltonian of a system symmetric under SU(2) must conserve all three angular momentum generators $\hat J^x,\,\hat J^y,\,\hat J^z$. Yet, energy eigenstates cannot be simultaneously eigenstates of all three charges as they do not commute. Energy eigenstates can generally be eigenstates of total spin $\hat J^2\equiv\hat J\cdot\hat J$, and given a $J$ quantum number, since this operator commutes with both the Hamiltonian and all three generators of the symmetry algebra, but $E$ and $J$ quantum numbers do not completely specify a state except in the $J=0$ sector. The crux of the difficulty is that $J$ sectors also carry multidimensional \textit{irreducible representations} or \textit{irreps} of the symmetry group. SU(2) representations are usually resolved with an eigenbasis of the $z$ angular momentum operator $\hat J^{z}$ labeled by the $M$ quantum number, but rotational invariance dictates that this is merely conventional. Thus, an application of the ETH to systems with non-Abelian symmetries must follow a correct formulation of non-Abelian symmetry resolution.

Recently, ref.~\cite{bianchi_non-abelian_2024} developed the theory of non-Abelian symmetry resolution for the purpose of computing a non-Abelian symmetry-resolved Page curve. Non-Abelian symmetry resolution contains, coarsely, two steps: (1) defining charge sectors and (2) tensor factoring an \textit{irreducible representation} or \textit{irrep} of the symmetry group. For SU(2), one decomposes the global Hilbert space into total spin $J$ sectors, and then factors those sectors into an irrep $\mathcal H^{(J)}_{\text{sym}}$ and an invariant multiplicity space $\mathcal H^{(J)}_{\text{inv}}$,
\begin{align}
\label{eq:introDecomp}
    \mathcal{H} = \bigoplus_{J}\left(\mathcal H^{(J)}_{\text{sym}}\otimes\mathcal H^{(J)}_{\text{inv}}\right).
\end{align}
We will build on their perspective to study quantum chaotic systems that carry non-Abelian symmetries. We argue that it is within narrow energy windows in $\mathcal H^{(J)}_{\text{inv}}$ that chaotic eigenstates can and do behave as random vectors. Separately, recent interest in the thermal properties of systems with non-Abelian symmetries has prompted the development of a non-Abelian ETH. For the special case of SU(2), ref.~\cite{murthy_non-abelian_2023} adapted the traditional statement of the ETH using the Wigner--Eckart theorem. Their result separates operator matrix elements into a kinematic part that is governed by the Clebsch--Gordan coefficients of SU(2) and a chaotic part that is governed by the ETH. As we discuss in appendix~\ref{app:background}, the Wigner--Eckart theorem and eq.~\eqref{eq:introDecomp} are both consequences of Schur's lemma. As such the approaches of both papers are fundamentally linked. 

Building on the ideas of refs.~\cite{bianchi_non-abelian_2024, murthy_non-abelian_2023} we develop a non-Abelian ergodic bipartition and use it to compute the Page curve of energy eigenstates for systems with non-Abelian symmetries. The basic conceptual purpose of this computation is to explicate a relationship between global microcanonical entropy and local entanglement entropy. The most general type of symmetry available to finite-dimensional quantum systems is compact Lie groups\footnote{The maximal group of unitary symmetries of a $d$-dimensional Hilbert space is the $d$-dimensional unitary group, $U(d)$, which is a compact Lie group and any subgroup of a compact Lie group is also a compact Lie group~\cite{koch_compact_nodate}.} and for pedagogical purposes, we split our treatment of compact Lie groups into three parts. In the main text, we establish the intuition using the case of SU(2). Then in appendices~\ref{app:WeylGeneral},~\ref{app:finiteDens},~\ref{app:zeroDens}, and~\ref{app:margDens} we generalize our results to the case of connected or at least semisimple, connected, compact Lie groups for which the core of our machinery applies without caveat. Then in appendix~\ref{app:generalCompact} we discuss the generalization to all compact Lie groups, which requires handling U(1) and finite subgroups. While our results generally hold for general compact Lie groups, the particular case of compact Lie groups with a non-central component subgroup, such as SU(3) with charge conjugation symmetry, allows a nuance that lies outside the scope of this work. After formulating the non-Abelian EB in section~\ref{sec:formulation}, we narrow our focus down to computing the thermodynamic and entanglement properties of local subsystems when the global system is prepared in an energy eigenstate in sections~\ref{sec:equi} and~\ref{sec:ent}. A tutorial of the relevant machinery of compact Lie groups is provided in appendix~\ref{app:background}.

Before stating our key results, we must discuss a core nuance of non-Abelian thermodynamics. In defining a spatially local subsystem, non-Abelian global symmetries provide us two choices: (I) all operators living within a spatial region or (II) only the symmetry-invariant operators, i.e. singlets, living within that region\footnote{For our purposes, to regard a set of operators as a subsystem, one merely needs algebraic closure~\cite{bianchi_non-abelian_2024}. In appendix~\ref{subapp:symm} we relate this algebraic notion of a subsystem to a more conventional Hilbert space notion but the operator notion is formally and conceptually cleaner.}. Definition (I) is more conventional and in a lattice system provides us with a simple tensor factorization of the Hilbert space. Definition (II) provides no such tensor factorization. However, there are some reasons one may choose to proceed with definition (II) over definition (I) anyway. First, the natural notion of subsystems associated withsubregions of spacetime in gauge theories and quantum gravity must be understood as a subsystem in the sense of definition (II) rather than definition (I)~\cite{donnelly_local_2016} essentially because they carry gauge symmetries. Second, we will see in the context of this paper that the thermodynamic consequences of non-Abelian symmetries even in non-relativistic systems, are best understood in a symmetry-resolved way in the sense of definition (II); we find that the concept is not optional. Lastly, both definitions of a subsystem carry respective notions of entanglement that have been used in the literature~\cite{patil_average_2023, bianchi_non-abelian_2024} and we will compute entanglement entropy for both cases.

\subsection{Summary of results}

The key results of our paper are as follows. In section~\ref{sec:formulation} we formulate the non-Abelian EB and provide numerical evidence for it with a Heisenberg spin chain. In section~\ref{sec:ent} we consider the thermodynamic consequences of restricting the randomness of eigenstates to the invariant multiplicity space and develop a non-Abelian microcanonical entropy, generalizing a result of~\cite{lasek_numerical_2024} from SU(2)\footnote{Our notion of a microcanonical subspace deviates in some respects from an existing notion of a non-Abelian microcanonical subspace~\cite{yunger_halpern_microcanonical_2016}. We will return to this point in the discussion.}. We find that non-Abelian microcanonical entropy carries a uniquely non-Abelian logarithmic reduction in vicinity of the zero charge sector. In section~\ref{sec:equi} we study the local thermodynamics for both definitions of a subsystem under a global eigenstate. For definition (II) we find that local symmetry invariant observables experience a Gibbs state that depends on the global energy and total spin but is blind to the polarization of the system along any axis. For definition (I), symmetry non-invariant observables, are sensitive to the global polarization of the system but experience it through raw Clebsch--Gordan coefficients, which do not approximate the exponential form of Gibbs state. This is one sense in which definition (II) exhibits more conventional thermal behavior of subsystems.

In section~\ref{sec:EE}, we compute the finite temperature Page curve of both notions of a subsystem. For definition (I) our Page curve is computed with the conventional definition of entanglement entropy. For definition (II) we compute symmetry-resolved entanglement entropy~\cite{bianchi_non-abelian_2024}. We find symmetry-resolved entropy to be closely related to the non-Abelian microcanonical entropy, while conventional entanglement entropy differs by a universal logarithmic enhancement. This enhancement was previously noticed in the comparison of both entropies for SU(2) systems at infinite temperature in refs.~\cite{bianchi_non-abelian_2024, patil_average_2023}. This enhancement was conjectured in both papers to be a universal feature of non-Abelian symmetry. We confirm this conjecture by computing this logarithmic correction for general compact Lie groups and find that it represents the intrinsic entanglement of kinematic degrees of freedom, encoded in the Clebsch--Gordan coefficients. We find no temperature dependence for the contribution. Interestingly, this contribution exactly cancels the aforementioned logarithmic reduction of the microcanonical entropy in the zero charge sector. For a general compact Lie group the contribution is $+\frac{d-r}{4}\ln N$, where $d$ is the dimension of the group, $r$ is its rank, and $N$ is the volume of the system. For SU(2) $d=3,\,r=1$ leaving a $+\frac{1}{2}\ln N$ contribution to entanglement entropy. 

Our results for entanglement entropy rely on a semiclassical analysis of recoupling coefficients which has been developed extensively for SU(2)~\cite{reinsch_asymptotics_1999, ponzano_semiclassical_1969} but to our knowledge has only been explored recently for higher symmetry groups~\cite{reshetikhin_semiclassical_2018, alekseev_tetrahedral_2025}. The results of appendices~\ref{app:finiteDens},~\ref{app:zeroDens}, and~\ref{app:margDens} then contain to our knowledge among the first extensions of the program started by Ponzano and Regge~\cite{ponzano_semiclassical_1969} to study the semiclassical limit of recoupling coefficients to arbitrary compact Lie groups. The techniques employed, particularly those of geometric quantization~\cite{alekseev_quantization_1988,reshetikhin_semiclassical_2018, alekseev_tetrahedral_2025}, may be particularly fruitful in improving and generalizing our results.

\section{Non-Abelian eigenstate thermalization and ergodic bipartition}
\label{sec:formulation}

In this section we state the core ansatze of the non-Abelian eigenstate thermalization and non-Abelian ergodic bipartition hypotheses. Our basic goal is to demonstrate that the constraints of the non-Abelian symmetry are encoded in full by a correct application of complete reducibility. We will focus on the simply reducible case of SU(2) for pedagogy, but all of our results extend to general compact Lie groups as discussed in the appendices.

A central tenet of quantum chaos is that high-energy eigenstates can be regarded as random vectors constrained by the symmetries of the system, a principle we refer to as the \textit{many-body Berry's conjecture} (MBBC). More precisely we assert that physical quantities, such as entropy or correlation functions, are invariant under random unitary\footnote{or orthogonal or symplectic} rotations of the Hamiltonian, if those rotations are restricted to a small enough microcanonical subspace~\cite{pappalardi_microcanonical_2023, richter_eigenstate_2020, foini_eigenstate_2019, foini_eigenstate_2019-1, wang_emergence_2023}. With a non-Abelian symmetry, the existence of non-commuting conserved charges, e.g. $\hat J^{x}$ and $\hat J^z$ for SU(2), implies one has to be careful in defining that microcanonical subspace. 

\subsection{Simple reducibility}
\label{subsec:SR}

Our setup begins with a finite-dimensional Hilbert space decomposed into spatial regions $A$ and $B$, $\mathcal{H} = \mathcal{H}_A\otimes\mathcal{H}_B$. On this Hilbert space acts a Hamiltonian $\hat H = \hat H_A + \hat H_B + \hat H_{AB}$, where $H_A$ acts on subsystem $A$, $\hat H_B$ acts on subsystem $B$, and $\hat H_{AB}$ connects the subsystems at their boundary. A compact, semisimple Lie group $G$ acts on the Hilbert space through a representation $U:G \rightarrow \mathcal{L}(\mathcal{H})$ that maps elements of $G$ to unitaries on $\mathcal H$. We assert that each term in the Hamiltonian is invariant under the action of $G$,
\begin{equation}
\label{eq:representation}
    g\in G \implies \hat U(g)\hat h\hat U(g)^{-1} = \hat h\text{,\quad   } \hat h=\hat H,\,\hat H_A,\,\hat H_B,\,\hat H_{AB}.
\end{equation}
We also require that $U$ acts independently on each subsystem, $\hat U(g)=\hat U^{(A)}(g)\otimes \hat U^{(B)}(g)$. 

A prototypical model we employ is the $N$-site next-neighbor Heisenberg chain~\cite{yunger_halpern_noncommuting_2020, majidy_noncommuting_2023, manzano_non-abelian_2022, lasek_numerical_2024, noh_kubo-martin-schwinger_2025},
\begin{equation}
\label{eq:heisenberg}
\begin{split}
    \hat H =  \hat H_A + \hat H_B + \hat H_{AB} &= -\sum_{r=1}^{N-2}{\hat \sigma}_r\cdot\left({\hat \sigma}_{r+1}+t{\hat \sigma}_{r+2}\right)- {\hat \sigma}_{N-1}\cdot{\hat \sigma}_{N} \\
    \hat H_A &= -\sum_{r=1}^{N_A-2}{\hat \sigma}_r\cdot\left({\hat \sigma}_{r+1}+t{\hat \sigma}_{r+2}\right) - {\hat \sigma}_{N_A-1}\cdot{\hat \sigma}_{N_A} \\
    \hat H_B &= -\sum_{r=N_A+1}^{N-2}{\hat \sigma}_r\cdot\left({\hat \sigma}_{r+1}+t{\hat \sigma}_{r+2}\right) - {\hat \sigma}_{N-1}\cdot{\hat \sigma}_{N} \\
    \hat H_{AB} &= - {\hat \sigma}_{N_A}\cdot{\hat \sigma}_{N_A+1} - t{\hat \sigma}_{N_A-1}\cdot{\hat \sigma}_{N_A+1} - t{\hat \sigma}_{N_A}\cdot{\hat \sigma}_{N_A+2} 
\end{split}
\end{equation}
with next-neighbor interaction $t$. $\hat H$ has been split into left and right halves $H_A$ and $H_B$ of sizes $N_A$ and $N_B=N-N_A$. $H,\,H_A,\,H_B$ are all invariant under a global SU(2) symmetry:
\begin{equation}
\label{eq:su2gen}
    {\hat J}\equiv{\hat J}_A+{\hat J}_B \equiv\frac{1}{2}\sum_{r=1}^{N_A}{\hat \sigma}_r+\frac{1}{2}\sum_{r=N_A+1}^{N}{\hat \sigma}_r,\quad\left[{\hat J},\hat H\right] = \left[{\hat J}_A,\hat H_A\right] = \left[{\hat J}_B,\hat H_B\right] = 0
\end{equation}
where $ \hat J = (\hat J^x,\hat J^y,\hat J^z)$ are the generators of the SU(2) algebra and their exponentiation is the SU(2) group.

The irreps of SU(2) carry an orthonormal basis $\ket{J,M}$ labeled by the eigenvalues of $\hat J^2\equiv \hat J\cdot \hat J$ and $\hat J^z$,
\begin{equation}
\label{eq:SakuraiMaterial}
\begin{split}
    \hat J^2\ket{J,M}\equiv J(J+1)\ket{J,M}\!,\,
    \, \hat J^{z}\ket{J,M}\equiv M\ket{J,M}.
\end{split}
\end{equation}
Eigenstates of $\hat H$ can be chosen such that
\begin{equation}
\label{SU2state}
    \ket{i}\equiv \ket{J_i,M_i,E_i}\! ,\,\quad  \hat H\ket{J_i,M_i,E_i}= E_i\ket{J_i,M_i,E_i}
\end{equation}
and similarly for subsystems $A$ and $B$,
\begin{equation}
\label{SU2states}
    \hat H_{A(B)}\ket{a(b)} \equiv H_{A(B)}\ket{J_{a(b)},M_{a(b)},E_{a(b)}} = E_{a(b)}\ket{J_{a(b)},M_{a(b)},E_{a(b)}}.
\end{equation}

As implied by eq.~\eqref{eq:introDecomp} and discussed in appendix~\ref{subapp:symm}, a state $\ket{JME}$ should be understood as a tensor product between a \textit{kinematic state} $\ket{JM}$ and an \textit{symmetry-invariant state} $\ket{JE}$,
\begin{align}
\label{eq:symmr}
    \ket{JME} \equiv \ket{JM}\otimes\ket{JE}.
\end{align}
In the context of eq.~\eqref{eq:introDecomp} $\ket{JM}$ and $\ket{JE}$ are respectively elements of the Hilbert spaces $\mathcal H^{(J)}_{\text{sym}}$ and $\mathcal H^{(J)}_{\text{inv}}$ which are tensor factors of the fixed-$J$ sector of $\mathcal H$. Our study will center around the overlap, 
\begin{equation}
\label{eq:overlap1}  
\begin{split}
    c^{i}_{ab} \equiv \bra{i}\ket{ab} &\equiv \bra{J_iM_iE_i}\ket{J_{a}M_{a}E_{a};J_{b}M_{b}E_{b}} \\
    &=\bra{J_iM_i}\ket{J_{a}M_{a};J_{b}M_{b}}\bra{J_iE_i}\ket{J_aE_{a};J_{b}E_{b}} \\
    &\equiv C^{i}_{ab}\tilde c^{i}_{ab}
\end{split}
\end{equation}
where the second line follows from the \textit{simple reducibility} of SU(2) and the third line introduces a shorthand we will use throughout this paper: $ C^i_{ab}=C^{J_iM_i}_{J_{a}M_{a};J_{b}M_{b}}\equiv\bra{J_iM_i}\ket{J_{a}M_{a};J_{b}M_{b}}
$ is the Clebsch--Gordan coefficient and $\tilde c^{i}_{ab} = \tilde c^{J_iE_i}_{J_{a}E_{a};J_{b}E_{b}}\equiv \bra{J_iE_i}\ket{J_aE_{a};J_{b}E_{b}}$ is the \textit{reduced overlap}. For both symbols we will use $\tilde c_{i}^{ab}\equiv\left(\tilde c^{i}_{ab}\right)^{*}$, $C_{i}^{ab}\equiv\left(C^{i}_{ab}\right)^{*}$. Simple reducibility and its generalization complete reducibility are the key property of compact Lie groups that enable a clean formulation of the EB and ETH in non-Abelian settings. We discuss it formally in appendix~\ref{subapp:comRed} but for the purpose of this section simple reducibility can be understood as the general factorizability of overlaps and correlation functions into kinematic components and symmetry-invariant components.

The Wigner--Eckart theorem tells us that a similar reduction applies to the matrix elements of \textit{spherical tensor operators}~\cite{sakurai_modern_2020}. A spherical tensor operator $\hat X^{(J,M)}$ is an eigenoperator of $\hat J^2$ and $\hat J^z$ under the adjoint action of the algebra, i.e. commutators, and as such carry $J$ and $M$ quantum numbers,
\begin{align}
\label{eq:adjointEigen}
     \hat J^2\left[\hat X^{(J,M)}\right] &\equiv \sum_{\alpha = x,y,z}\left[\hat J^\alpha,\left[\hat J^\alpha,\hat X^{(J,M)}\right]\right] = J(J+1)\hat X^{(J,M)}, \nonumber \\ \hat J^z\left[\hat X^{(J,M)}\right] &\equiv \left[\hat J^z,\hat X^{(J,M)}\right] = M\hat X^{(J,M)}.
\end{align}
Any operator in system that carries SU(2) symmetry will have a decomposition into spherical tensor components\footnote{In e.g. a qubit system the decomposition of a short product of local Pauli operators is generally quite simple.}. The Wigner--Eckart theorem is stated,
\begin{align}
\label{eq:su2WET}
    \bra{J_2M_2E_2}\hat X^{(J,M)}\ket{J_1M_1E_1} = \bra{J_2M_2}\ket{JM;J_1M_1}\bra{J_2E_2}\hat X^{(J)}\ket{J_1E_1}
\end{align}
where $\hat X^{(J)}$ is a \textit{reduced tensor operator} which acts only on the invariant states,
\begin{align}
\label{eq:RTO}
    \bra{J_2E_2}\hat X^{(J)}\ket{J_1E_1} \equiv \sum_{MM_1}\bra{JM;J_1M_1}\ket{J_2M_2}\bra{J_2M_2E_2}\hat X^{(J,M)}\ket{J_1M_1E_1}. 
\end{align}
It is to the reduced tensor operator and the reduced overlap that the ETH and the EB, respectively, can be naively applied. 

\subsection{The non-Abelian microcanonical subspace and its entropy}
\label{subsec:naMS}

A microcanonical subspace is a space of physical states that share a set of extensive labels. In our context, simple reducibility enabled us to define physical objects with $J$ and $E$ labels, but not $M$ labels. Specifically, we defined a symmetry invariant state $\ket{JE}$, which lives in the Hilbert space $\mathcal{H}_{\text{inv}}^{(J)}$. Our hypothesis is that the relevant microcanonical subspace of a non-Abelian system, for the purposes of the ETH and the EB, is the space of symmetry-invariant states with total spin $J$ and energy (approximately) $E$ and without $M$ labels. The entropy of this space is the logarithm of its dimensionality. Its dimension is what we refer to as the \textit{density of multiplets}\footnote{A multiplet is a set of states that transform into one another under a symmetry, i.e. $\mathcal{H}_{\text{sym}}^{(J)}\otimes\ket{JE}$. The density of symmetry-invariant states $\ket{JE}$ is then the density of multiplets $\mathcal{H}_{\text{sym}}^{(J)}\otimes\ket{JE}$.} $\tilde\rho(E, J)$. One can define $\tilde\rho(E, J)$ crudely, 
\begin{align}
\label{eq:sinjin}
    \tilde\rho(E, J) \equiv \sum_{J'}\sum_{E'} \delta_{JJ'}\delta(E-E')
\end{align}
where $\delta(E-E')$ is understood to be a smeared delta function\footnote{The precise width of the smearing can be generally taken as $N^{-\gamma}$ where $\gamma$ is an exponent that depends on the transport properties of the system~\cite{dymarsky_bound_2018, wang_emergence_2023, richter_eigenstate_2020}, but won't be relevant for our purposes in any case. Formally, this problem is wholly sidestepped by handling the heuristic derivation we give in section~\ref{subsec:SchurOrth} with the resolvent formalism~\cite{guhr_random-matrix_1998} and taking the thermodynamic limit.}. The corresponding non-Abelian microcanonical entropy is then $S(E,J)\equiv \ln\tilde\rho(E, J)$. We revisit $S(E,J)$ in section~\ref{sec:ent} and provide a more elegant view.

\subsection{Non-Abelian ergodic bipartition}
\label{subsec:naeb}

In section~\ref{subsec:SR} we described how eigenstates decompose into a kinematic part and an invariant part. It is to the invariant part that we can apply the MBBC. Concretely, the MBBC motivates the basic form of the non-Abelian ergodic bipartition hypothesis (EB)~\cite{jindal_generalized_2024, deutsch_thermodynamic_2010, murthy_structure_2019, huang_universal_2021, shi_local_2023, lu_renyi_2019},
\begin{equation}
\label{eq:eb}
    \tilde c^i_{ab} \equiv \bra{J_iE_i}\ket{J_aE_a;J_bE_b} = \sqrt{e^{-S(E_i,J_i)}F^{J_i}_{J_aJ_b}(E_i-E_a-E_b)}R^{i}_{ab}
\end{equation}
where $F$ is a smooth window function and $R$ are correlated Gaussian random numbers. The window function $F$ encodes the energy and spin dependence of correlations of $\tilde c^i_{ab}$ and its shape is dictated by the nature of the coupling between subsystems. For energy, the coupling Hamiltonian $H_{AB}$ is a small area-scaling perturbation to the bulk energies $E_a$ and $E_b$ if the subsystems are themselves thermodynamically large. Consequently, the total energy is very nearly additive, $E_i \approx E_a + E_b$, forcing the window function $F$ to be sharply peaked around $E_i - E_a - E_b = 0$.

The spin dependence of $F$ is more nuanced. The total overlap between the subsystem $J_a\otimes J_b$ sector and the global $J_i$ sector is fully set by the recoupling rules of SU(2), which are encoded in the Clebsch--Gordan coefficients. Thus, it must be understood $F$ says nothing about the fusion of kinematic labels. What the $J$ dependence of $F$ encodes is the how its energy dependence differs between sectors. The only rule $F$ must obey on $J$ is the triangle inequality:
\begin{equation}
\label{eq:triangle_inequality}
    |J_a - J_b| \le J_i \le J_a + J_b.
\end{equation}
We therefore expect $F$ to be a smooth function without sharp peaks across all values of $J_i$ permitted by Eq.~\eqref{eq:triangle_inequality} unless a nontrivial energy constraint imposes a strong selection rule for a particular system.

\begin{figure}[htbp]
    \centering
    \begin{subfigure}[b]{0.8\textwidth}
        \centering
        \includegraphics[width=\textwidth]{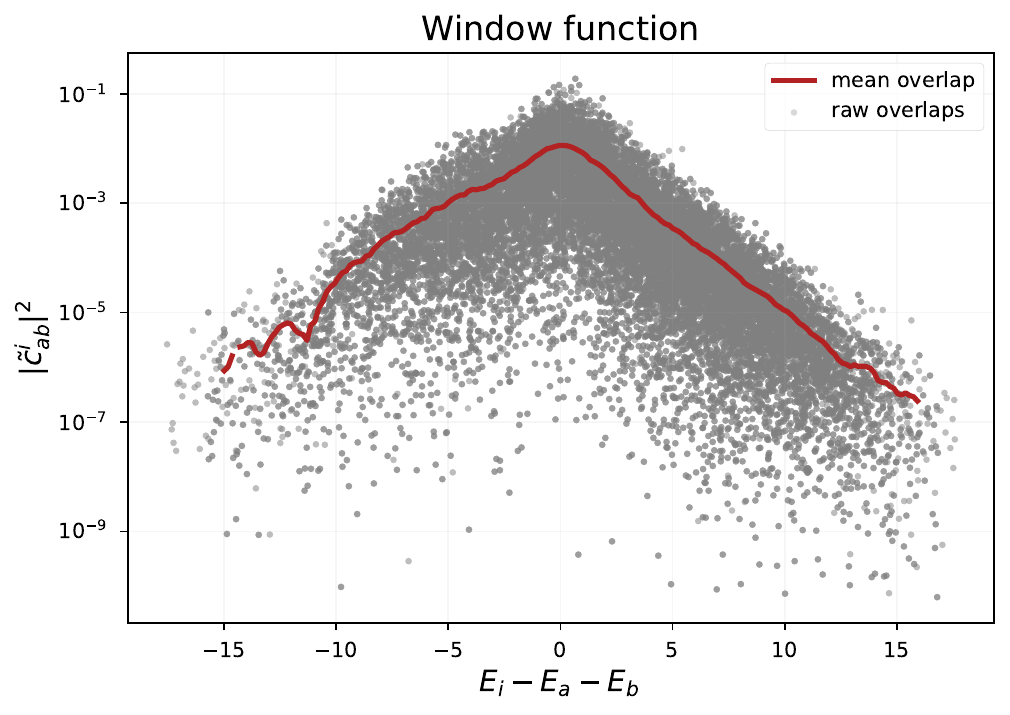}
        \caption{}
        \label{fig:a}
    \end{subfigure}
    \hfill 
    \begin{subfigure}[b]{0.8\textwidth}
        \centering
        \includegraphics[width=\textwidth]{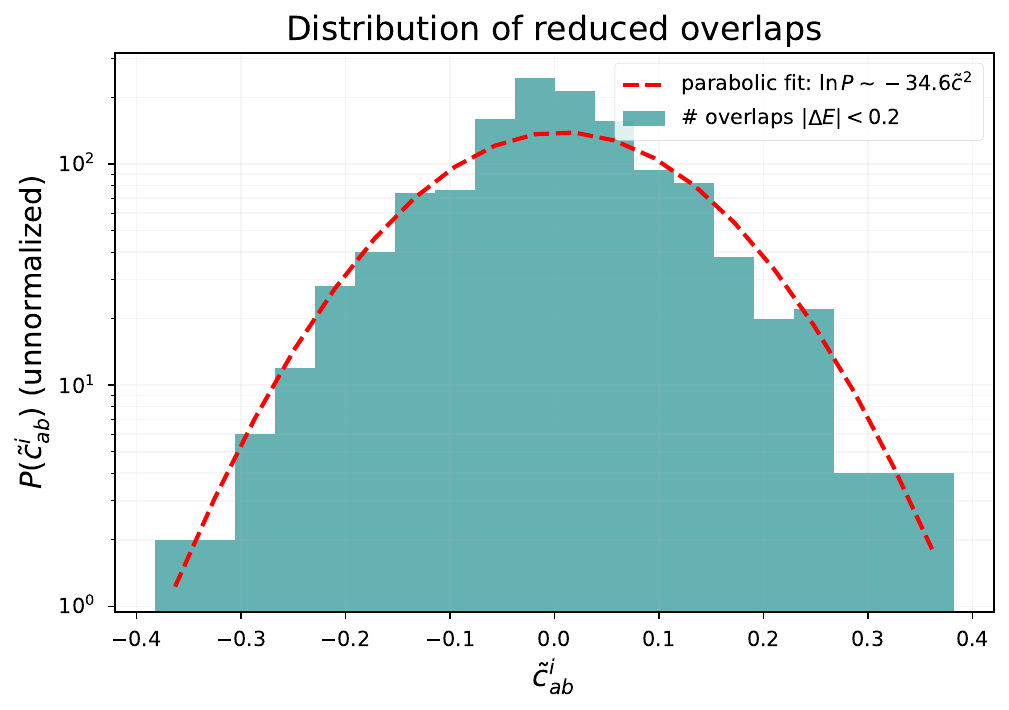}
        \caption{}
        \label{fig:b}
    \end{subfigure}

    \caption{Numerical evidence for the non-Abelian ergodic bipartition. We diagonalized a 14 site next-neighbor Heisenberg chain bipartitioned into 7 site subsystems with next neighbor interaction $t=0.5$ and open boundary conditions without resolving discrete symmetries. Nonzero values of the reduced overlap were computed by explicitly resolving the SU(2) symmetry and dividing out the appropriate Clebsch--Gordan coefficients and found to agree with eq.~\eqref{eq:eb}. We plot only states with $J_a=J_b=0.5,\, J_i=0$. (a) The modulus squared $\vert\tilde c^i_{ab}\vert^2$ versus energy. The window function is obtained by a Savitzky--Golay filter of degree 2 with bins of 2000 data points. (b) The distribution of $\tilde c^i_{ab}$ is found to be Gaussian within narrow energy windows as discussed in the text. }
    \label{fig:numerics}
\end{figure}

The utility of the EB is that it provides a conceptually clear ansatz for calculating state-based measures of chaos\footnote{See also the computation of distinguishability measures in~\cite{kudler-flam_distinguishing_2021} and~\cite{jindal_generalized_2024} and entanglement dynamics in~\cite{shi_local_2023} and~\cite{jindal_generalized_2024}.}, such as the Page curve of entanglement entropy. The basic technique for working with the EB is the replica trick, whose rules we summarize in appendix~\ref{app:replicas}. In essence, $\tilde c^{i}_{ab}$ is a statistical object whose properties are governed by its moments with respect to that Haar average of the unitary\footnote{or orthogonal or symplectic} group. Numerical evidence for the non-Abelian ergodic bipartition is provided in fig.~\ref{fig:numerics}.

\subsection{Non-Abelian ETH}
\label{subsec:naETH}

The many-body Berry's conjecture (MBBC) motivates the general form of the non-Abelian Eigenstate Thermalization Hypothesis (ETH). Traditionally~\cite{srednicki_approach_1999, murthy_non-abelian_2023},
\begin{equation}
\label{eq:su2_traditional}
\bra{J_i E_i} \hat{X}^{(k)} \ket{J_j E_j} = {X}^{(k)}(E, J) \delta_{ij} + e^{-S(E, J)/2} f^{(k)}(E, J; \omega,\Delta J) R_{ij}
\end{equation}
where $S(E, J)$ is the thermodynamic entropy at the average energy and spin density, $X^{(k)}$ and $f^{(k)}$ are smooth thermodynamic functions, and $R_{ij}$ are Gaussian random numbers with zero mean and unit variance. One also obtains an expression for the \textit{free cumulants}~\cite{pappalardi_eigenstate_2022},
\begin{equation}
\label{eq:su2_cumulant}
\overline{ \prod_{n=1}^q \bra{J_{i_{n+1}} E_{i_{n+1}}} \hat{X}_n^{(k)} \ket{J_{i_n} E_{i_n}} } = e^{-(q-1)S(E, J)} f_q^{(k)}(E, J; \vec{\omega}, \Delta \vec{J})
\end{equation}
where $i_{q+1} \equiv i_1$, $E$ and $J$ represent the average energy and spin of the ensemble, $\vec{\omega}$ denotes the $q-1$ independent energy differences, and $f_q$ is the free cumulant of order $q$. The non-Abelian ETH essentially states that reduced tensor operators obey a charged eigenstate thermalization~\cite{belin_charged_2022, patil_eigenstate_2025} without caveat; their frequency dependent correlation functions are stored in the empirical moments of their matrix elements. The correlation functions of other operators, however, must carry Clebsch--Gordan coefficients.

\section{Microcanonical Entropy}
\label{sec:ent}

In the last section we defined a non-Abelian microcanonical entropy by fixing the total energy and total spin, $S(E,J)$. However, total spin is quite different from fixing an Abelian charge. Thus it is necessary for us to demonstrate that the entropy, $S(E,J)$, has the expected scaling properties of a microcanonical entropy. In this section we demonstrate that $S$ is concave and extensive, but with a uniquely non-Abelian logarithmic correction in the zero spin density sector which will be later relevant to our calculation of entanglement entropy.

\subsection{The non-Abelian partition function}
\label{subsec:SchurOrth}

Usually, we define the density of states in energy alone as\footnote{Devotees of the resolvent formalism can follow along with their machinery instead~\cite{guhr_random-matrix_1998}.},
\begin{align}
\label{eq:DOS}
    \rho(E) \equiv \operatorname{Tr}\left[\delta\left(\hat H-E\right)\right]
\end{align}
which may be interpreted as the Fourier--Laplace transform of the canonical partition function,
\begin{align}
\label{eq:DOS2}
    \rho(E) = \int_\beta e^{\beta E}Z(\beta).
\end{align}
Applying Laplace's method to eq.~\eqref{eq:DOS2} one finds that the microcanonical entropy, $S$, defined as the logarithm of the density of states is given to volume order by the following optimization,
\begin{align}
\label{eq:Legendre}
    S(E) \equiv \ln\rho(E) = \sup_\beta\left[\beta E + \ln Z(\beta)\right].
\end{align}
Analogously, one can repeat the same analysis in multiple thermodynamic potentials for any number of Abelian charges and many standard results establish the properties of entropy and related quantities in such circumstances~\cite{pathria_statistical_2021}. 

However, to get $S(E,J)$ we want not a density of states $\rho(E,J,M)$, but the density of multiplets, $\tilde\rho(E,J)$ and we must apply the Fourier--Laplace analysis with respect to the group geometry. To arrive at our desired expression it is easier to work backwards. The SU(2) partition function is given by\footnote{For SU(2) $e^{-\beta \hat H}\hat U(g) = e^{-\beta \hat H - i\phi_x\hat J^x- i\phi_y\hat J^y- i\phi_z\hat J^z}$ which analytically continues the non-Abelian thermal state of refs.~\cite{guryanova_thermodynamics_2016, yunger_halpern_microcanonical_2016, lostaglio_thermodynamic_2017, yunger_halpern_noncommuting_2020, kranzl_experimental_2023} to imaginary chemical potential. The choice of imaginary over real chemical potential is nothing more than using a Fourier transform instead of a Laplace transform as it will provide us a more transparent derivation. The Fourier analysis is also automatic to generalize to the case of disconnected groups in a way the Laplace analysis is not.},
\begin{align}
\label{eq:su2part}
    Z(\beta,g) = \operatorname{Tr}\left[e^{-\beta \hat H}\hat U(g)\right].
\end{align}
Following the decomposition in eq.~\eqref{eq:introDecomp} and Schur's Lemma (see appendix~\ref{app:background} for details),
\begin{align}
\label{eq:charFind}
    Z(\beta,g) &= \sum_{J} \left( \sum_{E}\bra{JE}e^{-\beta \hat H} \ket{JE}\right)\times \left( \sum_{M}\bra{JM}\hat U(g)\ket{JM}\right) \nonumber \\ 
    &\quad\quad\equiv \sum_{J} \Tr_{\mathcal{H}^{(J)}_{\text{inv}}}\left[e^{-\beta \hat H}\right]\times \chi_J(g)
\end{align}
where $\chi_J$ is known as the \textit{character} of irrep $J$. It is essentially the partition function for the internal degrees of freedom of the representation. For us, its essential feature is given by the Schur orthogonality relation~\cite{fulton_representation_2004},
\begin{align}
\label{eq:schur}
    \int_{g\in SU(2)} \chi^*_{J}(g)\chi_{J'}(g) = \delta_{JJ'}
\end{align}
where $\int_g$ is taken with respect to the Haar measure on SU(2). Thus, the character provides precisely the term we need to count each J multiplet once as in eq.~\eqref{eq:sinjin}. It follows,
\begin{align}
\label{eq:rochester}
    \tilde\rho(E,J) \equiv \sum_{J'}\sum_J \delta_{JJ'}\delta(E-E') = \int_{\beta g} \chi^*_{J}(g) e^{\beta E} Z(\beta,g).
\end{align}
Eq.~\eqref{eq:rochester} provides a more elegant, and computationally useful, formulation of the density defined in eq.~\eqref{eq:sinjin} that we now apply.

\subsection{Logarithmic corrections}
\label{subsec:charInt}

Two more steps are required to compute the volume and logarithmic order contributions to eq.~\eqref{eq:rochester}. By applying the Weyl integral formula we can restrict the the integration to the U(1) subgroup of SU(2) generated by rotations about the $z$-axis. The Weyl integral formula states that when the integrand, as in eq.~\eqref{eq:rochester}, is invariant under group conjugation~\cite{koch_compact_nodate},
\begin{align}
\label{eq:WIF}
    \int_g = \int_\phi \sin(\phi/2)^2
\end{align}
where $\phi$ is the azimuthal angle integrated from $0$ to $4\pi$. This simplification is essentially a consequence of choosing a $z$-axis and integrating out the other Euler angles. Secondly, one needs the character formula. Following from the definition of the spin matrices,
\begin{align}
    \chi_J(\phi) \equiv \sum_{m = -J}^{J} e^{im\phi} = \frac{\sin((2J+1)\phi/2)}{\sin(\phi/2)}.
\end{align}
Then one has an expression for the density of multiplets in terms of the Abelian partition function,
\begin{align}
\label{eq:rochester2}
    \tilde\rho(E,J) &= \int_{\beta\phi}\sin(\phi/2)\sin\left[(2J+1)\phi/2\right]e^{\beta E}\operatorname{Tr}[e^{-\beta \hat H+i\phi \hat J_z}] \nonumber \\
    &= \int_{\beta\phi}\sin(\phi/2)\sin\left[(2J+1)\phi/2\right]e^{\beta E}Z(\beta,i\phi)
\end{align}

From $\sin(\phi/2)\sin\left[(2J+1)\phi/2\right]$ we need to pull out a factor $e^{\pm iJ\phi}$,
\begin{align}
\label{eq:sinAdd}
    \sin(\phi/2)\sin\left[(2J+1)\phi/2\right] = \frac{i}{2}\sin(\phi/2)e^{-i\phi/2}e^{-iJ\phi} +\text{c.c.}
\end{align}
and find the saddle points of
\begin{align}
\label{eq:saddleAction}
    \beta E\pm iJ\phi+\ln Z(\beta, i\phi)
\end{align}
where the $\pm$ comes from the complex conjugate term in eq.~\eqref{eq:sinAdd}. The saddle coordinate $\bar\phi$ will be complex, but the overall entropy will be real. Then the maximum of eq.~\eqref{eq:saddleAction} is the volume component of microcanonical entropy
\begin{align}
\label{eq:grandLegendre}
    \bar S \equiv \sup_{\beta,\phi} [\beta E\pm iJ\phi+\ln Z(\beta, i\phi)]
\end{align}
which provides saddle-points conditions for $\beta$ and $\phi$,
\begin{align}
\label{eq:saddlePoints}
    \expval{\hat J_z}_{\bar\beta,i\bar\phi} = \pm J \nonumber \\
    \expval{\hat H}_{\bar\beta,i\bar\phi} = E.
\end{align}
The $\pm$ in front of $J$ represents a reflection symmetry about the $x$-$y$ plane. For simplicity we will drop the $-J$ saddle\footnote{Multiple, equivalent saddles contribute only an $\mathcal{O}(1)$ term to entropy, i.e. $\ln(e^S+e^S) = S + \ln 2$.}. To obtain the leading logarithmic corrections we must apply Laplace's method to Gaussian order,
\begin{align}
\label{eq:Laplace}
    \tilde\rho(E,J) = \frac{i}{2}\int_{\beta\phi}\sin(\phi/2)e^{-i\phi/2}e^{\bar S-\frac{1}{2}(\delta\beta,\delta\phi)\cdot \bar \Sigma\cdot(\delta\beta,\delta\phi)}
\end{align}
where $\Sigma$ is the susceptibility matrix. 

From eq.~\eqref{eq:Laplace}, we will handle, for illustration, three scenarios for the prefactor $\frac{i}{2}\sin(\phi/2)e^{-i\phi/2}$: $J\sim\mathcal{O}(N)$, $J = 0$, and $J\sim\mathcal{O}(\sqrt N)$. First, by eq.~\eqref{eq:saddlePoints}, if $J\sim\mathcal{O}(N)$ then $\Im[\bar\phi]\sim\mathcal{O}(1)$. In this case, $\frac{i}{2}\sin(\phi/2)e^{-i\phi/2}$ is approximately a constant, yielding,
\begin{align}
\label{eq:jV}
    \tilde\rho(E,J) &= \frac{i}{2}\sin(\bar\phi/2)e^{-i\bar\phi/2}e^{\bar S} \sqrt{\frac{4\pi^2}{ \det \bar\Sigma}} \nonumber \\
    \implies  S(E,J)&= \ln\tilde\rho(E,J) = \bar S - \ln\sqrt{\det \bar\Sigma} + \mathcal{O}(1) \\
    &= \bar S - \ln N + \mathcal{O}(1). \nonumber
\end{align}
The logarithmic reduction in eq.~\eqref{eq:jV} is a physically unimportant and strictly Abelian consequence of defining $S(E,J)$ as the logarithm of a density, $\tilde\rho(E,J)$, and not a dimensionless, $\tilde\rho(E,J)\Delta E\Delta J$~\cite{pathria_statistical_2021}. In the second and third cases, we have $J\ll\mathcal{O}(N)$ which enforces  $\Im[\bar\phi]\ll\mathcal{O}(1)$. Here there are two saddles $\bar\phi \approx 0, 2\pi$ which apply respectively to integral and half-integral representations in the two reflection sectors. On each saddle, $\frac{i}{2}\sin(\phi/2)e^{-i\phi/2} $ $ \approx i\phi/4+\phi^2/8$, which yields for each saddle,
\begin{align}
\label{eq:j0m}
    \tilde\rho(E,J) &= \left(\frac{i\bar\phi}{4}+\frac{\bar\phi^2+\bar\Sigma_{\phi\phi}^{-1}}{8}\right) e^{\bar S} \sqrt{\frac{4\pi^2}{ \det \bar\Sigma}}.
\end{align}
The zero total spin sector has forced nontrivial polynomial corrections to the density of muliplets. Whether we are in case 2, $J =0$, or case 3, $J\sim\mathcal{O}(\sqrt N)$, will determine which polynomial factor is largest. For $J = 0$, $\Im[\bar\phi]=0$ and $\bar\Sigma_{\phi\phi}^{-1}\sim \mathcal{O}(N^{-1})$ and we find,
\begin{align}
\label{eq:j0}
    S(E,J)&= \ln\tilde\rho(E,J) = \bar S - \ln\bar\Sigma_{\phi\phi} - \ln\sqrt{\det \bar\Sigma} + \mathcal{O}(1) \nonumber \\
    &= \bar S - 2\ln N + \mathcal{O}(1). 
\end{align}
For the remaining case, $J\sim\mathcal{O}(\sqrt N)$, $\bar\phi \sim \mathcal{O}(N^{-1/2})\gg\bar\phi^2\sim \bar\Sigma_{\phi\phi}^{-1}\sim\mathcal{O}(N^{-1})$. Then,
\begin{align}
\label{eq:jM}
    S(E,J)&= \ln\tilde\rho(E,J) = \bar S - \ln\Im[\bar\phi] - \ln\sqrt{\det \bar\Sigma} + \mathcal{O}(1) \nonumber \\
    &= \bar S - \frac{3}{2}\ln N + \mathcal{O}(1). 
\end{align}
More generally we can consider $J\sim\mathcal O (N^p)$, $1>p>0$. In all cases, the logarithmic correction will come from $\bar \phi$ which scales $\sim\mathcal O (N^{p-1})$ and thus we expect $S(E,J)=\bar S - (2-p)\ln N +\mathcal O(1)$.

In eqs.~\eqref{eq:j0} and~\eqref{eq:jM}, we found extra reductions to entropy because of the Jacobian $2\sin(\phi/2)^2$ in the Weyl integral formula, eq.~\eqref{eq:WIF}. The physical origin of this reduction is that zero total spin is result of 3 restrictions, $\expval{\hat J_x} = \expval{\hat J_y} = \expval{\hat J_z} = 0$ rather than 1, $\expval{\hat J_z} = 0$ even though states are not labeled by eigenvalues of 2 of these operators. Consequently, one finds it is more challenging to construct such states, reflected in a logarithmic reduction of the microcanonical entropy. This reduction is $\frac{3-1}{2}\ln N = \ln N$ in the exactly zero total spin sector and by $\frac{1}{2}\frac{3-1}{2}\ln N = \frac{1}{2}\ln N$ in the $\sim\mathcal O(\sqrt N)$ sector. Indeed, in appendix~\ref{app:WeylGeneral} we derive semiclassically that in the general case of a compact, connected Lie group, the corrections are $-\frac{d-r}{2}\ln N$ and $-\frac{d-r}{4}\ln N$ where $d$ is the dimension of the group and $r$ is its rank.

\subsection{Concavity of microcanonical entropy}
\label{subsec:concave}

The fact that the non-Abelian microcanonical entropy is closely related to the Abelian partition function associated with its Cartan subalgebra implies that the usual rules of extensivity and concavity of Abelian microcanonical entropy apply equally well to the non-Abelian entropy. Then, it becomes possible to apply Jensen's inequality. Defining intensive densities, $s = S/N,\, \varepsilon  = E/N,\, j = J/N$,
\begin{align}
\label{eq:Jensen}
    \frac{N_A}{N}s(\varepsilon_A,j_A)+\frac{N_B}{N}s(\varepsilon_B,j_B)\leq s(\frac{N_A\varepsilon_A+N_B\varepsilon_B}{N},\frac{N_Aj_A+N_Bj_B}{N}) = s(\varepsilon,j)
\end{align}
The maximization of entropy by saddle-points in eq.~\eqref{eq:grandLegendre} established that the entropy is additive to volume order $S(E,J) \approx S_A(E_A,J_A) + S_B(E_B,J_B)$. The additivity implies the saturation of eq.~\eqref{eq:Jensen} which in turn implies equality of arguments. That is, in equilibrium, the energy and spin densities are the same across the whole system,
\begin{align}
\label{eq:densities}
    \varepsilon_A&=\varepsilon_B=\varepsilon \nonumber \\
    j_A&=j_B=j.
\end{align}
Essentially, to volume order the total spins of subsystems A and B are parallel and proportional to their respective volumes in equilibrium. As a consequence, total spin density $j$ is a valid intensive variable, i.e. stored locally, despite being composed of highly non-local terms in $\hat J\cdot \hat J$. The ability of a non-locally defined $J$ to have an intensive density $j$ is originates from the approximately Gaussian structure of eq.~\eqref{eq:rochester}.

However, the subsystem spin and energy densities can fluctuate away from the global spin and energy density. The degree of fluctuation is determined by $F^{J_i}_{J_aJ_b}(E_i-E_a-E_b)$ in eq.~\eqref{eq:eb}. The variance of this fluctuation was analyzed for energy alone in refs.~\cite{garrison_does_2018, jindal_generalized_2024}. Spin fluctuations only modify the analysis by the presence of hard walls imposed by the triangle inequality eq.~\eqref{eq:triangle_inequality}, so we just state our variances without elaborating on the calculation. Each subsystem has a susceptibility matrix $\Sigma_{A(B)}$. We define two effective susceptibilities $\Sigma_{1}\equiv \left(\Sigma_A^{-1}+\Sigma_B^{-1}\right)^{-1}$ and $\Sigma_{2}\equiv \Sigma_A+\Sigma_B$. Then for subsystem A,
\begin{align}
\label{eq:variances}
    \expval{\delta E_a^2} &\approx \left(\Sigma_1\right)_{\beta\beta} + \left(1-\frac{2}{\pi}\right) \frac{\left(\Sigma_A\Sigma_2^{-1}\right)_{\beta\phi}^2}{\left(\Sigma_2^{-1}\right)_{\phi\phi}} \nonumber \\
    \expval{\delta J_a^2} &\approx \left(\Sigma_1\right)_{\phi\phi} + \left(1-\frac{2}{\pi}\right) \frac{\left(\Sigma_A\Sigma_2^{-1}\right)_{\phi\phi}^2}{\left(\Sigma_2^{-1}\right)_{\phi\phi}}
\end{align}
where the second term in each expression is the influence of the upper wall imposed by the triangle inequality, $J_i \leq J_a +J_b$, and the approximation suppresses terms related to area order fluctuations and the lower wall of the triangle inequality $J_i\geq |J_a-J_b|$. The important takeaway from eq.~\eqref{eq:variances} is not the precise form of the variances but the general behavior: $\delta E_a,\,\delta J_a \sim\mathcal{O}(\sqrt {\min\left(N_A, N_B\right)})$. This fluctuation, along with the triangle inequality implies that $J_a + J_b$ typically exceeds $J_i$ by a small amount,
\begin{align}
\label{eq:toRef}
    J_a + J_b = J_i +\epsilon,\quad\epsilon\sim\mathcal{O}\left(\sqrt {\min\left(N_A, N_B\right)}\right).
\end{align}
If subsystems $A$ and $B$ are both finite fractions of the whole system $\delta E_a,\,\delta J_a \sim\mathcal{O}\left(\sqrt {\min\left(N_A, N_B\right)}\right)$ $\sim \mathcal{O}(\sqrt {N})$. Unless otherwise mentioned we will take this assumption and assume $\epsilon\sim\mathcal{O}(\sqrt {N})$.

\section{Equilibrium Thermodynamics}
\label{sec:equi}

Here we explore briefly the implication of choosing a symmetry-resolved global state, $\ket{JME}$, on local equilibrium thermodynamics. The principle of the MBBC, is that an eigenstate reproduces the predictions of the microcanonical ensemble. Relatedly, we expect local subsystems to experience a version of the canonical ensemble. We mentioned before two notions of a spatially local subsystem: (I) the (algebra of) all observables in a region and (II) the (algebra of) symmetry-invariant observables, i.e. spin singlets $\hat X^{(0,0)}$, in a region. We compute the equilibrium density matrix for each definition, (I) by a partial trace and (II) by a specific projection defined in appendix~\ref{subapp:symm}. Surprisingly, we will find that only the latter definition obtains a true canonical state. The issue is that only the $J$ and $E$ charges were used in the non-Abelian microcanonical entropy. The $M$ quantum numbers are fully blind in this set up to thermal fluctuations, except via their relationship to $J$. We first consider definition (II) and then return to definition (I).

\subsection{Invariant reduced density matrix}
\label{subsec:rdmInv}

In appendix~\ref{subapp:symm}, we show to define a density matrix for only symmetry-invariant observables $\hat{\rho}_{\text{invA}}$, via a specific projector applied to the global density matrix $\hat\rho$. Restricted to SU(2), the definition is written,
\begin{align}
\label{eq:projInvAJ}
    \hat{\rho}_{\text{invA}} = \hat{\Pi}_{\text{invA}}[\hat\rho] =\sum_{\substack{J_aM_aE_aE_a' \\J_bM_bE_b}}   \left(\bra{J_aM_aE_a;J_bM_bE_b}\hat\rho\ket{J_aM_aE_a';J_bM_bE_b}\right)\times\ket{J_aE_a}\!\bra{J_aE_a'}.
\end{align}
To study energy eigenstates, we choose $\hat\rho=\ket i \!\bra i$ and apply the non-Abelian EB to eq.~\eqref{eq:projInvAJ},
\begin{align}
\label{eq:pTha}
    \bra{J_aE_a}\hat{\rho}_{\text{invA}}\ket{J_aE_a'} &=\sum_{{M_aJ_bM_bE_b}}   \bra{ab}\ket{i} \!\bra i\ket{a'b}  \nonumber \\
    &= \sum_{{M_aJ_bM_bE_b}}   \vert C^{i}_{ab}\vert^2 \tilde c^{ab}_{i}\tilde c^{i}_{a'b}= \sum_{{J_bE_b}}  \tilde c^{ab}_{i}\tilde c^{i}_{a'b} \\
    &\approx \int_{J_bE_b}e^{S_B(E_b,J_b)}\left(e^{-S(E_i,J_i)}F^{J_i}_{J_aJ_b}(E_i-E_a-E_b)\delta_{aa'}\right) +\dots \nonumber
\end{align}
where the normalization rules of the Clebsch--Gordan coefficients caused them to sum out. Eq.~\eqref{eq:pTha} very closely resembles a partial trace of the norm-squared reduced overlap but with one extra restriction: $J_a = J_a'$; a singlet operator cannot change the total spin. The off-diagonal elements of $\hat{\rho}_{\text{invA}}$ encode non-trivial boundary interactions, but for the purposes of this section can be regarded as small as long as A is thermodynamically large. We will handle them implicitly in our calculation of entanglement entropy to come.

Evaluating eq.~\eqref{eq:pTha} follows a conventional application of Laplace's method. The window function $F$ enforces $E_i\approx E_a+E_b$ and $J_i \leq J_a + J_B$ which by the extensivity argument of section~\ref{subsec:concave} is optimized at equality for $J\sim\mathcal O (N)$. Thus,
\begin{align}
\label{eq:standard}
    \int_{J_bE_b}e^{S_B(E_b,J_b)}\left(e^{-S(E_i,J_i)}F^{J_i}_{J_aJ_b}(E_i-E_a-E_b)\right) \approx e^{S_B(E_i-E_a, J_i-J_a)-S(E_i,J_i)} \underset{N_A\ll N_B}{\longrightarrow} \frac{1}{Z_A}e^{-\beta E_a-\phi J_a}.
\end{align}
The Gibbs state in~\eqref{eq:standard} is a true non-Abelian Gibbs state. There is another notion of a non-Abelian thermal state is defined in refs.~\cite{guryanova_thermodynamics_2016, yunger_halpern_microcanonical_2016, lostaglio_thermodynamic_2017, yunger_halpern_noncommuting_2020, kranzl_experimental_2023} that we will return to in the discussion, sec.~\ref{sec:discussion}. We have again seen that $\hat J\cdot\hat J$ is a strictly non-local quantity, yet the local density matrix perceives an intensive $J$ density nevertheless. We return to this point in the discussion. 

\subsection{General reduced density matrix}
\label{subsec:generalRDM}

If one wishes to study operators $\hat X^{(J,M)}$ that carry non-zero $J$ and $M$ quantum numbers one must work within the conventional full algebra of local observables. Its density matrix is obtained by a partial trace,
\begin{equation}
\label{eq:pT}
\rho_A = \text{Tr}_B \big( |i\rangle \langle i| \big) = \sum_{b} |a\rangle \langle a b | i \rangle\langle i | a'b \rangle   \langle a'|
\end{equation}
We restrict our focus to the diagonal elements, 
\begin{equation}
\label{eq:diagonal}
\langle a | \rho_A | a \rangle \approx \sum_{J_B} \int dE_B \, e^{S_B(E_B, J_B)} |C_{J_A M_A; J_B M_B}^{J_i M_i}|^2 |\tilde{c}_{ab}^i|^2
\end{equation}
which, following section~\ref{subsec:rdmInv} but with the Clebsch--Gordan coefficients, evaluates to
\begin{equation}
\label{eq:fixx}
\rho_A(J_A, M_A, E_A) \approx e^{S_B(E_i - E_A, J_i - J_A)-S(E_i,J_i)} |C_{J_A M_A; \bar J_B \bar M_B}^{J_i M_i}|^2.
\end{equation}
Analogous to the restrictions imposed by $F$, the Clebsch--Gordan coefficient in eq.~\eqref{eq:fixx} enforces $M_i=M_A+M_B$ exactly. However, $S$ does not depend on $M$ and will not generate a conjugate potential for it. Instead, the reduced density matrix carries the raw Clebsch--Gordan coefficients. In appendices~\ref{app:finiteDens},~\ref{app:zeroDens}, and~\ref{app:margDens} we work out the asymptotics of Clebsch--Gordan coefficients in our regimes of interest. Restricting strictly to the case of SU(2), the squared coefficients take three precise forms depending on the polarization of the global state: $J_i = 0$, $J_i \sim \mathcal{O}(\sqrt{N})$, and $J_i \sim \mathcal{O}(N)$.

For a completely unpolarized global singlet state ($J_i = 0$), the squared Clebsch--Gordan coefficient is uniformly distributed across the allowed subsystem states,
\begin{equation}
\label{eq:cgc_unpolarized}
|C^{0 0}_{J_a M_a; J_b M_b}|^2 = \frac{1}{2J_a + 1} \delta_{J_a, J_b} \delta_{M_a, -M_b}.
\end{equation}
For a marginally polarized state ($J_i \sim \mathcal{O}(\sqrt{N})$), the coefficients are described by the semiclassical limit of geometric quantization,
\begin{equation}
\label{eq:cgc_marginal}
|C^{J_i M_i}_{J_a M_a; J_b M_b}|^2 \approx \frac{J_i}{2\pi A_{\text{xy}}}\delta_{i;ab},
\end{equation}
where $\delta_{i;ab}$ enforces that three vectors of $J$ lengths and $M$ heights can form a triangle and $A_{\text{xy}}$ is the area of that triangle when projected onto the $x$-$y$ plane. In the fully polarized, or stretched, regime ($J_i \sim \mathcal{O}(N)$), the coefficients take a tightly bound Gaussian form,
\begin{equation}
\label{eq:cgc_stretched}
    |C^{J_i M_i}_{J_a M_a; J_b M_b}|^2 \approx \frac{1}{\sqrt{2\pi\sigma^2}} \exp\left( -\frac{(M_a - \bar{M}_A)^2}{2\sigma^2} \right),
\end{equation}
with a mean subsystem magnetization $\bar{M}_A = M_i J_A / J_i$ and variance $\sigma^2 \approx \frac{J_A J_B}{J_i} \left(1 - \frac{M_i^2}{J_i^2}\right)$.

When completely unpolarized, the state has absolutely no $M$ dependence; the probability distribution over the magnetic quantum numbers is perfectly uniform across the subsystem multiplet. As the global state polarizes further into the marginal regime, the distribution concentrates onto an ever-smaller region of semiclassically allowed $M$ values, but remains approximately uniform within those boundaries. Once the system reaches the fully polarized regime, the $M$ values become tightly Gaussian distributed around their single classically and thermodynamically allowed value, $\bar{M}_A$. Naively, the Gaussian behavior of eq.~\eqref{eq:cgc_stretched} violates thermodynamic expectations of a Gibbs state/Boltzmann factor. However, because the subsystem magnetization $M_A$ is strictly pinned to a fixed ratio of $J_A$, the exponential distribution of $J_A$ implies an overall exponential distribution of $M_A$ nevertheless, but with a different chemical potential. We return to this point in the discussion.

\section{Entanglement Entropy}
\label{sec:EE}

We now elaborate algebraically on our two notions of a subsystem. As discussed in appendix~\ref{subapp:symm}, both definitions correspond to \textit{subalgebras of observables} of the Hilbert space. Further, following ref.~\cite{bianchi_non-abelian_2024}, we provide a definition of a reduced density matrix, $\hat\rho_*$, for an arbitrary observable subalgebra, $\mathcal A_*$. From a reduced density matrix, $\hat\rho_*$, one also obtains a notion of entanglement entropy, $S_*\equiv -\Tr\left[\hat\rho_*\ln\hat\rho_*\right]$. If it has an entanglement entropy, one then wonders what environment an arbitrary subalgebra is entangled with. In general, the (non-zero) spectrum of a reduced density matrix must be identical to the spectrum of the reduced density matrix of its environment. For the conventional notion of a subsystem, its environment is its complementary tensor product factor. For an arbitrary subalgebra, the environment is the \textit{commutant subalgebra}~\cite{bianchi_non-abelian_2024}. The commutant of an algebra $(\mathcal A_*)'$ is the subalgebra of all observables that commute with all elements of $\mathcal A_*$,
\begin{align}
\label{eq:commutantMain}
    (\mathcal A_*)' \equiv \left\{\hat X\mid \forall\hat X_*\in \mathcal A_*, \left[\hat X,\hat X_*\right]=0\right\}.
\end{align}
For the algebra of all observable in region A, $\mathcal A_A$, its commutant is simply $\mathcal A_B$, the algebra of all observables in its spatial complement region B. 

For our local invariant subalgebra $\mathcal A_{\text{inv}A}$, the commutant, $\mathcal A_{\text{inv}A}'$, the subalgebra of all observables which commute with all elements of $\mathcal A_{\text{inv}A}$. One can swiftly work out $\mathcal A_{\text{inv}A}'$ using the ordinary rules of operator subalgebras~\cite{bratteli_operator_1981},
\begin{align}
\label{eq:dM}
    \mathcal A_{\text{inv}A}' = \left(\mathcal A_A\cap \mathcal A_{\text{inv}} \right)' = \mathcal A_A'\mathcal A_{\text{inv}}' = \mathcal A_B\mathcal A_{\text{sym}}.
\end{align}
where $\mathcal{A}_{\text{sym}} = U(\mathfrak{su}(2))$ is the algebra generated by $\hat J^x, \hat J^y, \hat J^z$. Eq.~\eqref{eq:dM} needs substantial interpretation. The first equality encodes that the local symmetry invariant subalgebra is the intersection of the full local subalgebra with the global symmetry invariant subalgebra. The second statement is that the commutant of an intersection is the product of commutants and is essentially an algebraic generalization of de Morgan's laws. The product of two subalgebras is the smallest subalgebra that contains both factors\footnote{recognizing that a subalgebra demands closure}. Lastly, the commutant of operators in region A is the algebra of operators in region B, while the commutant of the symmetry invariant observables is $\mathcal A_{\text{sym}}$. 

Eq.~\eqref{eq:dM} then tells us that the algebra of symmetry invariant observables on A entangles not only to the subsystem B but also to the representations of the global Lie algebra. In other words, the local invariant density matrix composed of states $\ket{J_aE_a}$ can entangle with the global spin configuration composed of kinematic states $\ket{J_iM_i}$. The entanglement of $\mathcal A_{\text{inv}A}$ and $\mathcal A_{\text{inv}A}'$ is referred to symmetry-resolved entanglement entropy, while conventional entanglement entropy gives the entanglement between $\mathcal A_{A}$ and $\mathcal A_{B}$. The main result of this section is a logarithmic difference between the symmetry-resolved and conventional entanglement entropies.

\subsection{Symmetry-resolved entanglement entropy}
\label{subsec:symmEE}

In section~\ref{subsec:rdmInv}, we found that the invariant reduced density matrix is given by
\begin{align}
\label{eq:rdmInvAgain}
    \left(\hat \rho_{\text{invA}}\right)_{aa'} = \sum_b \tilde c_{i}^{ab}\tilde c^{i}_{a'b}
\end{align}
To compute the symmetry-resolved entanglement entropy we invoke the replica approach. We start with the $\alpha$-R\'enyi entropy,
\begin{equation}
\label{eq:Renyi_def}
  \tilde S_\alpha \equiv\frac{1}{1-\alpha}\ln\!\left(\Tr_{\text{inv}A}\left[\left(\hat \rho_{\text{invA}}\right)^\alpha\right]\right),
\end{equation}
for which we need to evaluate,
\begin{equation}
\label{eq:Renyi_product}
  \Tr_{\text{inv}A}\left[\left(\hat \rho_{\text{invA}}\right)^\alpha\right] = 
  \sum_{\{a_k b_k\}}\tilde c_{i}^{a_1 b_1}\tilde c^{i}_{a_2 b_1}\tilde c_{i}^{a_3 b_2}\cdots\tilde c_{i}^{a_\alpha b_\alpha}\tilde c^{i}_{a_1 b_\alpha}.
\end{equation}
This object has been studied extensively in past works~\cite{jindal_generalized_2024, murthy_structure_2019, lu_renyi_2019, shi_local_2023}. In fig.~\ref{fig:3replica} we provided an example of a leading contribution to 3-R\'enyi entropy. In it we find a suppression by 3 factors of $S_A$ and 5 factors of $S_B$ which can be collected into $-3S-2S_B$. Naturally there is a conjugate pairing of $i$'s that will yield a factor of $-3S-2S_A$. The larger term is simply given by the smaller of $S_A$ or $S_B$. For the $\alpha$-R\'enyi entropy, the overall weight of the summand is $e^{-\alpha S-(\alpha-1)S_{\text{min}}}$, $S_{\min} =  \min\left[S_A,S_B\right]$. Finally, the sum in ~\eqref{eq:Renyi_product}, when smoothed into an integral over the subsystem energies and spins, provides a factor of $e^{\alpha S_A+\alpha S_B}$, and the smooth $F$ function enforces the relevant saddle-point conditions. In summary, our expression becomes,
\begin{align}
\label{eq:RenyiEqResultInt}
    \Tr_{\text{inv}A}\left[\left(\hat \rho_{\text{invA}}\right)^\alpha\right] = \int_{E_{a_1} E_{b_1}J_{a_1}J_{b_1},\dots} e^{\alpha\left[S_A({E
    }_A,{J}_A) +S_B({E}_B,{J}_B)-S(E_i)\right]-(\alpha-1)S_{\text{min}}({E}_A,{J}_A,{E}_B,{J}_B)}\nonumber \\[-1em] \times F(E_{a_1}, E_{b_1},J_{a_1}, J_{b_1}, \dots).
\end{align}
The saddle-point entropy is given by,
\begin{align}
\label{eq:RenyiEqResult}
    \tilde S_\alpha = \frac{1}{1-\alpha}\left\{\alpha\left[S_A(\bar{E
    }_A,\bar{J}_A) +S_B(\bar{E}_B,\bar{J}_B)-S(E_i, J_i)\right]-(\alpha-1)S_{\text{min}}(\bar{E}_A,\bar{J}_A,\bar{E}_B,\bar{J}_B)\right\}.
\end{align}
The saddle-point $(\bar{E}_A,\bar{J}_A,\bar{E}_B,\bar{J}_B)$ is given by 
\begin{align}
\label{eq:saddlePoints2}
    &\text{(I)}\quad \bar{E}_{A}+\bar{E}_{B}=E_i,\,\bar{J}_{A}+\bar{J}_{B}=J_i \quad \& \quad
    \begin{cases}
      \text{(II.1)} \quad S_A' = \alpha S_B'\\
      \text{(II.2)} \quad S_B' = \alpha S_A'\\
      \text{(II.3)} \quad S_A=S_B
    \end{cases} 
\end{align}
where the second saddle-point condition (II) is given by which of the three choices minimizes the entropy.

The von Neumann entropy, given by $-\partial_\alpha\Tr[\hat\rho^\alpha]\big\rvert_{\alpha=1}$, becomes, following eq.~\eqref{eq:RenyiEqResultInt},
\begin{align}
\label{eq:vNint}
    \tilde S_1 = \int_{E_{A}, J_A} e^{S_A(E_A, J_A)+S_B(E_i-E_A,J_i-J_A)-S(E_i,J_i)}S_{\min}.
\end{align}
Evaluating by saddles yields the same saddle-point conditions~\eqref{eq:saddlePoints2} except (II.1) and (II.2) are equivalent and (II.3) is forbidden for $\alpha\geq 1$ unless the subsystems are identical in size. The entropy is,
\begin{equation}
\label{eq:Page_nonAbelian_sym}
  \tilde S_1 = S_{\text{min}}(\bar{E}_A,\bar{J}_A,\bar{E}_B,\bar{J}_B) + \Delta\tilde{S}_1,
\end{equation}
where $\Delta\tilde{S}_1$ represents corrections away from the saddle that we will now discuss. 

Physically, we consider corrections in eq.~\eqref{eq:vNint} from three sources: the thermal fluctuation of subsystem energies and spins from the width of $e^{S_A(E_A, J_A)+S_B(E_i-E_A,J_i-J_A)-S(E_i,J_i)}$, the perturbation of the saddle from the structure of $F$, and the perturbation of the saddle from $S_\text{min}$. First, since the thermal fluctuations are of order $\sim \mathcal O(\sqrt N)$, the integration measure in~\eqref{eq:vNint} contributes a factor of $1/2\ln N$ for each integration variable that produces an overall correction of $+\ln N$. Second, the perturbation of the saddle by $F$ depends on the non-universal structure of $F$ but will generally result in area order corrections since $F$ carries a sharp area-order profile in energy. Third, $S_{\text{min}}$, only linear in system size, is ordinarily too small to perturb the saddle governed by the exponential component. However, when $|N_B-N_A| \lesssim\mathcal{O}(\sqrt N)$, the saddles for both $S_A<S_B$ and $S_A>S_B$ contribute to eq.~\eqref{eq:vNint} and $S_{\text{min}}$ turns sharply at the critical point between them. Evaluating such a perturbation is analogous to taking the expectation of the absolute value of a Gaussian random variable, which pulls out a factor of the standard deviation: $\frac{1}{\sqrt{2\pi\sigma^2}}\int_{-\infty}^{+\infty}|x|e^{-x^2/2\sigma^2}dx = \sigma\sqrt{\frac{2}{\pi}}$. For us, the variance is given by susceptibility $\sqrt\Sigma\sim \mathcal{O}(\sqrt N)$. The detailed evaluation was performed in~\cite{murthy_structure_2019} in the context of Abelian charges. Within eq.~\eqref{eq:vNint}, $J$ behaves exactly as an ordinary Abelian charge thus we can directly summon their result.

For the saddle contribution itself, without loss of generality let us assume that $N_A\leq N_B$. First, we found not one, but three formulas for the microcanonical entropy $S$. However, only two of them can apply here. If $J_i\sim\mathcal {O}(N)$, then $\bar J_A\approx J_i\frac{N_A}{N}\sim\mathcal O(N)$ and $S_A = \bar S_A-\ln N$. If $J_i \lesssim \mathcal O(\sqrt N)$, however, thermal fluctuations place a floor on the value of $\bar J_A$ of $\sim \mathcal O(\sqrt N)$. Here $S_A = \bar S_A-3/2\ln N$. Putting together our result becomes,
\begin{align}
\label{eq:vNsym}
    \tilde S_1 = \min\left[\bar S_A,\bar S_B\right] + \mathcal O(\text{Area}) - \delta_{N_A,N_B}\sqrt{\frac{\sum_{mn}\bar\Sigma_{mn}}{2\pi}} 
        -\delta_{J_i,0}\frac{1}{2}\ln N
\end{align}
where $\bar \Sigma$ is the susceptibility matrix at the saddle. We remark, though we have simplified the expression using Kroenecker symbols, $\delta_{N_A,N_B}$ is specific Gaussian-tailed function with width $\sim\mathcal O(\sqrt N)$ and for $\delta_{J_i,0}$ we have not properly handled the case of $\mathcal O (N)\gg J_i \gg \mathcal O(\sqrt N)$. Following the arguments of section~\ref{subsec:charInt}, we expect that for $J\sim\mathcal O( N^p)$, $1>p>1/2$, the logarithmic correction will be $(1-p)\ln N$ which simply interpolates the two cases linearly. The analogous result is expected for the general compact semisimple Lie group case with global Casimir charge $\Lambda_i$, 
\begin{align}
\label{eq:vNsymGen}
    \tilde S_1 = \min\left[\bar S_A,\bar S_B\right] + \mathcal O(\text{Area}) - \delta_{N_A,N_B}\sqrt{\frac{\sum_{mn}\bar\Sigma_{mn}}{2\pi}} 
        -\delta_{\Lambda_i,0}\frac{d-r}{4}\ln N.
\end{align}
Again we expect intermediate scalings of $\Lambda_i$ to linearly interpolate between the two values of $\delta_{\Lambda_i,0}$. 

Ref.~\cite{bianchi_non-abelian_2024} found an asymmetry between subsystems $A$ and $B$ in the Page curve of symmetry resolved-entanglement that originates in the entanglement between $\mathcal A_{\text{invA}}$ and $\mathcal A_{\text{sym}}$\footnote{A similar asymmetry was found in ref.~\cite{yauk_typical_2026} in anyon chains.}. This asymmetry is evidently invisible to our analysis implying that the environment for a local singlet algebra is dominated by the singlet algebra of its spatial complement.

\subsection{Entanglement entropy}
\label{subsec:totalEE}

We now lift the symmetry resolution and consider the entanglement entropy of the full reduced density matrix on subsystem $A$. The full reduced density matrix restores the sums over the magnetic quantum numbers $M$ missing in eq.~\eqref{eq:Renyi_def} and the corresponding Clebsch--Gordan coefficients. The overall expression factors neatly into a kinematic and invariant part analogously to eq.~\eqref{eq:introDecomp},
\begin{equation}
\label{eq:full_renyi_trace}
\Tr_{A}\left[\left(\hat \rho_{A}\right)^\alpha\right] = \sum_{\{J_{a_k}, J_{b_k}\}} \left( \sum_{\{M_{a_k}, M_{b_k}\}} C_{i}^{a_1 b_1} C^{i}_{a_2 b_1} C_{i}^{a_3 b_2} \cdots C^{i}_{a_1 b_\alpha} \right) \left( \sum_{\{E_{a_k}, E_{b_k}\}} \tilde{c}_{i}^{a_1 b_1} \tilde{c}^{i}_{a_2 b_1} \tilde{c}_{i}^{a_3 b_2} \cdots \tilde{c}^{i}_{a_1 b_\alpha} \right).
\end{equation}
One should imagine working this expression from the outside in. As before, the discrete sums approximate continuous integrals to which Laplace's method can be applied. One first finds a saddle-line in the outer sum over $J$'s and then treats the kinematic part and the invariant part, separately. Although the expression is not identical, the evaluation of the invariant part is ultimately equal to that given in section~\ref{subsec:symmEE}. We study the kinematic part in appendices~\ref{app:finiteDens},~\ref{app:zeroDens},~\ref{app:margDens}. 

At a given saddle point, the total trace decomposes into a product of the contribution from the kinematic part (the Clebsch--Gordan coefficients) and the contribution from the invariant part (the reduced overlaps). Crucially, Clebsch--Gordan coefficients only carry, on average, a polynomial dependence on the system size. Therefore, the kinematic part of the expression cannot shift the saddle-point in $J$ of the invariant part, which is exponential in system size. As a result, taking the von Neumann limit ($\alpha \to 1$), the logarithm of this product splits into a simple sum:
\begin{equation}
\label{eq:entropy_split}
S_1 = S_{\text{inv}} + S_{\text{sym}}
\end{equation}
where $S_{\text{inv}}$ is the invariant entanglement entropy $\tilde{S}_1$ determined in Section~\ref{subsec:symmEE}, and $S_{\text{sym}}$ is the entanglement strictly originating from the Clebsch--Gordan kinematics. As derived in the appendices, the kinematic Clebsch--Gordan coefficient contribution $S_{\text{sym}} \approx +\frac{d-r}{4}\ln N = +\frac{1}{2}\ln N$ in all regimes.

By adding this kinematic contribution to the invariant results from the previous section, we obtain the full von Neumann entanglement entropy for the two primary regimes. For the completely unpolarized global state ($J_i = 0$),
\begin{equation}
\label{eq:S1_unpolarized}
S_1 = \min\left[\bar S_A, \bar S_B\right] + \mathcal{O}(\text{Area}) - \delta_{N_A,N_B}\sqrt{\frac{\sum_{mn}\bar\Sigma_{mn}}{2\pi}}.
\end{equation}
For the finitely polarized, or stretched, global state ($J_i \sim \mathcal{O}(N)$),
\begin{equation}
\label{eq:S1_polarized}
S_1 = \min\left[\bar S_A, \bar S_B\right] + \mathcal{O}(\text{Area}) - \delta_{N_A,N_B}\sqrt{\frac{\sum_{mn}\bar\Sigma_{mn}}{2\pi}} + \frac{1}{2}\ln N.
\end{equation}
Interestingly, the positive kinematic correction to entanglement entropy perfectly cancels the negative non-Abelian correction to microcanonical entropy in the zero total spin sector. Above zero total spin, it provides a strictly positive logarithmic correction. 

\section{Discussion}
\label{sec:discussion}

\subsection{Semiclassical spins}
\label{subsec:SSS}
Before reflecting on our results, we should discuss the calculations in appendices~\ref{app:finiteDens},~\ref{app:zeroDens}, and~\ref{app:margDens} which contain core technical material used in the analysis of Clebsch--Gordan coefficients that is not summoned in the main text. We stick to the language of SU(2) for clarity but our results generalize to all compact Lie groups. In section~\ref{subsec:concave} we found that in general 
\begin{align}
\label{eq:disc}
J_a+J_b=J_i+\epsilon
\end{align}
where $\epsilon$ is a strictly non-negative correction caused by thermal fluctuations of order $\mathcal{O}(\sqrt N)$. In fact, that is only true if the subsystems are finite fractions of the whole system. Strictly, $\epsilon$ is of order the square root of the volume of the smaller subsystem and we will return to this caveat shortly. One can imagine, semiclassically, the full system carrying an angular momentum $\vec J_i$ with length $J_i$. The subsystem angular momentum vectors add $\vec J_a + \vec J_b = \vec J_i$. Thermal equilibrium requires further that $\vec J_a, \vec J_b,\vec J_i$ are parallel and thus $J_a+J_b=J_i$, except that thermal fluctuations in the subsystems allow their angular momenta to be slightly oblique resulting in a non-negative correction $\epsilon$. The regime of validity of this semiclassical picture is simple to state: all three spins must be much larger than one\footnote{Or, with dimensions restored, much larger than $\hbar$.}, $\vec J_a, \vec J_b,\vec J_i\gg\mathcal{O}(1)$. The semiclassical behavior of large representations is a robust expectation Lie theory~\cite{perelomov_coherent_1972} notably applied fruitfully in the analysis of conformal field theories~\cite{komargodski_convexity_2013, hellerman_cft_2015} and string theories~\cite{berenstein_strings_2002}. 

A configuration of three large spins may be regarded, borrowing language from the CFT literature~\cite{di_francesco_conformal_1997}, as a \textit{heavy-heavy-heavy} (HHH) configuration\footnote{A large $J$ is heavy, a small $J$ is light.}. There are two other possible configurations: a \textit{light-heavy-heavy} (LHH) configuration of one small spin and two large spins, and a \textit{light-light-light} (LLL) configuration of three small spins. An HHH configuration is generally expected by eq.~\eqref{eq:disc} whenever $J_i\gg \mathcal{O}(1)$. However, if the total spin is very small $J_i\sim\mathcal{O}(1)$, one obtains an LHH configuration as the subsystem spins are no smaller than thermal fluctuations and remain heavy. Our analysis has till now presumed subsystems that are finite fractions of the whole system. One can, however, obtain an LLL configuration in a small subsystem size limit, since $\epsilon$ is is bounded by the size of the smaller subsystem per eq.~\eqref{eq:toRef}. Our analysis, based on coherent states~\cite{alekseev_quantization_1988} and limited to thermodynamically large subsystems, finds that semiclassicalization of heavy weights, i.e. large spin, provides a robust enough self-averaging of Clebsch--Gordan coefficients in both the HHH and LHH configurations to compute the relevant correction to entanglement entropy. This perspective on semiclassicalization suggests two mechanisms by which non-Abelian thermodynamics~\cite{murthy_non-abelian_2023, manzano_non-abelian_2022, yunger_halpern_microcanonical_2016, yunger_halpern_how_2022, yunger_halpern_noncommuting_2020, lasek_numerical_2024, noh_kubo-martin-schwinger_2025, lostaglio_thermodynamic_2017} may be considered truly quantum, and not semiclassical, thermodynamics~\cite{vinjanampathy_quantum_2016, campbell_roadmap_2026}: zero total charge and small (sub)system sizes.

\subsection{Local non-Abelian thermodynamics}
\label{subsec:LQP}
In this paper, as in ref.~\cite{bianchi_non-abelian_2024}, we focused on a symmetry-resolved approach to non-Abelian thermodynamics which we found to be a necessity of studying the non-Abelian ETH and EB. There is a more conventional approach to non-Abelian thermodynamics than the one presented in this paper where instead one focuses directly on the three charges defined by sums of local terms, $\hat J^x, \hat J^y, \hat J^z$~\cite{glorioso_hydrodynamics_2021, majidy_noncommuting_2023}. In this picture, one expects to study a (grand) canonical state, referred to as a non-Abelian thermal state, of the form~\cite{kranzl_experimental_2023, guryanova_thermodynamics_2016, yunger_halpern_microcanonical_2016, lostaglio_thermodynamic_2017, yunger_halpern_noncommuting_2020},
\begin{align}
\label{eq:GGE}
    \hat \rho = Z^{-1}e^{-\beta \hat H - \phi\cdot\hat J}
\end{align}
with $\hat J = (\hat J^x,\hat J^y,\hat J^z)$ and $\phi$ a vector of conjugate thermodynamic potentials. Another density operator was considered in ref.~\cite{noh_kubo-martin-schwinger_2025}, but assumed that $M$ quantum numbers also followed an exponential form,
\begin{align}
\label{eq:nonnGGE}
    \hat \rho = Z^{-1}\sum_a e^{-\beta E_a-\mu M_a-\gamma J_a}\ket{a}\!\bra{a}.
\end{align}
The basic assumption of eq.~\eqref{eq:GGE} is that the global system is an uncertainty-minimizing approximate microcanonical state in the spin components, but without fixing a total spin. Eq.~\eqref{eq:nonnGGE} assumes a system that is in a global state of fixed $J$ and $M$ quantum numbers instead. However, under fundamentally the same assumption we found in section~\ref{subsec:generalRDM} the density operator to be exponential in total spin $J$, but at finite spin density Gaussian in the $z$ spin component,
\begin{align}
\label{eq:nonGGE}
    \hat\rho = Z^{-1}\sum_a e^{ -\frac{(M_a - \bar{M}_A)^2}{2\sigma^2} -\beta E_a -\phi J_a}\ket{a}\!\bra{a}
\end{align}
contradicting the form in eq.~\eqref{eq:nonnGGE}. This does not, however, strictly contradict the information theoretic expectation of exponentiality~\cite{jaynes_information_1957} in $M_a$ overall. As discussed in section~\ref{subsec:generalRDM}, the marginal distribution of $M_a$ is still exponential even if its conditional distribution with respect to fixed $J_a$ and $E_a$ is Gaussian, because the Gaussian tightly concentrates $M_a$ to be a fixed ratio of $J_a$. The discrepancy between eqs.~\eqref{eq:nonnGGE} and~\eqref{eq:nonGGE} implies that $J_a$ and $M_a$ cannot be regarded as independent variables when defining an ensemble, even if they commute.

Nevertheless, the physical systems described in each perspective are the same. As discussed in the introduction, any conventional symmetry\footnote{i.e. represented by unitaries on the Hilbert space} of a finite dimensional system must be described by a compact Lie group. Thus, the symmetry-resolution approach and the component-wise approach to non-Abelian thermodynamics cannot be separated from one another. The difference comes from the choice of the global microcanonical configuration. Our symmetry resolution approach focuses on the eigenstates of commuting observables, $\hat J^2, \hat J^z, \hat H$. Whereas refs.~\cite{kranzl_experimental_2023, guryanova_thermodynamics_2016, yunger_halpern_microcanonical_2016, lostaglio_thermodynamic_2017, yunger_halpern_noncommuting_2020} focus on the quasi-local\footnote{sums of local terms} conserved charges $\hat J^x, \hat J^y, \hat J^z, \hat H$. Both are physically meaningful choices. One possible avenue for bridging the gap would be to repeat our analysis, but placing the system in a global spin coherent state of the kind mentioned in section~\ref{subsec:SSS}. It would interesting to explore more generally the state-dependence of non-Abelian thermodynamics.

There is another implication of density operators in eqs.~\eqref{eq:nonnGGE} and~\eqref{eq:nonGGE}: total spin $J$ is stored locally, despite being defined via sum of non-local terms in $\hat J\cdot\hat J$. Ref.~\cite{murthy_non-abelian_2023} provided a semiclassical argument for the local storage of $J$ quanta in trying to demonstrate that $X^{k}(E,J)$ in eq.~\eqref{eq:su2_traditional} for some $\hat X$ can have a linear dependence on $J$ near $J=0$. It would be interesting to connect the perspectives in future work.

\section{Conclusion}
\label{sec:conclusion}

In this work, we have considered the structure of chaotic eigenstates of a Hamiltonian that is composed of local terms and symmetric under a global compact Lie group. We constructed, via symmetry resolution, a microcanonical ensemble and computed its thermodynamic entropy, which we refer to as non-Abelian microcanonical entropy. We then computed the entanglement entropy and symmetry-resolved entanglement entropy of eigenstates for thermodynamically large subsystems. We found that symmetry-resolved entanglement entropy, via the replica approach, is closely related to the non-Abelian microcanonical entropy. The conventional entanglement entropy differs from the symmetry-resolved entanglement entropy by a universal enhancement logarithmic in volume, sourced in the intrinsic kinematic entanglement encoded in Clebsch--Gordan coefficients. This enhancement settles a conjecture of ref.~\cite{patil_average_2023} in which the authors suggested that non-Abelian symmetries carry a logarithmic signature in entanglement entropy.

The non-Abelian symmetry resolution approach developed in ref.~\cite{bianchi_non-abelian_2024} and utilized in this work is a necessity for defining and studying a non-Abelian ETH or EB; it is implicit in the invocation of the Wigner--Eckart theorem in ref.~\cite{murthy_non-abelian_2023}. Yet, as discussed in section~\ref{subsec:LQP}, there is another approach to non-Abelian thermodynamics~\cite{kranzl_experimental_2023, guryanova_thermodynamics_2016, yunger_halpern_microcanonical_2016, lostaglio_thermodynamic_2017, yunger_halpern_noncommuting_2020}. The difference originates in distinct definitions of the microcanonical ensemble. It would be interesting to bridge this gap in future work.

The picture we have painted of a chaotic system with non-Abelian symmetries largely relies on this semiclassical limit of kinematic degrees of freedom, as developed in appendices~\ref{app:finiteDens}, ~\ref{app:zeroDens}, and~\ref{app:margDens}. In some sense, the kinematic rules encoded by Clebsch--Gordan coefficients represent a small integrable piece imposed on a chaotic system. But in another sense, it obeys an analog of chaos. In the thermodynamic limit, the invariant part of our physical set-up, being described by the ETH, approaches a model of large random matrices~\cite{jindal_generalized_2024}, The kinematic part approaches a classical mechanical system. Both represent an exceptionally strong tendency of true quantum-ness to evaporate in a sufficiently complex system. The analogy between semiclassics and random matrices is very old~\cite{yaffe_large_1982} and of modern relevance, via the ETH, in the holographic literature~\cite{jafferis_matrix_2023, de_boer_principle_2023}.

\section*{Acknowledgements}
S.J. is grateful to Nicole Yunger Halpern and Simeon Hellerman for helpful discussions.
\paragraph{Funding information}
This work is supported by the National Science Foundation grant no. DMR-2047193.
% Authors are required to provide funding information, including relevant agencies and grant numbers with linked author's initials. Correctly-provided data will be linked to funders listed in the \href{https://www.crossref.org/services/funder-registry/}{\sf Fundref registry}.

\begin{appendix}
\numberwithin{equation}{section}

\section{Background: Lie groups}
\label{app:background}

In this appendix we provide a brief tutorial on the relevant basics of compact Lie groups needed to understand this paper. Proper treatment of the subject is available in ref.~\cite{koch_compact_nodate, fulton_representation_2004}. Our basic focus is the introduction and application of complete reducibility and of Schur's lemma. We sketch a proof of the Wigner--Eckart theorem for general compact Lie groups and discuss its relevance to the non-Abelian ETH.

In this appendix, we review complete reducibility in the context of unitary representations of compact Lie groups and use it to establish the general version of the Wigner--Eckart theorem for compact Lie groups. 

\subsection{Terminology: Casimirs, Cartans, and weights}
\label{subapp:terminology}

A Lie group $G$ is a group that is also a differentiable manifold. If one differentiates a Lie group in the neighborhood of the identity element one obtains a Lie algebra $\mathfrak g$ which is a vector space associated with a generally non-associative Lie bracket $[\cdot,\cdot]: \mathfrak g \times\mathfrak g \rightarrow \mathfrak g$. To interpret the elements of $\mathfrak g$ as observables in an associative quantum theory, we then embed $\mathfrak g$ in a universal enveloping algebra $U(\mathfrak g)$ in which the Lie bracket maps to the commutator $[X,Y] = XY-YX$. The core nuance that separates $\mathfrak g$ and $U(\mathfrak g)$ is that $\mathfrak g$ only contains linear combinations of its generators, e.g. $X+Y,\, X-Y,\,$ et cetera, whereas $U(\mathfrak g)$ contains powers of the generators, e.g. $X^2,\, Y^2,\,$ et cetera.

The {center} of an algebra is the set of operators within that algebra that commute with every element of the algebra. For a semi-simple Lie algebra $\mathfrak{g}$, the center of its universal enveloping algebra $U(\mathfrak g)$ is generated by a set of independent operators known as \textit{Casimir operators}. These operators are the direct generalization of the total angular momentum operator $\hat{J}^2$ in $SU(2)$. By Schur's lemma, developed below, the eigenvalues of the Casimir operators distinguish distinct irreps, just as the eigenvalue $J(J+1)$ labels representations of $SU(2)$.

Once a representation is fixed (by specifying the Casimir eigenvalues), we require the analog of the magnetic quantum number $M$ for $\hat{J}^z$ which labels the basis within the representation. To find these labels, we seek a maximal set of mutually commuting generators within the algebra. This abelian subalgebra is known as the \textit{Cartan subalgebra}, $\mathfrak{h}$. The dimension of $\mathfrak{h}$ is the rank of the group, $r$. We can simultaneously diagonalize these operators, and their simultaneous eigenvalues form a vector $\nu = (\nu_1, \dots, \nu_r)$ known as a \textit{weight}:
\begin{equation}
    H_\alpha \ket{\nu} = \nu_\alpha \ket{\nu}, \quad \text{for } H_\alpha \in \mathfrak{h}.
\end{equation}
Thus, a weight $\nu$ is the multidimensional generalization of $M$.

For SU(2) one usually uses the label $J$, rather $J(J+1)$ to denote irreps. This property is generalized by the theorem of highest weight. Every finite dimensional irrep has a \textit{highest weight}, $\Lambda$, which the vector of largest allowed values of $\nu$ in the irrep. The theorem of highest weight states that an irrep is uniquely determined by its highest weight. Thus we label states in an irrep by $\ket{\Lambda,\nu}$. However, we must caution that weights are usually not unique\footnote{Highest weights are always unique, however}. There may be multiple states $\ket{\Lambda,\nu,1} \neq \ket{\Lambda,\nu,2}$. In fact, the multiplicity is generically thermodynamically large. We will return to this point in the discussion of root systems in appendix~\ref{subapp:rootsLadders}.

\subsection{Complete reducibility}
\label{subapp:comRed}

A basic postulate of quantum mechanics is that a $d$-dimensional quantum system lives in a $d$-dimensional Hilbert space $\mathcal{H}$. 

First, one can always choose a unitary representation for a finite-dimensional representation of a compact Lie group, $\expval{\hat U(g^{-1})\hat U(g)} = \expval{\hat U(g)^{\dagger}\hat U(g)} = 1$. Then one considers a proper subspace of the Hilbert space It follows automatically that the orthogonal component of a symmetry-invariant subspace is also symmetry-invariant. That is, for some projector $\hat \Pi$, if $\hat U(g)^\dagger \hat \Pi \hat U(g) = \hat P$ then $\hat U(g)^\dagger \left(\hat 1-\hat \Pi\right) \hat U(g) = \hat 1-\hat \Pi$.

By repeatedly splitting the Hilbert space into symmetry invariant orthogonal components, one will eventually encounter units that contain no nontrivial symmetry invariant proper subspace. These are known as \textit{irreducible representations}, or \textit{irreps}. Then, one can always decompose a Hilbert space into a direct sum over irreps of the symmetry group,
\begin{align}
\label{eq:irreM}
    \mathcal H = \bigoplus_{\{\Lambda\}}\mathcal{H}_{\text{sym}}^{(\Lambda)}.
\end{align}
Generically, however, there will be multiple copies of each irrep with label $\Lambda$. Thus, collecting \textit{irrep multiplicities} we have,
\begin{equation}
\label{eq:hilbertOG}
    \mathcal H = \bigoplus_\Lambda\left(\mathcal H^{(\Lambda)}_{\text{sym}} \otimes \mathcal H^{(\Lambda)}_{\text{inv}}\right)
\end{equation}
where $\mathcal H^{(\Lambda)}_{\text{inv}}$ is the multiplicity space in which symmetry-invariant operators act. 

\subsection{Schur's lemma}
\label{subapp:schurs}

The two parts of Schur's lemma we need can be sketched very briefly. Consider an operator $\hat O$ that is invariant under the action of the symmetry group,
\begin{align}
\label{eq:xInv}
    \hat U(g)\hat O \hat U(g)^{\dagger} = \hat O.
\end{align}
First, consider $\bra{v}\hat O\ket{w}$ where $\ket{v}\in\mathcal H^{(\Lambda')}_{\text{sym}}$ and $\ket{w}\in\mathcal H^{(\Lambda)}_{\text{sym}}$. We can show that the kernel of $\hat O$ when mapping $\mathcal H^{(\Lambda)}_{\text{sym}}$ to $\mathcal H^{(\Lambda')}_{\text{sym}}$ is itself an irrep, 
\begin{align}
\label{eq:noKern}
    \text{if}&\quad \bra{v}\hat O\ket w = 0\quad\forall\ket{v}\in\mathcal H^{(\Lambda')}_{\text{sym}} \nonumber \\
    \text{then}&\quad \bra{v}\hat O \hat U(g)\ket w = \bra{v} \hat U(g)\hat O\ket w = \bra{U(g)^\dagger v}\hat O\ket w = 0.
\end{align}
That is if $\ket w$ gets mapped to $0$, then so does $\hat U(g)\ket w$, which for all $g$ generates an irrep. However, since $\mathcal H^{(\Lambda)}_{\text{sym}}$ is an irrep, the set $\left\{\hat U(g)\ket w\mid g\in G\right\}$ is either the zero vector, $\{0\}$, which implies $\Lambda = \Lambda'$, or the entire irrep $\mathcal H^{(\Lambda)}_{\text{sym}}$. Thus a symmetry invariant operator $\hat O$ cannot mix distinct irreps. Second, in a complex finite dimensional Hilbert space $\hat O$ has at least one eigenvector $\ket{o}$. Then it follows from
\begin{align}
\label{eq:irrepDef}
    \hat O \hat U(g)\ket{o} = \hat U(g)\hat O \ket{o} = o \left(\hat U(g)\ket{o}\right)
\end{align}
that $\hat U(g)\ket{o}$ is also an eigenstate of $\hat O$ with the same eigenvalue. If $\ket o$ lives within an irrep $\mathcal H^{(\Lambda)}_{\text{sym}}$ it follows that every state in $\mathcal H^{(\Lambda)}_{\text{sym}}$ has eigenvalue $o$. Thus, any operator that is invariant under the symmetry group must act as a multiple of the identity on an irrep. 

To summarize, the two parts of Schur's lemma are
\begin{subequations}
\label{eq:SchursLemma}
\begin{equation}
\label{eq:S1}
    \text{(I) A symmetry invariant operator }\hat O\text{ cannot mix distinct irreps}
\end{equation}
\begin{equation}
\label{eq:S2}
    \text{(II) A symmetry invariant operator }\hat O\text{ acts as a constant on an individual irrep}.
\end{equation}
\end{subequations}

\subsection{Wigner--Eckart theorem}
\label{subapp:WET}

We now apply the above tools to state and prove the Wigner--Eckart theorem. The Wigner--Eckart theorem is one of the most powerful theorems for working with angular momentum in quantum systems. Remarkably, it directly extends to arbitrary group representations that satisfy complete reducibility, including all finite dimensional representations of compact Lie groups. We review the theorem and its proof concretely for unitary representations. 

Consider an arbitrary operator living in the Hilbert space and its eigenbasis components,
\begin{equation}
    \hat X = \sum_{ij} \ket{i}X_{ij}\bra{j}.
\end{equation}
For notational purposes, it is convenient to use the canonical isomorphism to treat $\hat X$ as a state,
\begin{equation}
    \hat X \longrightarrow \ket{\ket{X}} \equiv \sum_{ij} X_{ij}\ket{i}\otimes\ket{j} \in \mathcal{H}\otimes\mathcal{H}.
\end{equation}
Just as $\mathcal{H}$ decomposes into irreducible representations of the symmetry group, complete reducibility means $\mathcal{H}\otimes\mathcal{H}$ must decompose as well,
\begin{equation}
    \mathcal{H}\otimes\mathcal{H} \equiv \mathcal H_2 =  \bigoplus_{\Lambda}\left(\mathcal H_2\right)^{(\Lambda)} = \bigoplus_{\Lambda}\left(\mathcal H^{(\Lambda)}_{\text{sym}}\otimes\left(\mathcal H_2\right)^{(\Lambda)}_{\text{inv}}\right)
\end{equation}
along with $\mathcal H_3 \equiv \mathcal{H}\otimes\mathcal{H}\otimes\mathcal{H}$ and all further tensor products of the Hilbert space. Then $\ket{\ket{X}}$ can be written,
\begin{equation}
    \ket{\ket{X}} = \sum_{\Lambda\nu}\ket{\ket{\Lambda \nu }}\ket{\ket{\Lambda X^{(\Lambda)}}} \equiv \sum_{\Lambda \nu}\ket{\ket{\Lambda \nu X^{(\Lambda)}}}.
\end{equation}
Reversing the canonical isomorphism we can define,
\begin{align}
    \ket{\ket{\Lambda X^{(\Lambda)}}} &\longrightarrow \hat X^{(\Lambda)} \nonumber \\
   \ket{\ket{\Lambda \nu X^{(\Lambda)}}} &\longrightarrow \hat X^{(\Lambda, \nu)} \\
   \ket{\ket{X}} =  \sum_{\Lambda\nu }\ket{\ket{\Lambda \nu X^{(\Lambda)}}} &\longrightarrow \hat X = \sum_{\Lambda \nu} \hat X^{(\Lambda, \nu)} \nonumber
\end{align}
where $\hat X^{(\Lambda, \nu)}$ is known as a \textit{tensor operator} and $\hat X^{(\Lambda)}$ is the \textit{reduced tensor operator}. Tensor operators live in irreducible representations of the symmetry group acting on the Hilbert space of operators, rather than the Hilbert space of states. It suffices then to study the properties of such tensor operators just as we focused on states with definite $\Lambda, \nu$ quantum numbers.

Agrawala's trick~\cite{agrawala_wignereckart_1980} to prove the theorem is to think of operator-vector multiplication as a tensor product subject to a particular homomorphism of vector spaces,
\begin{align}
\label{eq:agrawalaTrick}
    \hat X^{\left(\Lambda,\nu\right)}\ket{\Lambda_1\nu_1\Psi_1} \equiv \hat O \Bigl[\ket{\ket{\Lambda \nu X^{(\Lambda)}}}\otimes \ket{\Lambda_1\nu_1\Psi_1}\Bigr]
\end{align}
that maps $\mathcal H_3$ to $\mathcal H$\footnote{Explicitly, $\hat O = \sum_{i}\hat 1\otimes\bra{i}\otimes\bra i$}. We can show that $\hat O$ commutes with the group action\footnote{$\hat O$ is an intertwiner},
\begin{align}
\label{eq:intertwiner}
    \hat U(g)\hat O \Bigl[\ket{\ket{\Lambda \nu X^{(\Lambda)}}}\otimes \ket{\Lambda_1\nu_1\Psi_1}\Bigr] &= \hat U(g)\hat X^{\left(\Lambda, \nu\right)}\ket{\Lambda_1\nu_1\Psi_1} \nonumber \\
    &= \hat U(g)\hat X^{\left(\Lambda, \nu\right)}\hat U(g)^\dagger \hat U(g)\ket{\Lambda_1\nu_1\Psi_1} \\
    &= \hat O \Bigl[\left(\hat U(g)\otimes\hat U(g)^\dagger\right)\ket{\ket{\Lambda \nu X^{(\Lambda)}}}\otimes \hat U(g)\ket{\Lambda_1\nu_1\Psi_1}\Bigr]. \nonumber
\end{align}
Then it follows from the first part of Schur's lemma~\eqref{eq:S1} that for each irrep $\Lambda_2$ living in $\mathcal H_3$ $\hat O$ must map it only to the $\Lambda_2$ irrep living in $\mathcal H$ and from the second~\eqref{eq:S2} that it must be constant on each each irrep. Concretely,
\begin{align}
\label{eq:Odecomp} 
    \hat O = \bigoplus_{\Lambda'}\left(\hat 1^{(\Lambda')}_{\text{sym}}\otimes\hat O^{(\Lambda')}_{\text{inv}}\right).
\end{align}
Closing eq.~\eqref{eq:agrawalaTrick} with a state $\bra{\Lambda_2\nu_2\Psi_2}$,
\begin{align}
\label{eq:breakdown}
    \bra{\Lambda_2\nu_2\Psi_2}\hat X^{\left(\Lambda, \nu\right)}&\ket{\Lambda_1\nu_1\Psi_1} =  \bra{\Lambda_2\nu_2\Psi_2}\hat O \Bigl[\ket{\ket{\Lambda \nu X^{(\Lambda)}}}\otimes \ket{\Lambda_1\nu_1\Psi_1}\Bigr] \nonumber \\
    &= \sum_{\Lambda'\nu'\gamma'}\bra{\Lambda_2\nu_2\Psi_2}\hat O \Bigl[\ket{\Lambda'\nu'\gamma'}\!\bra{\Lambda'\nu'\gamma'}\ket{\Lambda_1\nu_1;\Lambda\nu}\ket{\ket{\Lambda X^{(\Lambda)}}} \ket{\Lambda_1\Psi_1}\Bigr] \\
    &=\sum_{\gamma_2} \bra{\Lambda_2\nu_2\gamma_2}\ket{\Lambda_1\nu_1;\Lambda\nu}\!\bra{\Lambda_2\Psi_2}\hat O^{(\Lambda_2)}_{\text{inv}} \Bigl[\hat\Pi^{(\Lambda_2)}_{\gamma_2}\ket{\ket{\Lambda X^{(\Lambda)}}} \ket{\Lambda_1\Psi_1}\Bigr] \nonumber
\end{align}
where $\gamma_2$ labels the independent copies of $\mathcal H_{\text{sym}}^{(\Lambda_2)}$ that appear in $\mathcal H_{\text{sym}}^{(\Lambda_1)}\otimes\mathcal H_{\text{sym}}^{(\Lambda)}$ and $\hat\Pi^{(\Lambda_2)}_{\gamma_2}$ is a projector onto the corresponding multiplicity space. Consistency with the definition~\eqref{eq:agrawalaTrick}, the action of $\hat O^{(\Lambda')}_{\text{inv}}$ is constrained to map the tensor product of multiplets into a matrix vector product,
\begin{align}
\label{eq:Oinv}
    \hat O^{(\Lambda_2)}_{\text{inv}} \Bigl[\hat\Pi^{(\Lambda_2)}_{\gamma_2}\ket{\ket{\Lambda X^{(\Lambda)}}} \ket{\Lambda_1\Psi_1}\Bigr] = \hat\Pi^{(\Lambda_2)}_{\gamma_2}\hat X^{(\Lambda)}\ket{\Lambda_1\Psi_1}
\end{align}
which immediately yields the general statement of the Wigner--Eckart Theorem,
\begin{align}
\label{eq:genWET}
    \bra{\Lambda_2\nu_2\Psi_2}\hat X^{\left(\Lambda, \nu\right)}\ket{\Lambda_1\nu_1\Psi_1} = \sum_{\gamma_2} \bra{\Lambda_2\nu_2\gamma_2}\ket{\Lambda_1\nu_1;\Lambda\nu}\!\bra{\Lambda_2\Psi_2}\hat\Pi^{(\Lambda_2)}_{\gamma_2}\hat X^{(\Lambda)}\ket{\Lambda_1\Psi_1}.
\end{align}

For the reader accustomed only to the statement for SU(2), we should discuss the role of the sum and the projector $\hat \Pi_\gamma$ in eq.~\eqref{eq:genWET}. In the tensor product of irreps $\mathcal H_{\text{sym}}^{(\Lambda_1)}\otimes\mathcal H_{\text{sym}}^{(\Lambda)}$ an irrep $\mathcal H_{\text{sym}}^{(\Lambda_2)}$ will appear multiple times in the direct sum unless the symmetry group is simply reducible as in the case of SU(2). This multiplicity creates an irrep or outer multiplicity space within the tensor product for which $\gamma_2$ labels an arbitrary basis. Elsewhere, unless otherwise mentioned, we will suppress the outer multiplicity as it does not impact our analysis. The dependence of the Clebsch--Gordan coefficients on $\gamma$ is nontrivial. Physically, $\gamma$ are eigenvalues of some operator(s) $\Gamma$ that live within $U(\mathfrak g)\otimes U(\mathfrak g)$ but not $U(\mathfrak g)$. However, the physical multiplicity space, labeled by $\Psi$, is much larger than elements of $U(\mathfrak g)\otimes U(\mathfrak g)$ and there is no reason that the labels $\Psi$ should carry eigenvalues of $\Gamma$. For example, if energy eigenvectors carried $\Gamma$ eigenvalues, the Hamiltonian would have a larger symmetry group than merely $G$, which we do not suppose. Thus, once we pull out a Clebsch--Gordan coefficient from eq.~\eqref{eq:breakdown} we obtain a label $\gamma$ onto which we must project $\bra{\Lambda_2\Psi_2}$.

\subsection{Symmetry resolution and entropy}
\label{subapp:symm}

The same ideas underlie non-Abelian symmetry resolution. A more thorough discussion of this topic is available in~\cite{bianchi_non-abelian_2024}. In general, to generate a complete basis of physical states, one needs a complete set of commuting operators whose eigenvalues may be used as labels. Then to get a complete basis of energy eigenstates we want a complete set of operators that commute with the Hamiltonian. Including a global SU(2) symmetry provides two labels, $J$ and $M$, for eigenstates of $\hat J^2$ and $\hat J^z$\footnote{The exact eigenvalue of $\hat J^2$ is $J(J+1)$ as usual.} indicating a complete labeling by $\ket{i} = \ket{J_iM_iE_i}$. 

The state $\ket{JME}$ is associated to the Hilbert space decomposition
\begin{equation}
\label{eq:hilbert}
    \mathcal H = \bigoplus_J\left(\mathcal H^{(J)}_{\text{sym}} \otimes \mathcal H^{(J)}_{\text{inv}}\right)
\end{equation}
where $\ket{JM}\in\mathcal H^{(J)}_{\text{sym}}$, $\ket{JE}\in\mathcal H^{(J)}_{\text{inv}}$ and $\ket{JM}\ket{JE}\equiv\ket{JME}\in \mathcal H$. The decomposition applies analogously to subsystems $A$ and $B$,
\begin{eqnarray}
 \mathcal H_A = \bigoplus_{J_A}\left(\mathcal H^{(J_A)}_{\text{symA}} \otimes \mathcal H^{(J_A)}_{\text{invA}}\right)\nonumber
\\
 \mathcal H_B = \bigoplus_{J_B}\left(\mathcal H^{(J_B)}_{\text{symB}} \otimes \mathcal H^{(J_B)}_{\text{invB}}\right). \label{eq:subhilbert}
\end{eqnarray}
Simple reduciblity of SU(2) decomposes the overlap of full system and subsystem eigenstates into Clebsch--Gordan coefficients and reduced overlaps,
\begin{eqnarray}
    &&\bra{i}\ket{ab}\equiv\bra{J_iM_iE_i}\ket{J_aM_aE_a;J_bM_bE_b} \nonumber \\ &&\quad= \bra{J_iM_i}\ket{J_aM_a;J_bM_b}\bra{J_iE_i}\ket{J_aE_a;J_bE_b} \nonumber \\ &&\quad\quad\equiv C^{J_iM_i}_{J_aM_a;J_bM_b}\tilde c^{J_iE_i}_{J_aE_a;J_bE_b} \equiv C^{i}_{ab}\tilde c^{i}_{ab}. \label{eq:simpleReducibility}
\end{eqnarray}

The mechanics of this decomposition follow from the structure of the $\mathfrak{su}$(2) algebra and Schur's lemma. The center of $U(\mathfrak{su}\text{(2)})$ is generated by the total spin $\hat J^2\equiv\hat J\cdot\hat J$, its only Casimir. Since $\hat J^2$ is invariant under the symmetry algebra, it obeys  eq.~\eqref{eq:Odecomp}. Consequently, the total Hilbert space $ \mathcal H $ decomposes into a direct sum of orthogonal subspaces, each labeled by a fixed quantum number $ J $. Each of these fixed-$J$ subspaces is itself a tensor product space, whose basis states are labeled by two sets of quantum numbers: the eigenvalue $ M $ of $ \hat{J}^z $, which specifies the state's orientation within the SU(2) representation, and additional labels that distinguish any remaining degenerate states, which in our case is just the energy $ E $.

To generalize, we swiftly introduce the language of operator algebras. All that follows is a consequence of repeated application of Schur's lemma. $\mathcal K$ is the set of bounded linear functions from $\mathcal H$ to itself. For any operator subalgebra\footnote{Operator subalgebras are closed under matrix addition and multiplication, scalar multiplication, and contain the additive and multiplicative identities.} $\mathcal A_*\subset \mathcal K$, one may define an associated restricted Hilbert space $\mathcal{H}_*$. For example, the SU(2) algebra $\mathcal A_\text{sym}$ generated by $ \hat J^x, \hat J^y, $ and $ \hat J^z $ is associated to the direct sum over irreducible representations $\mathcal H_{\text{sym}} = \bigoplus_J \mathcal H^{(J)}_{\text{sym}}$. We stress that $\mathcal{H}_*$ is not a subspace of $\mathcal H$; it is the space on which the irreducible representations of $\mathcal A_*$ act. The set of all operators in $\mathcal K$ that commute with all elements of a given subalgebra, $\mathcal A_*$, is known as its commutant\footnote{This is a reflexive relationship, $\mathcal A_*'' = \mathcal A_*$.}, $\mathcal A_*'$. The commutant of a subalgebra is also a subalgebra. The commutant of $\mathcal A_{\text{sym}}$ is $\mathcal A_{\text{sym}}'=\mathcal A_{\text{inv}}$, the set of all symmetry invariant operators. $\mathcal{A}_\text{inv}$ is associated to $\mathcal H_{\text{inv}} = \bigoplus_J \mathcal H^{(J)}_{\text{inv}}$. The center of a subalgebra is its intersection with its commutant, $\mathcal Z_* \equiv \mathcal A_*\cap\mathcal A_*'$.

Then from a given subalgebra, $\mathcal A_*$, a decomposition of $\mathcal H$ is given by first labeling a basis, $r$, for the action of its center, then fixed-$r$ bases labeled by $s$ and $t$, for the subalgebra and its commutant, respectively. In the SU(2) example, the center was given a basis $r\equiv J$, and $\mathcal A_{\text{sym}}$ and $\mathcal A_{\text{inv}}$ were given $s\equiv M$ and $t\equiv E$ bases, respectively. If $\mathcal Z_*$ contains only multiples of the identity, then the $r$ labels are empty and this decomposition reduces to a tensor product. This is the case for $\mathcal A_A$, the algebra of operators that live on subsystem $A$ and its commutant $\mathcal A_B$: $\mathcal H = \mathcal H_A \otimes \mathcal H_B$.

Starting from a (possibly mixed) state, represented by a density operator $\hat\rho$ on $\mathcal H$, one may define a reduced state $\hat\rho_*$ on $\mathcal{H}_*$ for any subalgebra of $\mathcal A_*\subset \mathcal K$ and a projector $\hat{\Pi}_*: \hat\rho \mapsto \hat\rho_*$ given by,
\begin{equation}
\label{eq:proj}
    \hat\rho_* \equiv \hat{\Pi}_*[\hat\rho] \equiv \sum_{rss't}{ \ket{rs} \left(\bra{rst}\!\hat\rho\!\ket{rs't}\right)\bra{rs'}}.
\end{equation}
For subalgebras such as $\mathcal{A}_A$ with a trivial center, $\hat{\Pi}$ reduces to the partial trace over the complementary subsystem,
\begin{equation}
\label{eq:projA}
    \hat\rho_A \equiv \hat{\Pi}_A[\hat\rho] \equiv \sum_{aa'b} \left(\bra{ab}\!\hat\rho\!\ket{a'b}\right)\ket{a}\!\bra{a'} \equiv \text{Tr}_B[\hat\rho].
\end{equation}

However, in many cases of physical interest the relevant subalgebra is not the full subsystem of operators local to $A$, but only those invariant under the global symmetry. For this purpose, one would study $\hat{\Pi}_{\text{invA}}$,
\begin{eqnarray}
\label{eq:projInvA}
    && \quad\hat{\rho}_{\text{invA}} = \hat{\Pi}_{\text{invA}}[\hat\rho] \nonumber \\
    && =\!\!\!\!\!\! \sum_{\substack{J_aM_aE_aE_a' \\J_bM_bE_b}} \!\!\!\!\!\!   \left(\bra{J_aM_aE_a;J_bM_bE_b}\!\hat\rho\!\ket{J_aM_aE_a';J_bM_bE_b}\right) \nonumber \\[-3.3ex] 
    && \hspace*{12em} 
    \times\ket{J_aE_a}\!\bra{J_aE_a'}
\end{eqnarray}
which merely strips out the physically irrelevant internal $M$ labels. 

This generalization of a reduced density matrix to general subalgebras extends the core properties of the usual definition in terms of the partial trace. Most crucially, $\hat\rho$ and $\hat\rho_*$ compute the same correlation functions for elements of $\mathcal{A}_*$. Remarkably, it also extends the equality of entanglement entropy of complementary tensor product factors to commutant subalgebras. If $\hat\rho$ is a pure state and $\mathcal{A}_*$ and $\mathcal{A}_{**}$ are commutants, then
\begin{equation}
\label{eq:complementarity}
    S_*\equiv -\text{Tr}\left[\hat\rho_*\text{ln}\hat\rho_*\right] = -\text{Tr}\left[\hat\rho_{**}\text{ln}\hat\rho_{**}\right] \equiv S_{**}.
\end{equation}
In fact, the much stronger statement that all non-zero eigenvalues of $\hat\rho_*$ and $\hat\rho_{**}$ are equal also holds. An external observer probing $\mathcal A_*$ has access to exactly the same information an observer probing $\mathcal{A}_{**}$. Thus, the entropies of the reduced states defined through the projector~\eqref{eq:proj} may be regarded as true entropies of entanglement. 

\section{The ETH replica trick}
\label{app:replicas}

The general postulate of the ETH, which we refer to as the MBBC, is that chaotic Hamiltonians behave, within small energy windows, as random matrices~\cite{deutsch_quantum_1991, pappalardi_microcanonical_2023, richter_eigenstate_2020, foini_eigenstate_2019, foini_eigenstate_2019-1, wang_emergence_2023} and consequently, their eigenstates are sampled from a corresponding eigenvector ensemble. The basic framework we employ is developed in ref.~\cite{jindal_generalized_2024}. In this section we summarize the basic calculation rule of this reference for the EB.

In general one will encounter a moment of the form
\begin{equation}
  \overline{
    \tilde c^{\,i_1}_{a_1 b_1}\,
     \tilde c^{\,i_2}_{a_2 b_2}\,\cdots\,
    \tilde c^{\,i_n}_{a_n b_n}
  }
  \label{eq:generic_product}
\end{equation}
where the overline denotes the expectation with respect to random unitary rotations within the microcanonical subspace implied by the MBBC. The MBBC fixes such quantities up to an overall exponential in the relevant entropies and a smooth function of the $E$'s and $J$'s. The
generic product \eqref{eq:generic_product} takes the form
\begin{equation}
\label{eq:product_scaling}
  \overline{ \prod_{v=1}^{n} \tilde c^{\,i_v}_{a_v b_v}}\;\sim\;e^{-n\,S(E,J) -n_{A(B)}S_{A(B)}(E_A,J_A)}F(E_{i_1},J_{i_1},\dots),
\end{equation}
where the integers $n,\,n_{A(B)}$ are determined entirely by combinatorial rules, and $F$ represents the leftover smooth dependence of the moment on the energies and spins. We now present the combinatorial rules.

Given a bare product of the form \eqref{eq:generic_product}, we first specify
which indices are equal and non-equal (i.e. which of the $i$'s, $a$'s, or $b$'s coincide). Then one computes a contribution from every partition of the full set of indices into identical pairs. If there are an odd number of a given index no such partition exists and the moment is always zero. Consider the following examples. The mean of a single overlap is always zero, $\overline{\tilde c^i_{ab}}=0$, because there is an odd number, one, of each index. The modulus squared overlap $\overline{\tilde c^i_{ab}\tilde c^{ab}_i}$ is non-zero and has exactly one contribution from the unique pairing of each $a$, $b$, and $i$ index. The covariance of two overlaps is 
\begin{align}
\label{eq:covri}
    \overline{\tilde c^i_{ab}\tilde c^{a'b'}_{i'}} - \overline{\tilde c^i_{ab}}\; \overline{\tilde c^{a'b'}_{i'}} = \delta_{aa'}\delta_{bb'}\delta_{ii'}\overline{\tilde c^i_{ab}\tilde c^{ab}_i} - 0\cdot0.
\end{align}
where we have seen that in $\overline{\tilde c^i_{ab}\tilde c^{a'b'}_{i'}}$ e.g. $a$ and $a'$ can only pair if they are equal, hence the Kronecker symbol $\delta_{aa'}$ and each mean overlap, e.g. $\overline{\tilde c^i_{ab}}$, is zero. We also should make a brief note about time reversal symmetry. To accommodate broken time reversal symmetry, one should keep bra indices upstairs, ket indices downstairs, and in the contraction of indices only count pairings between up- and downstairs indices. However, in a time reversal symmetric phase, one can pair bra with bra and ket with ket.

\begin{figure}[htbp]
    \centering
    \begin{subfigure}[b]{0.48\textwidth}
        \centering
        \includegraphics[trim={2cm 2.5cm 2.3cm 2cm}, clip, width=\textwidth]{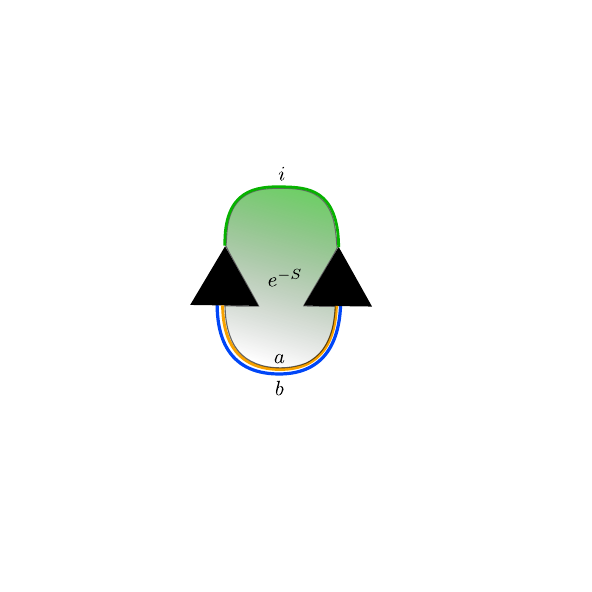}
        \caption{}
        \label{fig:1replica}
    \end{subfigure}
    \hfill 
    \begin{subfigure}[b]{0.48\textwidth}
        \centering
        \includegraphics[width=\textwidth]{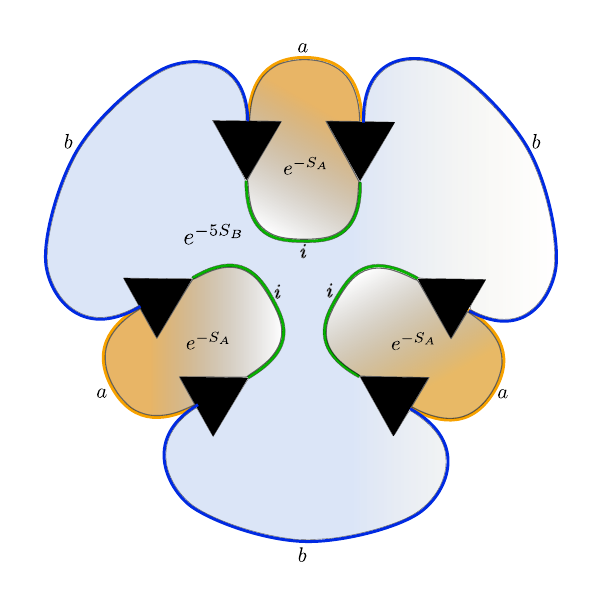}
        \caption{}
        \label{fig:3replica}
    \end{subfigure}

    \caption{Diagrammatic calculus of the ergodic bipartition. $\tilde c$ vertices are black triangles and eigenstate pairings are colored lines. The interior of each loop is shaded; leading order contributions to a monomial are always planar and admit a planar shading. (a) Visualization of the unique diagrammatic contribution to $\overline{\vert\tilde c^{i}_{ab}\vert^2}$. This diagram contains a single composite loop on both subsystems. Since the number of legs in the loop is $2$, the overall normalization gets $2-1=1$ factor of $S$. (b) Visualized is an example of one diagram that contributes to the calculation of the 3rd Renyi entropy of an eigenstate. Under the chosen pairing of indices there are 3 loops on subsystem $A$ of length $2$ and $1$ loop on subsystem $B$ of length $6$. The overall normalization is then $e^{-3S-2S_B}$.}
    \label{fig:replicas}
\end{figure}

Once a pair partition is determined, a graphical calculus is expedient to determine the contribution of that pairing. Treating each $\tilde c$ is a vertex and each pairing as a line that connects vertices, a given term decomposes into loops. There are two types of loops:
\begin{itemize}
  \item loops built from subsystem-$A$ indices $a$ and full system indices $i$,
  \item loops built from subsystem-$B$ indices $b$ and full system indices $i$.
\end{itemize}
A loop that lives in subsystem $A$ and passes through
$L_A$ factors contributes
\begin{equation}
  \sim \exp\!\big[-(L_A - 1)\,S_A\big]\,,
\end{equation}
and similarly for loops in $B$ with $S_B$. Lastly, by convention, factors of $S_A$ and $S_B$ are combined into $S_A + S_B \approx S$ and the approximation error is absorbed into the window function $F$. The evaluation of these loops is demonstrated in figure~\ref{fig:replicas}. We present numerical evidence for the non-Abelian ergodic bipartition in figure~\ref{fig:numerics}.

\section{Thermodynamic correction at finite and zero density}
\label{app:WeylGeneral}

Setting up the Weyl integral, Weyl dimension, and Weyl character formulas fully requires far more time than this appendix can justifiably demand of the reader\footnote{See refs.~\cite{koch_compact_nodate, fulton_representation_2004} for a proper treatment.}. In the interest of both comprehensibility and expedience we will state the formulas precisely and only remark on the definitions we need to understand the results. Let $G$ be a compact, connected, semisimple Lie group of rank $r$ and dimension $d$, with maximal torus $T$, Weyl group $W$, and a set of positive roots $\Delta_+$ (discussed in appendix~\ref{subapp:rootsLadders}). The Weyl denominator is defined as
\begin{align}
\label{eq:weylDenom}
    \Delta(\psi) \equiv \prod_{\alpha \in \Delta_+} \left(e^{i\alpha(\psi)/2} - e^{-i\alpha(\psi)/2}\right).
\end{align}
The Weyl integral formula for a class function $f(g)$ reduces integrals over the group to integrals over the maximal torus,
\begin{align}
\label{eq:weylIntGeneral}
    \int_G dg\, f(g) = \frac{1}{|W|} \int_T d\psi\, |\Delta(\psi)|^2 f(e^{i\psi}).
\end{align}
The Weyl dimension formula states that the dimension of an irrep $D_\Lambda$,
\begin{align}
\label{eq:weylDimGeneral}
    D_\Lambda = \prod_{\alpha\in\Delta_+}\frac{\left(\Lambda+\varrho\right)\cdot\alpha}{\varrho\cdot\alpha}
\end{align}
and the Weyl character formula for the irreducible representation of highest weight $\Lambda$ is,
\begin{align}
\label{eq:weylCharGeneral}
    \chi_\Lambda(e^{i\psi}) = \frac{\sum_{w \in W} \det(w) e^{iw(\Lambda+\varrho) \cdot \psi}}{\Delta(\psi)}
\end{align}
where $\varrho$ is the Weyl vector. The Weyl dimension formula implies $D_\Lambda\sim\Lambda^{|\Delta_+|}$.

Following the derivation of eq.~\eqref{eq:rochester2}, the density of multiplets $\tilde\rho(E, \Lambda)$ is found by integrating the Abelian partition function against the character. Using eq.~\eqref{eq:weylIntGeneral} and eq.~\eqref{eq:weylCharGeneral}, the integrand contains the combination,
\begin{align}
\label{eq:integrandWeyl}
    |\Delta(\psi)|^2 \chi_\Lambda^*(e^{i\psi}) = \Delta(\psi)^* \sum_{w \in W} \det(w) e^{-iw(\Lambda+\varrho) \cdot \psi}.
\end{align}
In eq.~\eqref{eq:sinAdd} applying sine addition rules to the character formula extracted a factor of $e^{-i\phi/2}$. The same logic applies here via the factor of $\varrho$ in eq.~\eqref{eq:integrandWeyl} and we find,
\begin{align}
\label{eq:integrandPart}
    |\Delta(\psi)|^2 \chi_\Lambda^*(e^{i\psi}) \sim \Delta(\psi)^*e^{-i\varrho \cdot \psi}e^{-i\Lambda \cdot \psi} + \text{reflections}.
\end{align}
Then we can focus on our three cases of interest.

At finite density, the saddle point occurs at $\bar{\psi} \neq 0$. The prefactor $\Delta(\psi)^*e^{-i\varrho \cdot \psi}$ evaluates to an $\mathcal{O}(1)$ constant. Accounting for the expected thermal fluctuation $\delta\psi \equiv \psi-\bar\psi \sim N^{-\frac{1}{2}}$, the Gaussian integration over the $r$ torus variables yields the standard Abelian suppression,
\begin{align}
\label{eq:finiteDensity}
    S(E, \Lambda) = \bar S - \frac{1}{2} \ln N - \frac{r}{2} \ln N + \mathcal{O}(1).
\end{align}
Then at exactly zero density the saddle point again occurs at the origin, $\bar{\psi} = 0$, or at any other zeros of the Weyl denominator. Near such a saddle point, the prefactor $\Delta(\psi)^*e^{-i\varrho \cdot \psi}$, once averaged over Weyl reflections, is a homogeneous polynomial of degree $2\vert\Delta_+\vert = d-r$ in the variable $\delta\psi$ which follows from the Weyl dimension formula. Then $\delta \psi \sim N^{-1/2}$ implies an additional polynomial suppression of $(N^{-1/2})^{d-r}$. Therefore, the non-Abelian microcanonical entropy at zero density receives an extra logarithmic correction,
\begin{align}
\label{eq:zeroDensity}
    S(E, 0) = \bar S - \frac{1}{2} \ln N  - \frac{r}{2}\ln N - \frac{d-r}{2} \ln N + \mathcal{O}(1) = \bar S - \frac{1}{2} \ln N  - \frac{d}{2} \ln N + \mathcal{O}(1).
\end{align}
When $\Lambda\sim\mathcal O (\sqrt N)$, $\bar\psi \sim\mathcal O (N^{-1/2})$ contributing a polynomial suppression of only $\vert\Delta_+\vert = (d-r)/2$, yielding,
\begin{align}
\label{eq:margDensity}
    S(E, 0) = \bar S - \frac{1}{2} \ln N  - \frac{r}{2}\ln N - \frac{d-r}{4} \ln N + \mathcal{O}(1) = \bar S - \frac{1}{2} \ln N  - \frac{d+r}{4} \ln N + \mathcal{O}(1).
\end{align}
Carefully reviewing the logic of the above integrals will reveal that intermediate scalings, $\Lambda\sim\mathcal{O}(N^p)$, interpolate the entropy linearly.
\section{Entanglement correction at $\Lambda\sim \mathcal O(N)$}
\label{app:finiteDens}

In this appendix, we calculate the necessary asymptotics for Clebsch--Gordan coefficients of general compact semi-simple Lie groups to compute the logarithmic correction to entanglement entropy at finite highest weight (i.e. spin) density. There is a physical expectation that highest weight states in the large representation limit for arbitrary Lie algebras map to a system of non-interacting bosons~\cite{inonu_contraction_1953}. We find that when the three representations are in the \textit{stretched} configuration (e.g. two parallel spins adding to a third spin) this expectation manifests in a Gaussian form for the Clebsch--Gordan coefficients. For our derivation we'll need to introduce the roots systems of compact semisimple Lie groups.
\subsection{Roots and Ladders}
\label{subapp:rootsLadders}

The core ingredient used to compute the required Clebsch--Gordan asymptotics are \textit{root systems}. Root systems are the direct generalization of the ladder operators, $\hat{J}^+$ and $\hat{J}^-$, that arise in the study of angular momentum.

\begin{figure}[htbp]
    \centering
    \includegraphics[width=0.6\linewidth]{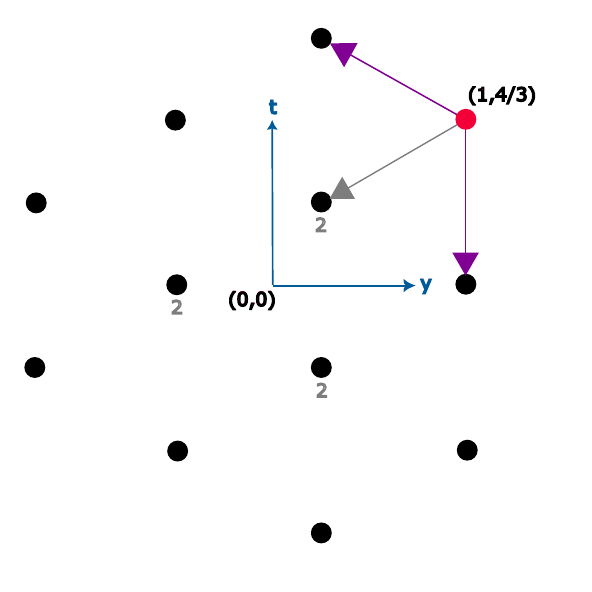}
    \caption{(2,1) irrep of SU(3) with highest weight isospin $T=1$ and hypercharge $Y=4/3$. The highest weight state is shown in red, but notice that a state with larger isospin than the highest weight state also exists in the irrep; a choice of highest weight state and associated roots is fundamentally conventional. Lattice sites indicate allowed weights. The three interior weights each have multiplicity 2. Exterior weights are always multiplicity one.  Roots are unit lattice translations and the 3 negative roots of SU(3) are drawn as arrows and the two simple roots are colored purple. Any site in the lattice can be reached by an application of simple negative roots onto the highest weight state. However, to probe the multiplicity space of the interior states, one needs to use the third negative root as well.}
    \label{fig:Dynkin}
\end{figure}

A root system is a set of non-commuting operators that generate translations along the lattice of weights in an irrep. Roots live in the complexification of the Lie algebra, $\mathfrak g_{\mathbb C} \equiv \mathbb C \otimes \mathfrak g$, i.e. elements of the Lie algebra but with complex coefficients. A \textit{positive root} (i.e. a raising operator) is a root that annihilates the highest weight state. A \textit{simple root} is a positive root that cannot be written as a sum of two other positive roots. In essence, simple roots correspond to a basis of translations along the weight lattice. Analogously, one defines negative roots (and negative simple roots) that move in the opposite directions of positive roots as the hermitian conjugate of the positive roots. Schematically,
\begin{align}
\label{eq:rootScheme}
    \hat E_{-\alpha} \ket{\Lambda,\nu}\propto\ket{\Lambda,\nu-\alpha}
\end{align}
where $\hat E_{-\alpha}$ is a negative root. The Poincar\'{e}--Birkhoff--Witt (PBW) theorem provides a precise procedure to generate the full basis of states in an irrep by applying a sequence lowering operators to the highest weight state. We assign an arbitrary ordering to the set of positive roots $\Delta_+$. Then, a basis for the representation can be labeled by a vector of root occupation numbers, $L = \{L_\alpha\}_{\alpha \in \Delta_+}$:
\begin{align}
\label{eq:pbwBasis}
    \ket{\Lambda L} \propto \left( \prod_{\alpha \in \Delta_+} (\hat E_{-\alpha})^{L_\alpha} \right) \ket{\Lambda\Lambda}
\end{align}
where the product enforces the chosen ordering and the weight is given by $\nu = \Lambda - \sum_{\alpha \in \Delta_+} L_\alpha \alpha$. This basis has many convenient properties we will exploit including that every element of this basis by construction has a definite weight. However, there are two issues that must be pointed out. First, the basis is highly redundant; we must arbitrarily discard paths that lead to linearly dependent states. Second, the basis is not orthogonal. This second problem, though apparently troublesome, we will see evaporate in the large representation limit. We illustrate the logic of root systems for an SU(3) irrep in fig.~\ref{fig:Dynkin}.

The algebraic relations of the root system are derived from those of the parent Lie algebra and can be summarized\footnote{Note that physicists usually include a factor of $2$ in front of $\hat H_\alpha$ in eq.~\eqref{eq:commOne} which allows weights such as $M$ to come in half-integral increments. Since the Clebsch--Gordan coefficients are the same either way we will omit this factor to simplify expressions.},
\begin{subequations}
\label{eq:commRoots}
\begin{equation}
\label{eq:commOne}
    [\hat E_{+\alpha},\hat E_{-\beta}] = \delta_{\alpha\beta}\hat H_{\alpha} + (1-\delta_{\alpha\beta})f_{\alpha,-\beta}\hat E_{\alpha-\beta}
\end{equation}
\begin{equation}
\label{eq:commTwo}
    [\hat E_{+\alpha},\hat E_{+\beta}] = f_{\alpha,\beta}\hat E_{\alpha+\beta}
\end{equation}
\end{subequations}
where $\hat H_{\alpha}$, known as the \textit{coroot}, is the element of the Cartan subalgebra parallel to the root $\alpha$ and the structure constants $f_{\alpha,\beta}$ evaluate to zero if their subscripts do not sum to a valid root in the algebra. It must be emphasized that the coroot $\hat H_{\alpha}$ is usually a superposition of basis elements of the Cartan subalgebra. That is a single lowering operator $\hat E_{-\alpha}$ can reduce multiple elements of the weight vector by either and integral or half-integral increment. The root system of a semisimple Lie algebra of dimension $d$ and rank $r$ has $\vert\Delta_+\vert = (d-r)/2$ positive roots and $r$ simple roots.

With the relations~\eqref{eq:commRoots} it becomes possible to compute the non-orthogonality of distinct PBW basis elements. Consider the overlap,
\begin{align}
\label{eq:gramEl}
    \bra{\Lambda L}\ket{\Lambda L'} \propto \bra{\Lambda\Lambda}\left( \prod_{\alpha \in \Delta_+} (\hat E_{+\alpha})^{L_\alpha} \right) \left( \prod_{\alpha \in \Delta_+} (\hat E_{-\alpha})^{L'_\alpha} \right) \ket{\Lambda\Lambda}
\end{align}
Our strategy is to annihilate positive roots $\hat E_{+\alpha}$ against $\ket{\Lambda\Lambda}$ collecting generated commutators along the way. From a commutation, either a coroot, $\hat H_{\alpha}$ may be generated or another root, $\hat E_{\alpha-\beta}$. Each root operator carries a weight $\sim\mathcal{O}(\sqrt{\Lambda})$ since a pair of them can generate a coroot whose highest eigenvalue is $\Lambda_\alpha$. Therefore, commuting two roots to create a single other root reduces the total weight of the term by $\sim \frac{1}{\sqrt{\Lambda}}$. However, any unpaired root operators will annihilate the highest weight state. Thus, the largest off-diagonal contribution comes from pairs of generated roots leading to a highest possible overlap of $\sim\mathcal{O}({\Lambda}^{-1})$. On its own, this could be sufficient to alter our result but, in fact, the condition is even stronger. Two distinct paths separated by $k(L,L')$ commutators will have overlap $\sim\mathcal{O}({\Lambda}^{-k(L,L')})$. Given there are only an $2\vert\Delta_+\vert \sim \mathcal{O}(1)$ number of commutator types, but $\text{poly}(\Lambda)$ states in the weight space, typically $k(L,L')\sim\text{poly}(\Lambda)$ and the basis is exponentially close to diagonal\footnote{at least up to blocks containing an $\mathcal{O}(1)$ number of a states each}.

\subsection{Asymptotics of stretched Clebsch--Gordan coefficients}
\label{subapp:calculation}

Now we are set up to calculate Clebsch--Gordan coefficients in the stretched irrep. Our starting point is the realization that the highest weight state always decomposes as a tensor product,
\begin{equation}
\label{eq:tensorFAX}
    \ket{\Lambda\Lambda} = \ket{\lambda\lambda}\otimes\ket{\mu\mu},\quad \mu + \lambda = \Lambda.
\end{equation}
We can interpret $\ket{\lambda\lambda}$ as living in subsystem A and $\ket{\mu\mu}$ in subsystem B. Then we wish to compute,
\begin{align}
\label{eq:stretchClebsch--Gordan coefficient}
    C^{\Lambda L}_{\lambda l;\mu m} \equiv \bra{\Lambda L}\ket{\lambda l;\mu m}.
\end{align}
Following eq.~\eqref{eq:pbwBasis}, we can write,
\begin{align}
\label{eq:pathOne}
    \ket{\Lambda L} = \frac{1}{\sqrt{N_{\Lambda L}}}\left(\prod_{\alpha \in \Delta_+}(\hat E_{-\alpha})^{L_\alpha}\right)\ket{\Lambda,\Lambda}.
\end{align}
Then,
\begin{align}
\label{eq:start}
    C^{\Lambda L}_{\lambda l;\mu m} = \frac{1}{\sqrt{N_{\lambda l}N_{\mu m}N_{\Lambda L}}}\bra{\Lambda,\Lambda}\left(\prod_{\alpha \in \Delta_+}(\hat E_{-\alpha})^{L_\alpha}\right)^\dagger\left(\prod_{\alpha \in \Delta_+}(\hat E_{-\alpha}^{(A)})^{l_\alpha}\right) \left(\prod_{\alpha \in \Delta_+}(\hat E_{-\alpha}^{(B)})^{m_\alpha}\right)\ket{\lambda,\lambda}\ket{\mu,\mu}.
\end{align}
The roots, being elements of the complexified lie algebra, are subject to the Leibniz rule,
\begin{align}
\label{eq:leibniz}
    \hat E_{\alpha} = \hat E_{\alpha}^{(A)}\otimes\mathbb{I}^{(B)} + \mathbb{I}^{(A)}\otimes\hat E_{\alpha}^{(B)}.
\end{align}
Therefore we can factorize the total lowering operators into binomial splits where $l_\alpha + m_\alpha = L_\alpha$,
\begin{align}
\label{eq:splits}
    \left(\prod_{\alpha \in \Delta_+}(\hat E_{-\alpha})^{L_\alpha}\right)^\dagger = \sum_{\text{splits}} \left(\prod_{\alpha \in \Delta_+}(\hat E_{-\alpha}^{(A)})^{l_\alpha}\right)^\dagger\otimes \left(\prod_{\alpha \in \Delta_+}(\hat E_{-\alpha}^{(B)})^{m_\alpha}\right)^\dagger.
\end{align}
Then we can rewrite eq.~\eqref{eq:start} as,
\begin{align}
\label{eq:restart}
    C^{\Lambda L}_{\lambda l;\mu m} = \frac{1}{\sqrt{N_{\lambda l}N_{\mu m}N_{\Lambda L}}}\sum_{\text{splits}}&\left(\bra{\lambda,\lambda}\left(\prod_{\alpha \in \Delta_+}(\hat E_{-\alpha}^{(A)})^{l_\alpha}\right)^\dagger\left(\prod_{\alpha \in \Delta_+}(\hat E_{-\alpha}^{(A)})^{l_\alpha}\right)\ket{\lambda,\lambda}\right) \nonumber \\    
    &\quad\times\left(\bra{\mu,\mu}\left(\prod_{\alpha \in \Delta_+}(\hat E_{-\alpha}^{(B)})^{m_\alpha}\right)^\dagger\left(\prod_{\alpha \in \Delta_+}(\hat E_{-\alpha}^{(B)})^{m_\alpha}\right)\ket{\mu,\mu}\right)
\end{align}
where we used $\bra{\Lambda\Lambda} = \bra{\lambda\lambda}\otimes\bra{\mu\mu}$ to split the inside of the sum into a product of two expectation values.

Our first order of business is to compute the normalization, $ N_{\Lambda L} $. We start from
\begin{align}
\label{eq:vev}
    N_{\Lambda L} &= \bra{\Lambda,\Lambda}\left(\prod_{\alpha \in \Delta_+}(\hat E_{-\alpha})^{L_\alpha}\right)^\dagger\left(\prod_{\alpha \in \Delta_+}(\hat E_{-\alpha})^{L_\alpha}\right)\ket{\Lambda,\Lambda}.
\end{align}
Our strategy once again will be to annihilate the positive roots on the left against the highest weight state on the right $\ket{\Lambda,\Lambda}$ and collect the factors generated by commuting the positive roots through the negative roots. There are two cases. First, a positive root, $\hat E_{+\alpha}$ commutes through its Hermitian conjugate $\hat E_{-\alpha}$ producing a coroot $H_\alpha$ which when measured against $\ket{\Lambda\Lambda}$ produces a factor of the highest weight vector along axis $\alpha$, $\Lambda_\alpha = \alpha\cdot\Lambda$. Second, a positive root, $\hat E_{+\alpha}$ commutes through a distinct negative root $\hat E_{-\alpha'}$ producing another root $\hat E_{+\alpha-\alpha'}$. Since this process reduces the number of root operators by one, it also reduces the number of coroots that generate factors of $\Lambda$ by one-half, suppressing the contributions of commutator terms. We will find in appendix~\ref{subapp:entStretch} that the suppression is qualitatively the same as that of the non-orthogonality found in~\ref{subapp:rootsLadders}.

Thus, at leading order, we only need to consider exact pairings. Schematically, from the commutation relations,
\begin{align}
\label{eq:appDer}
    \hat E_{+\alpha}\left(\hat E_{-\alpha}\right)^{L_\alpha}\ket{\Lambda\Lambda} = \left[\hat E_{+\alpha},\left(\hat E_{-\alpha}\right)^{L_\alpha}\right]\ket{\Lambda\Lambda} = \left(n\Lambda_\alpha\left(\hat E_{-\alpha}\right)^{(L_\alpha-1)} + \mathcal{O}(1)\right)\ket{\Lambda\Lambda}
\end{align}
where $\Lambda_\alpha$ is the value of $\Lambda$ in the $\alpha$ direction. Then commuting all $n$ factors of $\hat E_{+\alpha}$ yields,
\begin{align}
\label{eq:appPol}
    \left(\hat E_{+\alpha}\right)^{L_\alpha}\left(\hat E_{-\alpha}\right)^{L_\alpha}\ket{\Lambda\Lambda} = \left({L_\alpha}!\Lambda_\alpha^{L_\alpha} + \mathcal{O}(\Lambda_\alpha^{L_\alpha-1})\right)\ket{\Lambda\Lambda}.
\end{align}
When one reintroduces roots of several types in the ordered PBW basis, factors of lower order in $\Lambda$ will be generated by the cross-terms, but the leading order factor will only count the number of exact conjugate pairings. Thus,
\begin{align}
\label{eq:imNotWorried}
    N_{\Lambda L} \approx \prod_{\alpha \in \Delta_+} L_\alpha!\left(\Lambda_{\alpha}\right)^{L_\alpha}.
\end{align}

Then we are left to evaluate the terms in the sum in eq.~\eqref{eq:restart}. The result requires little further effort. The lowering operators applied to subsystem A and the operators in a given split have a non-zero overlap if and only if there is an exact pairing of roots between them. Then,
\begin{align}
\label{eq:imNotWorried2}
    \bra{\lambda,\lambda}\left(\prod_{\alpha \in \Delta_+}(\hat E_{-\alpha}^{(A)})^{l_\alpha}\right)^\dagger\left(\prod_{\alpha \in \Delta_+}(\hat E_{-\alpha}^{(A)})^{l_\alpha}\right)\ket{\lambda,\lambda}\approx \prod_{\alpha \in \Delta_+} l_\alpha!\left(\lambda_{\alpha}\right)^{l_\alpha}.
\end{align}
Then inserting eqs.~\eqref{eq:imNotWorried} and~\eqref{eq:imNotWorried2} into~\eqref{eq:restart} all the leading order terms fall out of the sum leaving,
\begin{align}
\label{eq:KrawtchouckBegins}
    C^{\Lambda L}_{\lambda l;\mu m} &\approx 
    \sqrt{\prod_{\alpha \in \Delta_+}
    \frac{l_\alpha!\left(\lambda_{\alpha}\right)^{l_\alpha}m_\alpha!\left(\mu_{\alpha}\right)^{m_\alpha}}{L_\alpha!\left(\Lambda_{\alpha}\right)^{L_\alpha}}}\sum_{\text{splits}}1 \nonumber \\
    &= \prod_{\alpha \in \Delta_+}\sqrt{
    \frac{l_\alpha!\left(\lambda_{\alpha}\right)^{l_\alpha}m_\alpha!\left(\mu_{\alpha}\right)^{m_\alpha}}{L_\alpha!\left(\Lambda_{\alpha}\right)^{L_\alpha}}} \binom{L_{\alpha}}{l_{\alpha}} \\
    & = \prod_{\alpha \in \Delta_+}\sqrt{\binom{L_{\alpha}}{l_{\alpha}}{\left(\frac{\lambda_{\alpha}}{\Lambda_{\alpha}}\right)^{{l_{\alpha}}}\left(\frac{\mu_{\alpha}}{\Lambda_{\alpha}}\right)^{{m_{\alpha}}}}} \nonumber 
\end{align}
which can be immediately identified as a product of the square root of independent binomial distributions for each positive root $\alpha$, with means $\bar l_\alpha = L_\alpha \lambda_{\alpha}/\Lambda_{\alpha}$ and variances $\sigma_{\alpha}^2 = L_\alpha \lambda_{\alpha}\mu_{\alpha}/\Lambda_{\alpha}^2$. In the asymptotic limit of large representations, the central limit theorem allows us to replace this product of binomials with a multivariate Gaussian spanning the $|\Delta_+|$-dimensional phase space,
\begin{align}
\label{eq:finalGaussianForm}
    \left\vert C^{\Lambda L}_{\lambda l;\mu m}\right\vert^2 \approx \frac{e^{-\frac{1}{2}(l-\bar{l})\cdot\Sigma^{-1}\cdot(l-\bar{l})}}{\sqrt{(2\pi)^{|\Delta_+|}\operatorname{det}[\Sigma]}}
\end{align}
where the elements of the diagonal covariance matrix are $\Sigma_{\alpha \beta} = \delta_{\alpha\beta} \sigma_\alpha^2$. 

\subsection{Neighborhood of the stretched irrep}
\label{subapp:neighborhood}

The major hole in the preceding argument is that it assumes that the stretched irrep is the only contribution in the thermodynamic limit just because it is the largest contribution. Here we show irreps in the neighborhood of the stretched irrep, $\Lambda' \lesssim \lambda + \mu = \Lambda$ retain the same overall Gaussian behavior in Clebsch--Gordan coefficients. In fact, we'll see that the behavior is precisely equivalent to the excited states of a harmonic oscillator.

We define the Inonu--Wigner contraction, a generalization of the Holstein--Primakoff transformation for large spins, of the Lie algebra generators in terms of $|\Delta_+|$ independent bosonic modes $[\hat a_\alpha, \hat a_{\beta}^\dagger] = \delta_{\alpha\beta}$. The idea is that at large representation the root operators $\hat E_{+\alpha}$ scale $\sqrt{\Lambda}$. However, since the commutator of distinct roots is at most another root per eq.~\eqref{eq:S2}, the ladder algebra approximately factorizes into indpendent bosonic modes for each root.

Expanding in inverse powers of the highest weight and retaining the leading order terms, the generators associated with the positive roots for the subsystems with highest weights $\lambda,\,\mu$ are given in terms of annihilation operators,
\begin{align}
\label{eq:apprHP}
    \hat E_{+\alpha}^{(A)} &= \sqrt{\lambda_{\alpha}}\left( \hat a_\alpha + \mathcal{O}(\lambda^{-1/2})\right), \nonumber \\
    \hat E_{+\alpha}^{(B)} &= \sqrt{\mu_{\alpha}}\left( \hat b_\alpha + \mathcal{O}(\mu^{-1/2})\right),
\end{align}
along with creation operators $\hat a_\alpha^\dagger,\,\hat b^\dagger_\alpha$ corresponding to negative roots. Then one defines a symmetric annihilation operator and an antisymmetric creation operator,
\begin{align}
\label{eq:HolsteinPrimakoffTrick}
    \hat u_\alpha &= \frac{1}{\sqrt{\Lambda_{\alpha}}}\left(\sqrt{\lambda_{\alpha}}\, \hat a_\alpha + \sqrt{\mu_{\alpha}}\, \hat b_\alpha\right), \nonumber \\
    \hat v_\alpha^\dagger &= \frac{1}{\sqrt{\Lambda_{\alpha}}}\left(\sqrt{\mu_{\alpha}}\, \hat a_\alpha^\dagger - \sqrt{\lambda_{\alpha}}\, \hat b_\alpha^\dagger\right). 
\end{align}
It follows from its definition that $\hat u_\alpha \propto \hat E_{+\alpha}$ to leading order. By construction $\hat v_\alpha^\dagger$ is also a lowering operator on each subsystem, and thus a lowering operator on the whole system. But because it commutes with the positive root it is not precisely a negative root. Furthermore,
\begin{align}
\label{eq:fockRealize}
    \hat u_\alpha \hat v_\alpha^\dagger \ket{\Lambda,\Lambda} = \hat v_\alpha^\dagger \hat u_\alpha \ket{\Lambda,\Lambda} = 0.
\end{align}
Since positive roots annihilate the highest weight state, $\hat v_\alpha^\dagger \ket{\Lambda,\Lambda}$ is a highest weight state but with weight $\Lambda$ reduced by one root to $\Lambda-\alpha$. More generally,
\begin{align}
\label{eq:believe}
    \ket{\Lambda-n_\alpha\alpha,\Lambda-n_\alpha\alpha} \propto \prod_{\alpha \in \Delta_+} \frac{1}{\sqrt{n_\alpha!}} (\hat v_\alpha^\dagger)^{n_\alpha} \ket{\Lambda,\Lambda}.
\end{align}
To reiterate, we have shown that $\hat v_\alpha^\dagger$ reduces the representation as well as the weight of the state it acts on.

To find the Clebsch--Gordan coefficient $C^{\Lambda' L}_{\lambda l;\mu m}$, we can repeat the analysis of subappendix \ref{subapp:calculation} but replacing some of the $\hat E_{+\alpha}$ in the normalization overlap with $\hat v_\alpha$. Recall that the normalization factor $N_{\Lambda L}$ was computed by commuting raising operators through lowering operators. For the excited irreps, we instead compute matrix elements of the form,
\begin{align}
\label{eq:matrixElem}
     \bra{\Lambda,\Lambda}  \cdots(\hat v_\alpha^\dagger)^{n_\alpha} (\hat E_{-\alpha}^{(A)})^{l_\alpha} (\hat E_{-\alpha}^{(B)})^{m_\alpha}\ket{\lambda\lambda}\ket{\mu\mu}.
\end{align}
Using the definition of $\hat v_\alpha$ and the binomial theorem, the $n_\alpha$-th power of the excitation operator expands as,
\begin{align}
\label{eq:krawtchoukExpand}
    (\hat v_\alpha)^{n_\alpha} = \Lambda_{\alpha}^{-n_\alpha/2} \sum_{j=0}^{n_\alpha} \binom{n_\alpha}{j} (-1)^j \left(\sqrt{\mu_{\alpha}} \hat a_\alpha\right)^{n_\alpha-j} \left(\sqrt{\lambda_{\alpha}} \hat b_\alpha\right)^j.
\end{align}
When this operator acts in the overlap, the term with index $j$ contributes only if it matches the specific partitioning of lowering operators between subsystems A and B determined by the target occupations $l_\alpha$ and $m_\alpha$. The resulting sum over binomial weights modulated by the phase $(-1)^j$ is precisely the definition of the Krawtchouk polynomials, $K_{n_\alpha}(x; p, N)$, which are the orthogonal polynomials of the binomial distribution.

In our asymptotic limit of large representations, we can summon the standard result that the discrete Krawtchouk polynomials converge to the continuous Hermite polynomials, $H_{n_\alpha}(x)$. This is a corollary of the central limit theorem for the binomial distribution. Consequently, for an irrep displaced from the stretched limit by $\Lambda - \Lambda' = \sum_\alpha n_\alpha \alpha$, the squared Clebsch--Gordan coefficient is given by modulating the universal Gaussian envelope derived in eq.~\eqref{eq:finalGaussianForm} with the corresponding Hermite polynomials:
\begin{align}
\label{eq:hermiteFinal}
    \left\vert C^{\Lambda' L}_{\lambda l;\mu m}\right\vert^2 \approx \left[ \prod_{\alpha \in \Delta_+} \frac{1}{2^{n_\alpha} n_\alpha!} H_{n_\alpha}\left( \frac{l_{\alpha} - \bar{l}_{\alpha}}{\sigma_\alpha} \right)^2 \right] \frac{e^{-\frac{1}{2}(l-\bar{l})\cdot\Sigma^{-1}\cdot(l-\bar{l})}}{\sqrt{(2\pi)^{|\Delta_+|}\operatorname{det}[\Sigma]}}
\end{align}
where $\sigma_\alpha$ is the Gaussian width for the $\alpha$-th root direction derived in the previous section. This result confirms an expectation that large representations of compact Lie groups behave like non-interacting oscillators along each root.
\subsection{The entropy of a stretched Clebsch--Gordan coefficient}
\label{subapp:entStretch}

Since we have described the Clebsch--Gordan coefficient in the stretched irrep as a product of $(d-r)/2$ independent Gaussians with variances $\sigma_{\alpha}^2 = L_\alpha \lambda_{\alpha}\mu_{\alpha}/\Lambda_{\alpha}^2 \sim \mathcal{O}(N)$ one can guess that the entropy should be $\frac{d-r}{2}\ln\left(\sqrt{N}\right) = \frac{d-r}{4}\ln N$. After setting up the problem properly and resolving possible off-diagonal correlations we will confirm this guess.

We are calculating the entropy of $\ket{\Lambda L}$ is a product of representations $\lambda$ and $\mu$. We can obtain a density matrix on $\mathcal H^{(\lambda)}_{\text{sym}}$ by tracing out the $\mu$ representation,
\begin{align}
    \hat\rho = \sum_{ll'm} \ket{l}\!\bra{\lambda l;\mu m}\ket{\Lambda L}\!\bra{\Lambda L}\ket{\lambda l';\mu m}\!\bra{l'}.
\end{align}
In appendix~\ref{subapp:calculation} we found that $C^{\Lambda L}_{\lambda l;\mu m} = \bra{\Lambda L}\ket{\lambda l;\mu m}$ is only large if $l$ and $m$ split the path that leads to $L$ in the PBW basis. This leads to a diagonal approximation of $\rho$,
\begin{align}
\label{eq:diagEnt}
    \hat\rho_{\text{diag}} = \sum_{l} \vert C^{\Lambda L}_{\lambda l;\mu l^c}\vert^2 \ket{l}\!\bra{l} \implies S_{\text{sym}} = -\sum_{l} \vert C^{\Lambda L}_{\lambda l;\mu l^c}\vert^2 \ln\vert C^{\Lambda L}_{\lambda l;\mu l^c}\vert^2 \approx \frac{d-r}{4}\ln N
\end{align}
where we have used $l^c$ to denote the complement of $l$ with respect to $L$.

To make sure the diagonal approximation is sufficient we have to show that the non-zero off-diagonal elements of $\hat\rho$ are neither large enough nor numerous enough to change its overall spectrum. First, if $l$ and $l'$ do not have the same weight $\hat\rho_{ll'}$ is automatically zero. Next, within the weight multiplicity space $\hat\rho_{ll'}/\hat\rho_{ll} \sim N^{-k(l,l')}$ where $k(l,l')$ is again the minimum number of commutators needed to deform the path to $l$ into a path to $l'$. Since generically $k(l,l')\sim\text{poly}(N)$, we conclude that the off-diagonal elements are too small to alter the overall spectrum of $\hat\rho$.

Lastly, we should define the domain of validity of our result. First, the above derivation relies on the assumption that $d,r \sim \mathcal{O}(1)$ but it would be interesting to test a double scaling limit of a large rank group $d,r \sim \text{poly}(N)$. Second, the group contraction eq.~\eqref{eq:apprHP} implies that the approximate commutation of roots holds so long as $\Lambda - \lambda -\mu \ll \sqrt N $. However, thermal fluctuations of $\lambda$ and $\mu$ are expected to be of order $\sqrt{N}$ by eq.~\eqref{eq:toRef} thus the inequality cannot strictly hold in any regime. Nevertheless, while $\Lambda \sim \mathcal{O}(N)$, the commutation works as mean-field model that correctly extracts the logarithmic scaling of entropy and higher-order terms should only modify the $\mathcal O(1)$ prefactor.

\section{Entanglement correction at $\Lambda\approx 0$}
\label{app:zeroDens}

We now must comment on the zero density sector. In fact, this is insufficiently precise. We assume $\Lambda \lesssim \mathcal O(1)$. Physically, in this limit thermal fluctuations, per eq.~\eqref{eq:toRef}, of the subsystems place a floor on the typical value $\lambda$ and $\mu$ of $\sim \mathcal O \left(\sqrt N\right)$. Thus we must consider $\lambda \approx \mu \gg \Lambda$\footnote{$\lambda \not\approx \mu$ cannot contain $\Lambda \ll \mu,\,\lambda$}. We will find that this irrep is essentially the infinite temperature sector of the algebra with maximal or near-maximal entanglement. 

\subsection{The identity irrep}
\label{subapp:pureGNS}
Our strategy is simplest starting from the cases of $\lambda = \mu$ and $\Lambda = 0$. Here there is only one global state, $\ket{00}$ and we want its Clebsch--Gordan coefficients,
\begin{align}
\label{eq:MESsetup}
    \ket{00} = \sum_{ll'} C^{00}_{\lambda l;\lambda l'}\ket{\lambda l} \otimes \ket{\lambda l'}.
\end{align}
Invariance of the state under group action fixes $C^{00}_{\lambda l;\lambda l'}$ to be proportional to $\delta_{ll'}$\footnote{This is strictly untrue and one should have instead $C^{00}_{\lambda l;\lambda^* l'} \propto \delta_{ll'}$ where $\lambda^*$ denotes the conjugate representation. For SU(2) for example, the results is $C^{00}_{J M;JM'} \propto \delta_{M,-M'} \neq \delta_{M,M'}$. Here and below, for notational simplicity, we suppress conjugation of the representation.}. To see this take
\begin{align}
\label{eq:MESinv}
    \ket{00} &= \hat U(g)\ket{00} \nonumber \\
    \sum_{ll'} C^{00}_{\lambda l;\lambda l'}\ket{\lambda l} \otimes \ket{\lambda l'}&= \sum_{ll'}\sum_{mm'} \left(U(g)\right)_{lm} C^{00}_{\lambda m;\lambda m'} \left(U(g)\right)_{l'm'} \ket{\lambda l} \otimes \ket{\lambda l'} \\
    C^{00}_{\lambda l;\lambda l'}\ket{\lambda l} &= \left(U(g)\right)_{lm} C^{00}_{\lambda m;\lambda m'} \left(U(g)\right)_{l'm'}
\end{align}
from which we find that $C^{00}_{\lambda l;\lambda l'}$, when interpreted as matrix elements of an operator $\hat C$, is an intertwiner and Schur's lemma ~\eqref{eq:S1} guarantees proportionality to the identity. Consequently,
\begin{align}
\label{eq:MEScgc}
    C^{00}_{\lambda l;\lambda l'} = \frac{1}{\sqrt {D_\lambda}}\delta_{ll'}
\end{align}
and
\begin{align}
\label{eq:MES}
    \ket{00} = \frac{1}{\sqrt {D_\lambda}}\sum_l \ket{\lambda l}\otimes\ket{\lambda l}
\end{align}
where $D_\lambda \sim \mathcal{O}(\lambda^{\vert\Delta_+\vert})$ is the dimension of the $\lambda$ irrep. One can readily show that the reduced density matrix $\hat\rho_{00} = \hat C_{00}\hat C_{00}^\dagger$ has entropy $\ln D_\lambda$.

In eq.~\eqref{eq:MES}, we have found that $\ket{00}$ is essentially the maximally entangled state of the tensor product. Such a state can be used to construct a useful operator-state duality. Consider the action of a generator $\hat J$ of the Lie algebra upon the state. By construction,
\begin{align}
\label{eq:GNSstart}
    \hat J \ket{00} &= 0 \nonumber \\
    \left(\hat J_A + \hat J_B\right) \ket{00} &= 0 \\
    \hat J_A\ket{00} &= - \hat J_B\ket{00}. \nonumber
\end{align}
Then consider the action of the same generator concatenated with an operator $\hat T$ living only on subsystem A,
\begin{align}
\label{eq:GNSpush}
    \hat J \hat T\ket{00} &= \left(\hat J_A \hat T + \hat J_B \hat T\right) \ket{00} \nonumber \\
     &= \left(\hat J_A \hat T + \hat T\hat J_B \right) \ket{00} \\
     &= \left(\hat J_A \hat T - \hat T\hat J_A \right) \ket{00} =\left[\hat J_A, \hat T\right] \ket{00}. \nonumber
\end{align}
Thus, as vector spaces, the tensor product of irreps is isomorphic\footnote{to prove isomorphism, one just needs to observe that the kernel of the homomorphism is trivial} to the operator space of a subsystem. The consequence of this construction is that every state in the tensor product maps one-to-one to an operator in a subsystem component,
\begin{align}
\label{eq:GNScon}
    \ket{T} \equiv \hat T\ket{00}.
\end{align}
This is the Gelfand-Naimark-Segal (GNS) construction applied to a tensor product of Lie groups with $\ket{00}$ as the GNS state. This isomorphism can be realized more trivially by noting that $\hat T$ and $\ket{T}$ both live in $\mathcal{H}^{(\lambda)}_{\text{sym}}\otimes\mathcal{H}^{(\lambda)}_{\text{sym}}$.

The utility of the GNS construction for our purposes, however, is that the transformation rule in eq.~\eqref{eq:GNSpush} implies that a state that carries labels $\Lambda,\nu$ can be constructed by directly applying a corresponding spherical tensor operator $\hat T^{(\Lambda, \nu)}$ to the GNS state (suppressing multiplicities),
\begin{align}
\label{eq:osc}
    \ket{\Lambda\nu} \equiv \hat T^{(\Lambda, \nu)}\ket{00},
\end{align}
that the Clebsch--Gordan coefficients are just the matrix elements of $\hat T^{(\Lambda, \nu)}$,
\begin{align}
\label{eq:Clebsch--Gordan coefficientelement}
    C^{\Lambda \nu}_{\lambda l;\lambda l'} \propto \left(\hat T^{(\Lambda, \nu)} \right)_{ll'}
\end{align}
and consequently that the reduced density matrix on each tensor product factor is simply,
\begin{align}
\label{eq:rdmGNS}
    \hat\rho_{00} = \frac{\hat T^{(\Lambda)\dagger}_{(\nu)}\hat T^{(\Lambda, \nu)}}{\operatorname{Tr}\left[\hat T^{(\Lambda)\dagger}_{(\nu)}\hat T^{(\Lambda, \nu)}\right]}.
\end{align}
Now the problem of studying entanglement in the identity irrep reduces to constraining the spectrum of $T^\dagger T$. We return to this in the next part of the appendix.

\subsection{Quasi-identity irreps}
\label{subapp:impureGNS}
Now we must set up the case of $\lambda \approx \mu$. Here our vacuum is no longer a GNS state but instead the highest weight state of a minimal weight irrep. Take wlog $\lambda > \mu$ and define $\Lambda_0 = \lambda-\mu$. We take as our vacuum $\ket{\Lambda_0\Lambda_0}$. We'd like a canonical way to associate to the Clebsch--Gordan coefficients,
\begin{align}
\label{eq:Clebsch--Gordan coefficientquasi}
    \ket{\Lambda_0\Lambda_0} = \sum_{lm} C^{\Lambda_0\Lambda_0}_{\lambda l;\mu m}\ket{\lambda l} \otimes \ket{\mu m}
\end{align}
an operator,
\begin{align}
\label{eq:vacOp}
    \left(\hat C_0\right)_{lm} = C^{\Lambda_0\Lambda_0}_{\lambda l;\mu m}.
\end{align}
Unlike $\hat T^{(0, 0)}$, $\hat C_0$ is not an intertwiner. It does, however, anticommute with positive roots,
\begin{align}
\label{eq:posInt}
    \hat{E}_{+\alpha}\ket{\Lambda_0\Lambda_0} &= \left(\hat{E}_{+\alpha}^{(\lambda)} + \hat{E}_{+\alpha}^{(\mu)}\right)\ket{\Lambda_0\Lambda_0} = 0 \nonumber \\ 
    &= \sum_{lm} (\hat{E}_{+\alpha}^{(\lambda)}\hat C_0 + \hat C_0\hat{E}_{+\alpha}^{(\mu)})_{lm}\ket{\lambda l} \otimes \ket{\mu m} = 0 \\
    &\implies \hat{E}_{+\alpha}^{(\lambda)}\hat C_0 + \hat C_0\hat{E}_{+\alpha}^{(\mu)} = 0. \nonumber
\end{align}
This property allows us to exploit the PBW basis introduced in appendix~\ref{subapp:rootsLadders}. 

The Clebsch--Gordan coefficient may be written,
\begin{align}
\label{eq:PBWagain}
    \bra{\lambda l}\hat C_0\ket{\mu m} &= \frac{1}{\sqrt{N_{\lambda l}N_{\mu m}}}\bra{\lambda \lambda}\left(\prod_{\alpha}\left(E^{(\lambda)}_{+\alpha}\right)^{l_\alpha}\right)\hat C_0\left(\prod_{\alpha}\left(E^{(\mu)}_{-\alpha}\right)^{m_\alpha}\right)\ket{\mu \mu}.
\end{align}
Eq.~\eqref{eq:posInt} implies that $\hat C_0$ can anti-commute through the positive roots to its left leaving behind positive roots on $\mu$ and at most an overall sign flip. Additionally, since the usual properties of highest weight states enforce $\ket{\lambda\lambda} = \ket{\Lambda_0\Lambda_0}\otimes\ket{\mu\mu}$, by construction $\bra{\lambda\lambda}\hat C_0 =\bra{\mu\mu}c_0$ with $c_0 = \bra{\lambda\lambda}\hat C_0\ket{\mu\mu}$. Thus we obtain a surprisingly simple expression for the matrix elements of $\hat C_0$,
\begin{align}
\label{eq:hereAgain}
    \bra{\lambda l}\hat C_0\ket{\mu m} &= \frac{1}{\sqrt{N_{\Lambda l}N_{\mu m}}}\bra{\lambda \lambda}\left(\prod_{\alpha}\left(E^{(\lambda)}_{+\alpha}\right)^{l_\alpha}\right)\hat C_0\left(\prod_{\alpha}\left(E^{(\mu)}_{-\alpha}\right)^{m_\alpha}\right)\ket{\mu \mu} \nonumber \\
    & = \pm\frac{c_0}{\sqrt{N_{\Lambda l}N_{\mu m}}}\bra{\mu\mu}\left(\prod_{\alpha}\left(E^{(\mu)}_{+\alpha}\right)^{l_\alpha}\right)\left(\prod_{\alpha}\left(E^{(\mu)}_{-\alpha}\right)^{m_\alpha}\right)\ket{\mu \mu} \\
    &= \pm\delta_{lm}c_0\sqrt\frac{N_{\mu m}}{N_{\lambda l}}(1+\mathcal{O}(\lambda^{-1})). \nonumber
\end{align}
The Kronecker $\delta$ in eq.~\eqref{eq:hereAgain} needs some interpretation. It merely declares that the sequence of roots operations are the same on both sides and does not indicate a one-to-one mapping of states. However, since $\lambda$ is the larger irrep, it should be cautioned that some paths that are unique within it will be redundant in $\mu$, i.e. a rectangular $\hat C_0$ cannot have a full set of linearly independent left eigenvectors. Separately, for the same reasons established in appendices~\ref{subapp:rootsLadders},~\ref{subapp:entStretch} the non-orthogonality of the basis and the off-diagonal elements are too small to modify the diagonal properties. Then, tracing out the $\lambda$ irrep,
\begin{align}
\label{eq:thermalBosons}
    \left(\hat\rho_0\right)_{mm'} = \left(\hat C_0^\dagger\hat C_0\right)_{mm'} &\approx c_0^2\delta_{mm'}\frac{N_{\mu m}}{N_{\lambda m}} \\
    &= \delta_{mm'} \prod_{\alpha\in\Delta_+} \left(1-\frac{\mu_\alpha}{\lambda_\alpha}\right)\left(\frac{\mu_\alpha}{\lambda_\alpha}\right)^{m_\alpha}
\end{align}
The density matrix $\hat{\rho}$ is found to be exactly the canonical thermal state for an oscillator along each root axis $\alpha$, with an effective Boltzmann weight $e^{-\beta\omega_\alpha} \equiv x_\alpha = \mu_\alpha/\lambda_\alpha$. For $\mu\approx\lambda$ it is straightforward to verify that the entropy of $\hat\rho_0$ is $\ln D_\mu$. We will consider to $\mu\not\approx\lambda$ in the next section.

We must now extend our analysis for $\ket{\Lambda_0\Lambda_0}$ to other states. Following the operator-state logic established in appendix~\ref{subapp:pureGNS}, we construct states carrying labels $\Lambda,\nu$ by acting on our new vacuum with a spherical tensor operator localized on subsystem A (the $\lambda$ irrep),
\begin{align}
\label{eq:genState}
    \ket{\Lambda\nu} = \left(\hat T^{(\Lambda, \nu)} \otimes \mathbb{I}_B\right) \ket{\Lambda_0\Lambda_0}.
\end{align}
which corresponds directly to matrix left-multiplication, yielding
\begin{align}
\label{eq:genCoeff}
    \ket{\Lambda\nu} = \sum_{lm} \left(\hat T^{(\Lambda, \nu)} \hat C_0\right)_{lm} \ket{\lambda l} \otimes \ket{\mu m}.
\end{align}
Tracing out the $\lambda$ irrep to obtain the reduced density matrix for subsystem B,
\begin{align}
\label{eq:genRDM}
    \hat\rho = \frac{\hat C_0^\dagger \hat T^{(\Lambda, \nu)\dagger} \hat T^{(\Lambda, \nu)} \hat C_0}{\operatorname{Tr}\left[\hat C_0^\dagger \hat T^{(\Lambda, \nu)\dagger} \hat T^{(\Lambda, \nu)} \hat C_0\right]}.
\end{align}

Our argument proceeds identically to previous arguments in appendix~\ref{app:finiteDens} so we elaborate in words only. By construction, $\hat T^{(\Lambda, \nu)\dagger}$, $\hat T^{(\Lambda, \nu)}$ are polynomials of root operators, $\hat E_{\pm\alpha}$, of degree $K \sim \text{poly}(\Lambda)$. As discovered in appendix~\ref{subapp:calculation} overlap of a sequence of root operators with its adjoint is to leading order a simple polynomial or zero. These two cases correspond to the diagonal and off-diagonal elements of $\hat\rho$ in the PBW basis. Then $\hat T^{(\Lambda, \nu)}$ at most modifies the spectrum of $\hat C_0^\dagger \hat C_0$ by a smooth polynomial envelope of low degree. Thus, to within a logarithm we expect no modification of the entropy as long as $\Lambda$ is very small.

All cases in appendices~\ref{subapp:pureGNS} and~\ref{subapp:impureGNS} we have predicted maximal entanglement entropy, precisely,
\begin{align}
\label{eq:highEnt}
    S_{\text{sym}} \approx \ln\left[\min\left(D_\lambda,D_\mu\right)\right] \approx \frac{d-r}{4}\ln N.
\end{align}

\section{Entanglement correction for $\Lambda\sim \mathcal O(N^p)$, $p\in(0,1)$}
\label{app:margDens}

In the last two appendices we demonstrated that the PBW basis has an incredibly strong tendency to orthogonalize in the large representation limit. In this appendix we interpret this result physically in terms of semiclassicalization and exploit it to compute the intrinsic entanglement of a generic fusion of irreps. The key insight is that PBW basis states are, semiclassically, angular momentum eigenstates. The framework we require is \textit{geometric quantization} which understands representations of Lie groups as the quantization of a classical phase space. This appendix relies on techniques of symplectic geometry that sharply diverge from all other sections of this paper. However, as a reward for working through the general case, we will grant the reader the opportunity to do the calculation themselves for SU(2) which requires no knowledge of symplectic geometry, but realizes all of the abstract steps clearly. We will also relegate the some mathematical aspects of the calculation to its own subappendix~\ref{subapp:gq}.

\subsection{Semiclassical geometry}
\label{subapp:sg}

First, let us properly exploit the quasi-orthogonality of the PBW basis. Ordinarily, the largest Abelian subalgebra of a Lie algebra is its Cartan subalgebra, which is generated by the coroots of simple roots. However, the physical content of the group contraction~\eqref{eq:apprHP} and of the quasi-orthogonality of the PBW basis is that the coroots of the positive, but not simple, roots become very strongly approximately commutative as well in the large irrep limit. Thus instead of working with a $r$-dimensional basis of weights, in practice we get a full $\frac{d-r}{2}$ dimensional basis of weights. Whereas for small irreps one can only demand additivity of the Cartan weights for large irreps one can also demand additivity of all coroots, $\hat H^{(\Lambda)}_\alpha \approx \hat H^{(\lambda)}_\alpha+\hat H^{(\mu)}_\alpha,\,\alpha \in\Delta_{+}$. The consequence of this fact for Clebsch--Gordan coefficients of 3 large representations is that their values in the PBW basis are essentially Schmidt coefficients,
\begin{align}
\label{eq:Schmidt}
    \left(\hat \rho_{\text{sym}A}\right)_{ll'} &= \sum_{m} C^{\Lambda L}_{\lambda l;\mu m}\left(C^{\Lambda L}_{\lambda l';\mu m}\right)^* \approx \sum_{m} C^{\Lambda L}_{\lambda l;\mu m}\left(C^{\Lambda L}_{\lambda l';\mu m}\right)^*\delta_{l,(L-m)}\delta_{l',(L-m)} \nonumber \\
    &\approx\left\vert C^{\Lambda L}_{\lambda l;\mu (L-l)}\right\vert^2\delta_{ll'}.
\end{align}
The entropy of this object is pleasingly simple to define,
\begin{align}
\label{eq:easyEnt}
    S_{\text{sym}} = -\sum_l \left\vert C^{\Lambda L}_{\lambda l;\mu (L-l)}\right\vert^2\ln\left\vert C^{\Lambda L}_{\lambda l;\mu (L-l)}\right\vert^2.
\end{align}
Next we will see that semiclassics will allow us to precisely estimate $\left\vert C^{\Lambda L}_{\lambda l;\mu (L-l)}\right\vert^2$.

A Lie algebra with $d$ generators is a $d$-dimensional Euclidean manifold, $\mathbb R^d$. Then one considers the \textit{adjoint orbit}\footnote{Since we are dealing with finite dimensional unitary representations, the nuances of coadjoint representations are unnecessary for our purposes and we can work purely within the adjoint representation.} of a Lie algebra element, $\hat H_0$, which is the submanifold of $\mathbb R^d$ obtained by acting on $\hat H_0$ by every element of the group,
\begin{align}
\label{eq:adjointOrbit}
    \mathcal O(\hat H) = \{\hat U(g)\hat H_0 \hat U(g^{-1}),\,g\in G\}.
\end{align}
$\mathcal O(\hat H_0)$ is naturally $d-r$ dimensional, since $\hat H_0$ carries $r$ Casimir invariants $\Lambda_0$ while acted on by a $d$-dimensional group. We can uniquely identify orbits by their Casimir invariants, $\mathcal O(\hat H_0) = \mathcal O_{\Lambda_0}$. The orbit $\mathcal O_{\Lambda_0}$ also naturally carries a symplectic structure~\cite{alekseev_quantization_1988}, the details of which we will summarize at the end of this appendix. What we learn from the mere existence of a symplectic structure is that the dimension of a Hilbert space on this adjoint orbit is given by $D_{\Lambda_0} = \hbar^{-\frac{d-r}{2}}\text{Vol}[\mathcal O_{\Lambda_0}]$ since volume is measured in units of $\hbar$ for each canonical conjugate pair\footnote{The precise relationship between irreps and the coadjoint orbit is given by the Borel--Weil theorem\cite{fulton_representation_2004,lurie_proof_nodate}. Formally, a wavefunction is a section of a particular complex line bundle on the orbit. The line bundle must have an integral Chern number on any 2-cycle. The line bundles that satisfy this condition correspond precisely to the irreps of the group and their holomorphic sections are elements of the irrep.}. While we will naturally take $\hbar = 1$, it is easy to see in this perspective how a large charge limit is physically a semiclassical limit.

Exploiting the semiclassical limit of the irrep, it becomes possible to construct coherent states that are tightly concentrated on individual points of the orbit. We compute the Clebsch--Gordan coefficient in this basis,
\begin{align}
\label{eq:coherentState}
    \left|\bra{\Lambda L}\ket{\lambda l; \mu m}\right|^2 \approx \int_{v_1'v_2'v_3'}\int_{v_1v_2v_3}&\bra{\Lambda L}\ket{\Lambda v_3}\bra{\lambda v_1}\ket{\lambda l}\bra{\mu v_2}\ket{\mu m} \bigg(\bra{\Lambda L}\ket{\Lambda v_3'}\bra{\lambda v_1'}\ket{\lambda l}\bra{\mu v_2'}\ket{\mu m}\bigg)^* \nonumber \\
    &\times\bra{v_3}\ket{v_1;v_2}\bra{v_1';v_2'}\ket{v_3'}
\end{align}
First, $\bra{v_3}\ket{v_1;v_2}$ enforces a sharp kinematic constraint $v_1+v_2=v_3$. Second, we can exploit a WKB type analysis for each PBW state, 
\begin{align}
\label{eq:AharonovBohm}
    \bra{\Lambda L}\ket{\Lambda v_3}\sim \sqrt{L(v_3)}e^{iS(v_3)}.
\end{align}
The normalization ${L(v_3)}$ is tightly concentrated on the submanifold of points $v_3$ that are consistent with the coroot charges of $L$ and roughly uniform on that manifold, $L(v_3)\approx \frac{\delta(v_3\in\mathcal O_{\Lambda\rvert L})}{\text{Vol}[\mathcal O_{\Lambda\rvert L}]}$. The phase $S(v_3)$ is given by the symplectic structure and is generally rapidly oscillating~\cite{alekseev_quantization_1988}. We will return to these points in the next subappendix, but for now we can use these facts to finish our computation. 

Unless $v_3=v_3'$ and similarly for $v_1$ and $v_2$ the classical phases will interfere destructively, which yields a saddle at the stationary phase. Then the integrand is constant and merely computes the volume of coherent states that are consistent with the PBW states,
\begin{align}
\label{eq:coherentStateEval}
    \overline{\left\vert C^{\Lambda L}_{\lambda l;\mu (L-l)}\right\vert^2} &\approx \int_{v_1'v_2'v_3'}\int_{v_1v_2v_3}\bra{\Lambda L}\ket{\Lambda v_3}\bra{\lambda v_1}\ket{\lambda l}\bra{\mu v_2}\ket{\mu m} \bigg(\bra{\Lambda L}\ket{\Lambda v_3'}\bra{\lambda v_1'}\ket{\lambda l}\bra{\mu v_2'}\ket{\mu m}\bigg)^* \nonumber \\
    &\quad\quad\quad\quad\quad\quad\quad\times\bra{v_3}\ket{v_1;v_2}\bra{v_1';v_2'}\ket{v_3'} \nonumber \\[1em]
    &\approx \int_{v_1v_2v_3}L(v_3)l(v_1)m(v_2) \vert\bra{v_3}\ket{v_1;v_2}\vert^2  \\[1em]
    &\approx \int_{v_1v_2v_3}\frac{\text{Vol}\left[\mathcal O_{\Lambda\rvert L}\right]\delta^{(d-r)}(v_1+v_2-v_3)}{\text{Vol}\left[\mathcal O_{\Lambda\rvert L}\times\mathcal O_{\lambda\rvert l}\times\mathcal O_{\mu\rvert m}\right]} = \frac{\text{Vol}\left[\mathcal O_{\Lambda\rvert L}\cap\left(\mathcal O_{\lambda\rvert l}\times\mathcal O_{\mu\rvert m}\right)\right]}{\text{Vol}\left[\mathcal O_{\lambda\rvert l}\times\mathcal O_{\mu\rvert m}\right]}.  \nonumber
\end{align}
The ratio of volumes in~\ref{eq:coherentStateEval} has a semiclassical interpretation: what is the probability that an element of $\mathcal O_{\lambda\rvert l}$ added to an element of $\mathcal O_{\mu\rvert m}$, each drawn uniformly, produces a particular element of $\mathcal O_{\Lambda\rvert L}$? The three $v$'s are restricted to respective, concentric $\frac{d-r}{2}$ dimensional tori so the expression is actually a product of $\frac{d-r}{2}$ integrals over circles. The radius of each circle is merely the coroot charge $\Lambda_\alpha$, etc. Then the evaluation simplifies enormously and the result is
\begin{align}
\label{eq:coherentStateResult}
    \overline{\left\vert C^{\Lambda L}_{\lambda l;\mu (L-l)}\right\vert^2}=\prod_{\alpha}\frac{\Lambda_{\alpha}}{2\pi A_\alpha}
\end{align}
where $A_{\alpha}$ is the area of the orthogonal projection of the triangle with sides $\mu$, $\lambda$, and $\Lambda$ onto the plane of the $\alpha$ circle. We will guide the reader through the calculation concretely for SU(2) shortly. First we comment that the non-trivial area of $A_\alpha$ is supported by thermal fluctuations, and without it the expression would appear singular and one could worry that the contributions of such configurations shift the thermal saddle-points. Thankfully, this does not occur because the oscillating component of the Clebsch--Gordan coefficient suppresses contributions without non-trivial area. This is concretely worked out for SU(2)\footnote{It is also possible to verify directly from the explicit formula for SU(2) coefficients.} in~\cite{reinsch_asymptotics_1999} but the physical principle is general; such configurations are not semiclassically permitted. Now this estimate can be employed to evaluate entropy.

Assuming that each subsystem is a finite fraction of the whole system, thermal fluctuations will ensure that $\lambda + \mu = \Lambda + \epsilon$ with $0\leq \epsilon\sim\mathcal{O}(\sqrt N)$. This implies scaling for each $A_\alpha\sim\mathcal{O}(N^{p+\frac{1}{2}})$. However, $A_\alpha$ divides $\Lambda_\alpha\sim\mathcal{O}(N^{p})$ so the overall expression scales
\begin{align}
\label{eq:finalEntResultResult}
    &\overline{\left\vert C^{\Lambda L}_{\lambda l;\mu (L-l)}\right\vert^2} \sim \mathcal{O}(({1}/{\sqrt N})^{-\frac{d-r}{2}})=\mathcal{O}(N^{-\frac{d-r}{4}})\\
    &\implies S_{\text{sym}} = -\expval{\ln\left\vert C^{\Lambda L}_{\lambda l;\mu (L-l)}\right\vert^2} \approx +\frac{d-r}{4}\ln N
\end{align}
yielding our main result of the section. Here we have seen the remarkable cancellation that results in a universal contribution to entanglement entropy from the Clebsch--Gordan coefficients. Physically, the stretched case of appendix~\ref{app:finiteDens} is analogous to a low entanglement ground state, the quasi-identity case of appendix~\ref{app:zeroDens} to a highly entangled infinite temperature state, and the present case to the intermediate finite temperature regime. However, the reduction in entanglement from polarizing a representation from the infinite temperature state towards the ground state is exactly compensated with the growth in the size of the effective representation. 

\begin{figure}[htbp]
    \centering
    \includegraphics[trim={1.5cm 3.5cm 1.5cm 0.9cm}, clip, width=0.5\linewidth]{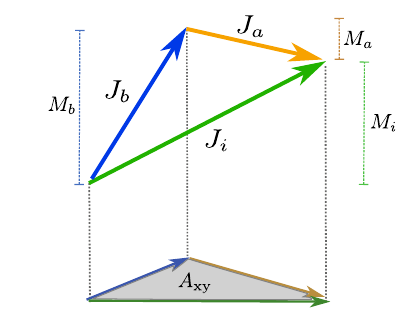}
    \caption{The asymptotic squared SU(2) Clebsch--Gordan coefficient is given by $\frac{J_i}{2\pi A_{\text{xy}}}$~\cite{reinsch_asymptotics_1999}. $A_{\text{xy}}$ is the area formed by the triangle with sides given by the 3 vectors projected onto the $x$-$y$ plane.}
    \label{fig:coadjoint}
\end{figure}

Now as promised, we describe the SU(2) calculation. First, a spin coherent state is created by rotating the highest weight state and obtains a particular form in terms of  $\theta,\phi$~\cite{fradkin_field_2013},
\begin{align}
\label{eq:hw1}
    \ket{Jv} &\equiv \ket{\theta, \phi} = e^{i\theta\left(v_x\hat J^y-v_y\hat J^x\right)}\ket{JJ} = \sum_M \sqrt{{2J}\choose{J+M}}\cos(\theta/2)^{J+M}\sin(\theta/2)^{J-M}e^{-iM\phi}\ket{JM}\nonumber \\
    &\implies \bra{\theta,\phi}\ket{JM} = \sqrt{{2J}\choose{J+M}}\cos(\theta/2)^{J+M}\sin(\theta/2)^{J-M}e^{+iM\phi}
\end{align}
First, one should show that eq.~\eqref{eq:hw1} agrees with eq.~\eqref{eq:AharonovBohm} in the semiclassical limit. Specifically, show that the wavefunction concentrates on one $\theta$ and is rapidly oscillating as a function of $\phi$. Then, since
Eq.~\eqref{eq:coherentStateEval} implicitly reduces the Clebsch--Gordan coefficient to a conditional probability, our question is: what is the probability that vectors of lengths $J_a$, $J_b$ and heights $M_a$, $M_b$ add to a vector of length and height $J_i$, $M_i$ if uniformly distributed? The answer, but not the solution, and a visual aid are given in fig.~\ref{fig:coadjoint}.

\subsection{Geometric quantization}
\label{subapp:gq}

Two key points in appendix~\ref{subapp:sg} were brushed over: the precise symplectic structure of the adjoint orbit and why it provides a precise form for the PBW wavefunction. For clarity and replicability we bring the necessary background out of our main technical reference for this appendix~\cite{alekseev_quantization_1988}. The definition of the symplectic form is simple. For a point $\hat M$ (remember that our coordinates are actually elements of the Lie algebra), a tangent vector is generated by commuting $\hat H$ with another Lie algebra generator $\hat X_M \equiv [\hat X, \hat M]$. The symplectic form is then,
\begin{align}
\label{eq:symplectic}
    \Omega_M(\hat X_M, \hat Y_M) = \langle \hat M, [\hat X, \hat Y]\rangle
\end{align}
where the bracket $\langle,\rangle$ is the vector inner product of the Lie algebra. Alekseev et al. then prove that $\Omega$ is an exact form $\Omega = -d\omega$ where $\omega$ is given by,
\begin{align}
\label{eq:gauge}
    \omega = \sum_{k}\mathcal L_kd\phi^k
\end{align}
where $\mathcal L,\,\phi$ are canonical momenta and positions and $\omega$ serves as a gauge potential for our classical action. 

Our form of the PBW wavefunction~\eqref{eq:AharonovBohm} implies that PBW states are in fact, momentum eigenstates, and thus plane waves. This is only true in the semiclassical limit. The true momentum eigenstates are given by the Gelfand-Tsetlin basis~\cite{alekseev_quantization_1988} which\footnote{for the unitary groups. The general case is also worked out in~\cite{molev_gelfand-tsetlin_2002} follows analogously} can be expressed in terms of Mickelsson lowering operators~\cite{molev_gelfand-tsetlin_2002},
\begin{align}
\label{eq:molev}
    \ket{\Lambda \mathcal L} \propto \prod_{\alpha}\hat Z_{-\alpha}^{\mathcal L_\alpha} \ket{\Lambda\Lambda}
\end{align}
where the product in~\eqref{eq:molev} is ordered analogously to the product in~\eqref{eq:pbwBasis}. The Mickelsson operators share an explicit but nuanced relationship with the usual negative roots. Explicitly,
\begin{align}
\label{eq:rootTrans}
    \hat Z_{-\alpha} = \sum_{\text{paths} \, p} \left(\prod_{\alpha'\in p}\hat E_{-\alpha'}\right)\prod_{\alpha'' \not\in D_\alpha/p}\left(\hat H_{\alpha''}+\left(\varrho\cdot{\alpha''}\right)\hat 1)\right).
\end{align}
There are two parts of eq.~\eqref{eq:rootTrans} that are opaque without elaboration. The paths $p$ are products of other roots that produce a change in weight equal to $-\alpha$. For example in fig.~\ref{fig:Dynkin} an application of the two simple roots is equal to the application of the one non-simple root. The second part is the term $D_\alpha$ which is a set of dimensions of submanifolds that arises in a dimensional reduction procedure we do not discuss technically. For each dimension in the dimensional reduction procedure that is skipped in the product of negative roots, $D_\alpha/p$, one gets a factor of an associated coroot $\hat H_{\alpha''}$. Physically, each $\hat Z_{-\alpha}$ isolates one momentum coordinate by finding a U(1) submanifold of the adjoint orbit that is generated by that momentum. Successive applications of negative roots steps through lower dimensional submanifolds until the target submanifold is reached. 

We do not need the mechanics of dimensional reduction, however, to recognize that there is a unique leading order term in~\eqref{eq:rootTrans}. Each factor of $\hat H_{\alpha''}$ carries weight $\sim\Lambda$. Therefore the largest contribution is the one that skips every intermediate submanifold to go right to the target weight. That is precisely the term associated with $\hat E_{-\alpha}$. Essentially,
\begin{align}
\label{eq:justified}
    \hat Z_{-\alpha} \approx \mathcal O(\Lambda^{|D_\alpha|})\left(\hat E_{-\alpha} + \mathcal O(\Lambda^{-1})\right)
\end{align}
which justifies the treatment of PBW states as momentum eigenstates and our approximate wavefunction on the adjoint orbit. 

\section{General compact Lie groups}
\label{app:generalCompact}

In the main text and preceding appendices, we established the microcanonical entropy using the Weyl integration formula, which assumes a \textit{connected} Lie group and the entanglement corrections using the Clebsch--Gordan coefficients of the PBW basis, which assumes a \textit{semisimple} Lie \textit{algebra}. To apply our results to general (conventional) symmetries in finite dimensional systems, we must consider arbitrary compact Lie groups. 

The structure of an arbitrary compact Lie group $G$ can be understood in two steps. First, the component of the group that contains the identity element, $G_0$, is a connected compact Lie group. The full group $G$ is formed by adjoining a finite number of disconnected components, such that $G/G_0 \cong D$, where $D$ is a finite discrete group. Second, the connected component $G_0$ itself is isomorphic to a quotient,
\begin{align}
\label{eq:lieStructure}
    G_0 \cong \frac{U(1)^k \times K}{Z}
\end{align}
where $K$ is a (simply) connected, compact, semisimple Lie group, $U(1)^k$ represents $k$ Abelian phase factors (a $k$-dimensional torus), and $Z$ is a finite central subgroup. The preceding sections and appendices of the paper should be understood to have studied systems symmetric under $K$ only. In this section we unravel the consequences of the finite groups $Z$ and $D$, and the torus $U(1)^k$. At the logarithmic precision of our analysis, $G_0$ symmetry and $K$ symmetry only differ by the presence of Abelian charges provided by the torus $U(1)^k$. $G$ symmetry and $G_0$ symmetry differ if $D$ is non-central in a subtle way whose consequences lie beyond the scope of this work. In the future, it would be interesting to revisit the difference between compact, compact connected, and compact, connected, semisimple Lie groups in the context of $\mathcal O(1)$ corrections.

\subsection{The component subgroup}
\label{subapp:components}
The basic subtlety of the component subgroup $D$ is that it does not \textit{necessarily} commute with the generators of the Lie algebra within the universal enveloping algebra. This non-centrality means that choosing a Casimir and Cartan basis can by default \textit{break} a symmetry of the system. In the commutative case the generators of $D$ live within the center of the algebra and merely contribute new quantum numbers next to $\Lambda$ that define Abelian sectors. A simple example is the parity subgroup of O(3) symmetry; eigenstates in the multiplicity space merely carry an additional $\pm$ quantum number to denote parity sector. 

In the non-commuting case, resolving the symmetry entails mixing Casimir labels. A simple example is SU(3) symmetry with charge conjugation, $G = SU(3)\rtimes \mathbb Z_2^{\text{C}}$. SU(3) has a quadratic and a cubic Casimir $\hat C_2$ and $\hat C_3$. Charge conjugation $\hat C$ inverts all generators thus, $\hat C\hat C_3\hat C^{-1} = -\hat C_3$ since $\hat C_3$ is a cubic polynomial of the generators. Thus, to resolve $\hat C$ we must (anti-)symmetrize eigenstates of $\hat\Lambda_3$. Using the standard Dynkin labels of SU(3) and suppressing weight multiplicities,
\begin{align}
\label{eq:chargeRes}
    \ket{E, C_2, |\nu|, \pm} = \frac{1}{\sqrt{2}}\left[\ket{(\Lambda_1,\Lambda_2),\nu,E}\pm\ket{(\Lambda_2,\Lambda_1),-\nu,E}\right].
\end{align}
Thus, in order to define the ETH when the symmetry algebra has a component subgroup which does not live in its center, one has to choose whether the operator(s) and states in questions carry a full set of Casimir weights or weights of the component subgroup before applying an ansatz. For our analysis we have stuck to the resolving by the full set of Casimirs, but it may be interesting to revisit our results within e.g. charge conjugation sectors.

\subsection{Thermodynamic and entanglement corrections}
\label{subapp:genCorrectPart}
Let the semisimple group $K$ have dimension $d_K$ and rank $r_K$. The total group $G$ then has dimension $d = k + d_K$ and rank $r = k + r_K$. Because the $U(1)$ generators commute with the entire algebra, the highest weight vectors simply concatenate, $\Lambda_G = ({q}, \Lambda_K)$, where ${q}$ is a $k$-dimensional vector of $U(1)$ charges. The character of the tensor product representation perfectly factorizes as
\begin{align}
\label{eq:charRel}
    \chi_{\Lambda_G}({\theta}, \psi) = e^{i {q} \cdot {\theta}} \chi_{\Lambda_K}(\psi)
\end{align}
where ${\theta}$ are the $k$ angles of the $U(1)$ factors and $\psi$ are the coordinates on the maximal torus of $K$. When we evaluate the density of multiplets $\tilde\rho(E, \Lambda_G)$ as in eq.~\eqref{eq:rochester2}, the integral over the group manifold splits. The $U(1)$ sector contributes a standard Abelian saddle point integral over $k$ variables. Regardless of the charge density, these $k$ variables experience thermal fluctuations of order $\delta\theta \sim \mathcal{O}(N^{-1/2})$, contributing a logarithmic volume penalty of $-\frac{k}{2}\ln N$. 

The semisimple sector $K$ behaves exactly as derived in eq.~\eqref{eq:finiteDensity} and eq.~\eqref{eq:zeroDensity}. Summing the Abelian and non-Abelian contributions, we find the total microcanonical entropy at finite density to be
\begin{align}
    S(E, \Lambda_G) &= \bar{S} - \frac{k}{2}\ln N - \frac{r_K}{2}\ln N + \mathcal{O}(1) \nonumber \\
    &= \bar{S} - \frac{r}{2}\ln N + \mathcal{O}(1).
\end{align}
At zero density (the identity representation), the non-Abelian sector contributes its additional polynomial root suppression. The Abelian sector, lacking roots, does not. The zero-density entropy is therefore,
\begin{align}
    S(E, 0) &= \bar{S} - \frac{k}{2}\ln N - \frac{d_K}{2}\ln N + \mathcal{O}(1) \nonumber \\
    &= \bar{S} - \frac{d}{2}\ln N + \mathcal{O}(1).
\end{align}
Conveniently, expressing the final correction in terms of the total group dimension $d$ and total rank $r$ absorbs the Abelian parameters entirely, leaving the formulas from appendix~\ref{app:WeylGeneral} structurally unchanged.

To discuss entanglement entropy, we must evaluate the effect of the $U(1)^k$ factors on the Clebsch--Gordan coefficients. When tracing out subsystem A to find the reduced density matrix, the Abelian charges merely impose a superselection rule: $q_A + q_B = q$. The Clebsch--Gordan coefficient for the Abelian sector is simply a Kronecker delta, $\delta_{q_A + q_B, q}$, which contributes no spreading and no Gaussian envelope. Consequently, the logarithmic entropy correction derived in eq.~\eqref{eq:diagEnt} is generated solely by the $|\Delta_+|$ non-interacting harmonic oscillators of the semisimple root space and is unchanged from our preceding results, but restated in terms of $K$. Thus,
\begin{align}
\label{eq:last}
    S_{\text{sym}} &= \frac{d_K-r_K}{4}\ln N .
\end{align}

\end{appendix}


% Provide your bibliography here. You have two options:

%%% FIRST OPTION
% Write your entries here directly, following the example below, including:
% Author(s), Title, Journal Ref. with year in parentheses at the end, followed by the DOI number.

%\begin{thebibliography}{99}
%\bibitem{1931_Bethe_ZP_71} H. A. Bethe, {\it Zur Theorie der Metalle. i. Eigenwerte und Eigenfunktionen der linearen Atomkette}, Zeit. f{\"u}r Phys. {\bf 71}, 205 (1931), \doi{10.1007\%2FBF01341708}.
%\bibitem{arXiv:1108.2700} P. Ginsparg, {\it It was twenty years ago today... }, \url{http://arxiv.org/abs/1108.2700}.
%\end{thebibliography}

% Use your bibtex library, formatted by the SciPost style file.
\bibliography{naeb.bib}

%%%%%%%%%% END TODO: BIBLIOGRAPHY

\end{document}